\documentclass[12pt]{article}
\usepackage{jheppub}

\pdfoutput=1

\usepackage{amsmath,bbm,array,amsfonts,graphicx,wrapfig,lscape,float,mathtools,multirow,longtable}
\usepackage[dvipsnames]{xcolor}
\usepackage{stackrel}
\usepackage[all]{xy}

\newcommand{\be}{\begin{equation}}
\newcommand{\ee}{\end{equation}}
\newcommand{\beq}{\begin{equation}}
\newcommand{\beql}[1]{\begin{equation}\label{#1}}
\newcommand{\eeq}{\end{equation}}
\newcommand{\ba}{\begin{array}}
\newcommand{\ea}{\end{array}}
\newcommand{\bea}{\begin{eqnarray}}
\newcommand{\beal}[1]{\begin{eqnarray}\label{#1}}
\newcommand{\eea}{\end{eqnarray}}
\newcommand{\ben}{\begin{enumerate}}
\newcommand{\een}{\end{enumerate}}
\newcommand{\bean}{\begin{eqnarray*}}
\newcommand{\eean}{\end{eqnarray*}}

\newcommand{\fref}[1]{Figure \ref{#1}}
\newcommand{\btab}[1]{\begin{tabular}{#1}}
\newcommand{\etab}{\end{tabular}}

\def \Black {\pdfsetcolor{0 0 0 1}}

\newcommand{\comment}[1]{}

\newcommand{\qed}{\nobreak \ifvmode \relax \else
      \ifdim\lastskip<1.5em \hskip-\lastskip
      \hskip1.5em plus0em minus0.5em \fi \nobreak
      \vrule height0.75em width0.5em depth0.25em\fi}

\newcommand{\old}[1]{}

\usepackage{array}
\newcolumntype{C}[1]{>{\centering\let\newline\\\arraybackslash\hspace{0pt}}m{#1}}

\usepackage{amsmath,amssymb,amsfonts,mathtools,mathdots}
\usepackage{graphicx}
\usepackage{caption}
\usepackage[bottom]{footmisc}
\usepackage{multirow}
\usepackage{mathrsfs}
\usepackage{tikz}
\usepackage{tkz-graph}
\usepackage{bm}
\usepackage{color}
\usepackage{enumerate}
\usetikzlibrary{decorations.pathreplacing}
\usepackage{diagbox}
\numberwithin{equation}{section}
\numberwithin{figure}{section}
\numberwithin{table}{section}
\usepackage{float}
\usepackage{pgfplots}
\pgfplotsset{compat=1.14}
\usepackage[autostyle=true]{csquotes}
\usepackage[toc,page]{appendix}
\usepackage{longtable}

\title{Quiver Mutations, Seiberg Duality and Machine Learning} 

\author[a]{Jiakang Bao,}
\author[b,c,d]{Sebasti\'an Franco,}
\author[a,e,f]{Yang-Hui He,}
\author[a]{Edward Hirst,}
\author[g]{Gregg Musiker,}
\author[a,h]{Yan Xiao}

\affiliation[a]{
Department of Mathematics, City, University of London, EC1V 0HB, UK}

\affiliation[b]{
Physics Department, The City College of the CUNY \\
160 Convent Avenue, New York, NY 10031, USA}

\affiliation[c]{Physics Program and $^d$Initiative for the Theoretical Sciences \\
The Graduate School and University Center, The City University of New York  \\
365 Fifth Avenue, New York NY 10016, USA}

\affiliation[e]{
Merton College, University of Oxford, OX14JD, UK}

\affiliation[f]{
School of Physics, NanKai University, Tianjin, 300071, P.R. China}

\affiliation[g]{
School of Mathematics, University of Minnesota, Minneapolis, MN 55455, USA}

\affiliation[h]{Department of Physics, Tsinghua University
Beijing 100084, China}

\emailAdd{jiakang.bao@city.ac.uk}
\emailAdd{sfranco@ccny.cuny.edu}
\emailAdd{hey@maths.ox.ac.uk}
\emailAdd{edward.hirst@city.ac.uk}
\emailAdd{musiker@math.umn.edu}
\emailAdd{steven1025xiao@gmail.com}

\preprint{
\begin{flushright}

\end{flushright}
}

\abstract{We initiate the study of applications of machine learning to Seiberg duality, focusing on the case of quiver gauge theories, a problem also of interest in mathematics in the context of cluster algebras. Within the general theme of Seiberg duality, we define and explore a variety of interesting questions, broadly divided into the binary determination of whether a pair of theories picked from a series of duality classes are dual to each other, as well as  the multi-class determination of the duality class to which a given theory belongs.
We study how the performance of machine learning depends on several variables, including number of classes and mutation type (finite or infinite). In addition, we evaluate the relative advantages of Naive Bayes classifiers versus Convolutional Neural Networks.
Finally, we also investigate how the results are affected by the inclusion of additional data, such as ranks of gauge/flavor groups and certain variables motivated by the existence of underlying Diophantine equations. In all questions considered, high accuracy and confidence can be achieved.
}

\begin{document}

\maketitle

\section{Introduction}\label{intro}
\subsection{Preface}
Seiberg duality \cite{Seiberg:1994pq} for supersymmetric quantum field theories is one of the most fundamental concepts in modern physics, generalizing the classical electro-magnetic duality of the Maxwell equations.
In parallel, cluster algebras \cite{MR1887642,MR2004457} have become a widely pursued topic in modern mathematics, interlacing structures from geometry, combinatorics and number theory.
These seemingly unrelated subjects were brought together in \cite{Feng:2000mi,Feng:2001bn,Cachazo:2001sg} in the context of quiver gauge theories realized as world-volume theories of D-brane probing Calabi-Yau singularities.
Interestingly, the common theme - quiver Seiberg duality in physics and mutations of cluster algebras in mathematics - emerged almost simultaneously around 1995, completely unbeknownst to the authors of each. It was not until almost a decade later that a proper dialogue was initiated.

Meanwhile, \cite{Hanany:2005ve,Franco:2005rj,Franco:2005sm,Feng:2005gw,Benvenuti:2006qr} placed the study of quiver gauge theories and toric Calabi-Yau spaces on a firm footing via brane tilings, or dimer models, which are bipartite tilings of the torus.
In the mathematics community, cluster algebras have taken a life of their own \cite{lauren}.
Seiberg duality for quiver gauge theories and cluster mutations for quivers have thus allianced a fruitful matrimony.
Continued and often surprising interactions between the physics and mathematics have persisted, ranging from QFT amplitudes \cite{Bourjaily:2016mnp,Arkani-Hamed:2019plo}, to quantization \cite{MR2567745}, to dualities \cite{Franco:2017lpa}.

Recently, a program of using the latest technology of machine learning and data science to study mathematical structures was launched \cite{He:2017aed,He:2017set,He:2018jtw}.
Indeed, \cite{He:2017aed,He:2020lcy,Krefl:2017yox,Ruehle:2017mzq,Carifio:2017bov} introduced the machine learning paradigm to string theory; and \cite{Betzler:2020rfg,Krippendorf:2020gny} to symmetries and dualities.
Methods in neural networks and classifiers have been applied to study diverse problems in physics and mathematics ranging from triangulations in Calabi-Yau hypersurfaces in toric varieties \cite{Altman:2018zlc,Demirtas:2018akl,He:2015fif}, to flux compactifications in string theory \cite{Cole:2019enn}, to AdS/CFT \cite{Hashimoto:2018ftp}, to distinguishing elliptic fibrations \cite{Anderson:2017aux,He:2019vsj} and classifications of Calabi-Yau threefolds \cite{Grimm:2019bey}, to finding bundle cohomology on varieties \cite{Ruehle:2017mzq,Brodie:2019dfx}, to knot hyperbolic volumes \cite{Jejjala:2019kio}, to distinguishing standard models properties \cite{Mutter:2018sra,Deen:2020dlf,Gal:2020dyc},
to machine learning the Donaldson algorithm for numerical Calabi-Yau metrics \cite{Ashmore:2019wzb}, to the algebraic structures of groups and rings \cite{He:2019nzx}, to dessin d'enfants \cite{He:2020eva}, and to the Birch-Swinnerton-Dyer conjecture in number theory \cite{Alessandretti:2019jbs}, etc.

Given the highly combinatorial nature of quivers and cluster algebras, it is natural to ask whether the machine learning program could be applied to this context.
Specifically, one could wonder where in the hierarchy of difficulty, from the least amenable numerical analysis to the most resilient number theory, would quivers and mutations reside.
This is thus the motivation of our current work.
The paper is organized as follows.
After a rapid parallel introduction to Seiberg duality in quiver gauge theories and cluster mutation, from the physics and mathematics point of view in Section \S\ref{seiberg} and \S\ref{clusters}, we proceed in Sections \S\ref{recogmut} to \S\ref{rankinfo} to study a host of pertinent problems which we will summarize shortly. We conclude in Section \S\ref{outlook} and present some details of the neural networks, and their performances over training in the appendices.

\subsection{Summary of Results}\label{summary}
To provide the readers with an idea of the machine learning performance at a glance, we provide here: a brief description of the problem-styles addressed in this paper; a list of the quivers used to generate the mutation classes examined in the investigations; and a table summarizing the investigations' key results.

\paragraph{Data Format}
The datasets used in these investigations represent each quiver in consideration by its graph-theoretic adjacency matrix (in some investigations with an additional vector structure augmented on).
Each investigation has its own dataset of quivers, generated using the Sage software \cite{sagemath}, such that each full dataset is the union of mutually exclusive sets of quiver matrices, where all quivers in each set belong to the same duality class.

Two styles of classification problem are addressed in this paper, and each processes the input quiver data in a different format. The first is binary classification on pairwise data inputs. Here each data input is a pair of matrices, and each pair can be classified as having its two constituent quivers in the same class, or not in the same class. On these problems the Naive Bayes (NB) classification method, as described in appendix \ref{NB}, performed best and was hence used. The second problem style is multiclassification directly on the matrices. Here each data input is a matrix, and the matrix is classified into one of the duality classes the classifier is trained on. On these problems Convolutional Neural Networks (NN), as described in appendix \ref{python_appendix}, performed best and were hence used.

Within each investigation 5-fold cross validation was used to produce a statistical dataset of measures for the analysis of the classifier's performance. In 5-fold cross validation, 5 independent classifiers are each trained on 80\% of the data, and validated on the remaining 20\%, such that the union of the validation sets gives the full dataset for the investigation. Measures of the classifiers' performance are calculated for each classifier and averaged. In addition, the investigations were also run for varying training/validation \% splits, with results plotted as 'learning curves', shown in appendix \ref{learning_curves}.

\paragraph{Quivers considered}
Here we list the quivers used to generate the duality classes making up the datasets of the investigations considered in this paper. They are listed with an adjacency matrix representation and are labelled in the form: \textbf{Q\textit{i}}. Different combinations of these quivers (with further Dynkin type examples) were used in each investigation, as listed in the following table. 

The first 3 quivers, \textbf{Q1, Q2, Q3}, as well as \textbf{Q12, Q13, Q15}, are finite mutation type under the duality, whilst the remaining listed here are infinite mutation type. Additionally other Dynkin and finite mutation types were used in investigations, labelled in the standard Sage quiver package format \cite{2011arXiv1102.4844M}. These additional quivers considered were either Dynkin type of various sizes, labelled by the letter and rank of the Dynkin diagram they are equivalent to (with direction added to the edges); or affine type which correspond to affine Dynkin diagrams, and are labelled using Kac's notation with Dynkin letter, rank, and an optional twist. In the case of affine A, rank is given by a pair of integers for the number of clockwise/anticlockwise edges respectively. The specific affine quiver used to generate a mutation class used in an investigation is the choice auto-generated by the Sage package for the input label information. Finally, 'T' type are so named for being shaped like a letter 'T', their three integer entries give the number of nodes in each of the branches from the branch point (inclusive). These quivers are described further as they are introduced with each investigation.
\break

\begin{figure}[H]
	\centering
	\begin{tikzpicture}
	[round/.style={circle, draw=black, fill=white, minimum size=5mm},]
	\tikzstyle{arrow} = [,->,>=stealth]
	\node[round] (a) at (-5,0){};
	\node[] (b) at (-3,0){$\overbrace{\cdots}^\text{n-3 nodes}$};
	\node[round] (c) at (-3,-2){};
	\node[round] (d) at (-5,-2){};
	\draw[arrow] (a)--(b);
	\draw[arrow] (c)--(b);
	\draw[arrow] (c)--(d);
	\node[] at (-2.2,-3){\textbf{Q1}: ['A', $n$]};
	\node[] at (-1,-1){$\begin{pmatrix}
		0&\cdots &0&0\\
        \vdots & \ddots &\vdots &\vdots \\
        0&\cdots &0&1\\
        0&\cdots &-1&0\\
		\end{pmatrix}$};
	\node[round] (a) at (2,-1){};
	\node[] (e) at (3.5,-1){$\overbrace{\cdots}^\text{n-4 nodes}$};
	\node[round] (b) at (5,-1){};
	\node[round] (c) at (6,-2){};
	\node[round] (d) at (6,0){};
	\draw[arrow] (a)--(e);
	\draw[arrow] (e)--(b);
	\draw[arrow] (c)--(b);
	\draw[arrow] (d)--(b);
	\node[] at (5.8,-3){\textbf{Q2}: ['D', $n$]};
	\node[] at (9,-1){$\begin{pmatrix}
		0&\cdots &0&0&0\\
		\vdots & \ddots & \vdots &\vdots &\vdots \\
        0&\cdots &0&-1&-1\\
        0&\cdots &1&0&0\\
        0&\cdots &1&0&0\\
		\end{pmatrix}$};
	\end{tikzpicture}
\end{figure}
\begin{figure}[H]
    \vspace{-0.75cm}
	\centering
	\begin{tikzpicture}
	[round/.style={circle, draw=black, fill=white, minimum size=5mm},]
	\tikzstyle{arrow1} = [,->,>=stealth]
	\tikzstyle{arrow2} = [,->,>=stealth]
	\node[round] (a) at (0,-2){};
	\node[round] (b) at (1.5,-2){};
	\node[round] (c) at (3,-2){};
	\node[round] (d) at (4.5,-2){};
	\node[round] (e) at (6,-2){};
	\node[round] (f) at (3,-0.5){};
	\node[] at (5.5,-3.5){\textbf{Q3}: ['E', 6]}; 
	\draw[arrow2] (a)--(b);
	\draw[arrow1] (b)--(c);
	\draw[arrow1] (c)--(d);
	\draw[arrow1] (d)--(e);
	\draw[arrow1] (c)--(f);
	\node[] at (9,-1){$\begin{pmatrix}
		0&1&0&0&0&0\\
		-1&0&1&0&0&0\\
		0&-1&0&1&0&1\\
		0&0&-1&0&1&0\\
		0&0&0&-1&0&0\\
		0&0&-1&0&0&0\\
		\end{pmatrix}$};
	\end{tikzpicture}
\end{figure}
\begin{figure}[H]
	\centering
	\begin{tikzpicture}
	[round/.style={circle, draw=black, fill=white, minimum size=5mm},]
	\tikzstyle{arrow1} = [,->,>=stealth]
	\tikzstyle{arrow2} = [,->>,>=stealth]
	\tikzstyle{arrow3} = [,->>>,>=stealth]
	\node[round] (a) at (-5,0){};
	\node[round] (b) at (-3,0){};
	\node[round] (c) at (-3,-2){};
	\node[round] (d) at (-5,-2){};
	\node[] at (-2.5,-3){\textbf{Q4}}; 
	\draw[arrow2] (a)--(b);
	\draw[arrow2] (b)--(c);
	\draw[arrow2] (c)--(d);
	\draw[arrow2] (d)--(a);
	\node[] at (-1,-1){$\begin{pmatrix}
		0&2&0&-2\\
		-2&0&2&0\\
		0&-2&0&2\\
		2&0&-2&0\\
		\end{pmatrix}$};
	\node[round] (a) at (3,0){};
	\node[round] (b) at (5,0){};
	\node[round] (c) at (5,-2){};
	\node[round] (d) at (3,-2){};
	\node[] at (5.5,-3){\textbf{Q5}}; 
	\draw[arrow1] (a)--(b);
	\draw[arrow3] (b)--(c);
	\draw[arrow3] (c)--(d);
	\draw[arrow1] (d)--(a);
	\draw[arrow2] (d)--(b);
	\node[] at (7,-1){$\begin{pmatrix}
		0&1&0&-1\\
		-1&0&3&-2\\
		0&-3&0&3\\
		1&2&-3&0\\
		\end{pmatrix}$};
	\end{tikzpicture}
\end{figure}
\begin{figure}[H]
	\centering
	\begin{tikzpicture}
	[round/.style={circle, draw=black, fill=white, minimum size=5mm},]
	\tikzstyle{arrow1} = [,->,>=stealth]
	\tikzstyle{arrow2} = [,->>,>=stealth]
	\tikzstyle{arrow3} = [,->>>,>=stealth]
	\tikzstyle{arrow4} = [,->>>>,>=stealth]
	\node[round] (a) at (-5,0){};
	\node[round] (b) at (-3,0){};
	\node[round] (c) at (-3,-2){};
	\node[round] (d) at (-5,-2){};
	\node[] at (-2.5,-3){\textbf{Q6}}; 
	\draw[arrow1] (b)--(a);
	\draw[arrow2] (c)--(a);
	\draw[arrow3] (a)--(d);
	\draw[arrow2] (c)--(b);
	\draw[arrow1] (b)--(d);
	\draw[arrow4] (d)--(c);
	\node[] at (-1,-1){$\begin{pmatrix}
		0&-1&-2&3\\
		1&0&-2&1\\
		2&2&0&-4\\
		-3&-1&4&0\\
		\end{pmatrix}$};
	\node[round] (a) at (3,0){};
	\node[round] (b) at (5,0){};
	\node[round] (c) at (5,-2){};
	\node[round] (d) at (3,-2){};
	\node[] at (5.5,-3){\textbf{Q7}}; 
	\draw[arrow3] (a)--(b);
	\draw[arrow3] (b)--(c);
	\draw[arrow3] (c)--(d);
	\draw[arrow3] (d)--(a);
	\node[] at (7,-1){$\begin{pmatrix}
		0&3&0&-3\\
		-3&0&3&0\\
		0&-3&0&3\\
		3&0&-3&0\\
		\end{pmatrix}$};
	\end{tikzpicture}
\end{figure}
\begin{figure}[H]
	\centering
	\begin{tikzpicture}
	[round/.style={circle, draw=black, fill=white, minimum size=5mm},]
	\tikzstyle{arrow1} = [,->,>=stealth]
	\tikzstyle{arrow2} = [,->>,>=stealth]
	\tikzstyle{arrow3} = [,->>>,>=stealth]
	\tikzstyle{arrow5} = [,->>>>>,>=stealth]
	\node[round] (a) at (1,0){};
	\node[round] (b) at (3,0){};
	\node[round] (c) at (3,-2){};
	\node[round] (d) at (1,-2){};
	\draw[arrow2] (a)--(b);
	\draw[arrow3] (a)--(c);
	\draw[arrow3] (b)--(d);
	\draw[arrow1] (c)--(b);
	\draw[arrow5] (d)--(a);
	\draw[arrow2] (c)--(d);
	\node[] at (4,-3){\textbf{Q8}}; 
	\node[] at (6,-1){$\begin{pmatrix}
		0&2&3&-5\\
		-2&0&-1&3\\
		-3&1&0&2\\
		5&-3&-2&0\\
		\end{pmatrix}$};
	\end{tikzpicture}
\end{figure}
\begin{figure}[H]
	\centering
	\begin{tikzpicture}
	[round/.style={circle, draw=black, fill=white, minimum size=5mm},]
	\tikzstyle{arrow1} = [,->,>=stealth]
	\tikzstyle{arrow2} = [,->>,>=stealth]
	\tikzstyle{arrow3} = [,->>>,>=stealth]
	\tikzstyle{arrow4} = [,->>>>,>=stealth]
	\node[round] (a) at (-7.5,0){};
	\node[round] (b) at (-6.5,-1.4){};
	\node[round] (c) at (-8.5,-1.4){};
	\node[] at (-6.2,-2.5){\textbf{Q9}}; 
	\draw[arrow3] (a)--(b);
	\draw[arrow3] (b)--(c);
	\draw[arrow3] (c)--(a);
	\node[] at (-5,-0.8){$\begin{pmatrix}
		0&3&-3\\
		-3&0&3\\
		3&-3&0\\
		\end{pmatrix}$};
	\node[round] (a) at (-2,0){};
	\node[round] (b) at (-1,-1.4){};
	\node[round] (c) at (-3,-1.4){};
	\node[] at (-0.8,-2.5){\textbf{Q10}}; 
	\draw[arrow2] (a)--(b);
	\draw[arrow2] (b)--(c);
	\draw[arrow4] (c)--(a);
	\node[] at (0.5,-0.8){$\begin{pmatrix}
		0&2&-4\\
		-2&0&2\\
		4&-2&0\\
		\end{pmatrix}$};
	\node[round] (a) at (3.5,0){};
	\node[round] (b) at (4.5,-1.4){};
	\node[round] (c) at (2.5,-1.4){};
	\node[] at (4.8,-2.5){\textbf{Q11}}; 
	\draw[arrow1] (a)--(b);
	\draw[arrow2] (b)--(c);
	\draw[arrow3] (c)--(a);
	\node[] at (6,-0.8){$\begin{pmatrix}
		0&1&-3\\
		-1&0&2\\
		3&-2&0\\
		\end{pmatrix}$};
	\end{tikzpicture}
\end{figure}
\begin{figure}[H]
	\centering
	\begin{tikzpicture}
	[round/.style={circle, draw=black, fill=white, minimum size=5mm},]
	\tikzstyle{arrow} = [,->,>=stealth]
	\node[round] (a) at (-0.2,-1){};
	\node[round] (b) at (0,-2){};
	\node[round] (c) at (-2,-1.8){};
	\node[round] (d) at (-1,0){};
	\node[round] (e) at (1,0){};
	\node[round] (f) at (2,-1.8){};
	\node[round] (g) at (0.2,-3){};
    \node[] at (3,-4.2){\textbf{Q12}: triangulated 10-gon in the mutation class ['A', 7]}; 
	\draw[arrow] (d)--(e);
	\draw[arrow] (a)--(d);
	\draw[arrow] (e)--(a);
	\draw[arrow] (a)--(c);
	\draw[arrow] (c)--(b);
	\draw[arrow] (b)--(a);
	\draw[arrow] (b)--(g);
	\draw[arrow] (g)--(f);
	\draw[arrow] (f)--(b);
	\node[] at (6,-1.4){$\begin{pmatrix}
		0&1&-1&0&0&0&0\\
		-1&0&1&0&0&0&0\\
		1&-1&0&1&-1&0&0\\
		0&0&-1&0&1&0&0\\
		0&0&1&-1&0&1&-1\\
		0&0&0&0&-1&0&1\\
		0&0&0&0&1&-1&0\\
		\end{pmatrix}$};
	\end{tikzpicture}
\end{figure}
\begin{figure}[H]
	\centering
	\begin{tikzpicture}
	[round/.style={circle, draw=black, fill=white, minimum size=5mm},]
	\tikzstyle{arrow1} = [,->,>=stealth]
	\tikzstyle{arrow2} = [,->>,>=stealth]
	\node[round] (a) at (0,0){};
	\node[round] (b) at (2,0){};
	\node[round] (c) at (1,-1.4){};
	\node[round] (d) at (-1,-1.4){};
	\node[round] (e) at (3,-1.4){};
	\node[round] (f) at (0,-2.8){};
	\node[round] (g) at (2,-2.8){};
	\node[] at (3.5,-4.2){\textbf{Q13}: ['X', 7]}; 
	\draw[arrow2] (a)--(b);
	\draw[arrow1] (b)--(c);
	\draw[arrow1] (c)--(a);
	\draw[arrow1] (d)--(c);
	\draw[arrow1] (c)--(f);
	\draw[arrow2] (f)--(d);
	\draw[arrow1] (c)--(e);
	\draw[arrow1] (g)--(c);
	\draw[arrow2] (e)--(g);
	\node[] at (6.5,-1.4){$\begin{pmatrix}
		0&2&-1&0&0&0&0\\
		-2&0&1&0&0&0&0\\
		1&-1&0&-1&1&1&-1\\
		0&0&1&0&-2&0&0\\
		0&0&-1&2&0&0&0\\
		0&0&-1&0&0&0&2\\
		0&0&1&0&0&-2&0\\
		\end{pmatrix}$};
	\end{tikzpicture}
\end{figure}
\begin{figure}[H]
	\centering
	\begin{tikzpicture}
	[round/.style={circle, draw=black, fill=white, minimum size=5mm},]
	\tikzstyle{arrow1} = [,->>,>=stealth]
	\tikzstyle{arrow2} = [,->>,>=stealth]
	\node[round] (a) at (0,0){};
	\node[round] (b) at (2,-1){};
	\node[round] (c) at (2.4,-3){};
	\node[round] (d) at (1.2,-4.6){};
	\node[round] (e) at (-1.2,-4.6){};
	\node[round] (f) at (-2.4,-3){};
	\node[round] (g) at (-2,-1){};
	\node[] at (2.8,-5.2){\textbf{Q14}}; 
	\draw[arrow2] (a)--(b);
	\draw[arrow1] (b)--(c);
	\draw[arrow1] (c)--(d);
	\draw[arrow1] (d)--(e);
	\draw[arrow1] (e)--(f);
	\draw[arrow2] (f)--(g);
	\draw[arrow1] (g)--(a);
	\node[] at (6.5,-2.3){$\begin{pmatrix}
		0&2&0&0&0&0&-2\\
		-2&0&2&0&0&0&0\\
		0&-2&0&2&0&0&0\\
		0&0&-2&0&2&0&0\\
		0&0&0&-2&0&2&0\\
		0&0&0&0&-2&0&2\\
		2&0&0&0&0&-2&0\\
		\end{pmatrix}$};
	\end{tikzpicture}
\end{figure}
\begin{figure}[H]
	\centering
	\begin{tikzpicture}
	[round/.style={circle, draw=black, fill=white, minimum size=5mm},]
	\tikzstyle{arrow1} = [,->,>=stealth]
	\tikzstyle{arrow2} = [,->,>=stealth]
	\node[round] (a) at (0,0){};
	\node[round] (b) at (1,0){};
	\node[round] (c) at (2,0){};
	\node[round] (d) at (3,0){};
	\node[round] (e) at (4,0){};
	\node[round] (f) at (2,-1){};
	\node[round] (g) at (2,-2){};
	\node[] at (4.5,-3.5){\textbf{Q15}: T-type, ['T', [3,3,3]], which is also of type affine $E_6$, ['E', 6, 1] }; 
	\draw[arrow2] (a)--(b);
	\draw[arrow1] (b)--(c);
	\draw[arrow1] (c)--(d);
	\draw[arrow1] (d)--(e);
	\draw[arrow1] (c)--(f);
	\draw[arrow2] (f)--(g);
	\node[] at (8,-1){$\begin{pmatrix}
		0&1&0&0&0&0&0\\
		-1&0&1&0&0&0&0\\
		0&-1&0&1&0&1&0\\
		0&0&-1&0&1&0&0\\
		0&0&0&-1&0&0&0\\
		0&0&-1&0&0&0&1\\
		0&0&0&0&0&-1&0\\
		\end{pmatrix}$};
	\end{tikzpicture}
\end{figure}

\paragraph{Investigation Results}
Here we tabulate each investigation with a brief description, a list of the quivers used to generate the duality classes in the dataset, and the measures of learning performance. The measures of performance (as described in appendix \ref{measures}) are presented as a pair: $(acc,\phi)$, consisting of accuracy of agreement, $acc$, and Matthews' correlation coefficient, $\phi$, where calculated. Both evaluate to 1 for perfect learning, results are shown to 2 decimal places.

Dynkin and T type quivers are denoted using the Sage quiver package convention, other infinite mutation type quivers are denoted using the label assigned in the preceding 'Quivers considered' list.

NB classifier results showed perfect classification between 2 mutation type classes. Classifying classes of different quiver sizes was trivial and did not reduce performance as expected. Where classification was between more than 2 classes the performance was lower but still very good. Enhancing the datasets with rank information, or Diophantine-inspired variables, did not improve NB classification.

NNs required rank information in their dataset to classify well, but with this included NNs outperformed the NB classifier, particularly when classifying quivers at unseen mutation depths, and when classifying against random antisymmetric matrices.

We should also mention that we are using the word ``depth'' throughout the paper. Starting with a quiver (at depth 0) having $n$ nodes, we have $n$ choices of dualizing one node. These newly generated quivers are said to be at depth 1. We can then apply mutations to these depth-one quivers again by choosing one node to dualize. Such quivers obtained are at depth 2 (except the quiver at depth 0 we start with, i.e., dualizing the same node twice). Hence, when we say a quiver is at depth $k$, the (shortest) distance would be $k$ from this quiver to our starting quiver under mutations.

\newpage
\thispagestyle{empty}
\begin{table}[H]
\vspace*{-1.75cm}
\makebox[1 \textwidth][c]{
\begin{tabular}{|l|c|c|}
\hline
Investigation Description & Quivers & \begin{tabular}[c]{@{}c@{}}Results\\ $(acc,\phi)$\end{tabular} \\ \hline
\begin{tabular}[c]{@{}l@{}}NB classification between \\ 2 mutation classes\end{tabular}  & \begin{tabular}[c]{@{}c@{}}{[}'A',4{]} - {[}'D',4{]}\\ Q4 - Q5\\ {[}'D',4{]} - {[}'A',(3,1),1{]}\\ {[}'D',4{]} - Q4\end{tabular} & \begin{tabular}[c]{@{}c@{}}(1.00,1.00)\\ (1.00,1.00)\\ (1.00,1.00)\\ (1.00,1.00)\end{tabular} \\ \hline
\begin{tabular}[c]{@{}l@{}}NB classification on datasets\\ with varying quiver sizes\end{tabular} & Q4 - Q5 - Q9 - Q10 & (1.00,1.00) \\ \hline
\begin{tabular}[c]{@{}l@{}}NB classification on datasets \\ with more than 2 mutation classes\end{tabular}  & \begin{tabular}[c]{@{}c@{}}{[}'A',6{]} - {[}'D',6{]} - {[}'E',6{]}\\ {[}'A',4{]} - {[}'D',4{]} - {[}'A',(3,1),1{]} - {[}'A',(2,2),1{]}\\ {[}'A',6{]} - {[}'D',6{]} - {[}'E',6{]} - {[}'A',4{]} -\\ - {[}'D',4{]} - {[}'A',(3,1),1{]} - {[}'A',(2,2),1{]}\\ Q4 - Q5 - Q6\\ Q4 - Q5 - Q6 - Q7\\ Q4 - Q5 - Q6 - Q7 - Q8\\ {[}'T',(4,4,4){]} - {[}'T',(4,5,3){]} - {[}'T',(4,6,2){]}\end{tabular} & \begin{tabular}[c]{@{}c@{}}(0.90,0.82)\\ (0.85,0.70)\\ (0.75,$\sim$)\\ \\ (0.91,0.82)\\ (0.86,0.72)\\ (0.84,0.67)\\ (0.89,0.78)\end{tabular} \\ \hline
\begin{tabular}[c]{@{}l@{}}NB extrapolation predictions:\\ validating on different classes / \\ mutation depths to training\end{tabular}  & \begin{tabular}[c]{@{}c@{}}Train {[}'A',6{]},{[}'D',6{]} - Valid {[}'E',6{]}\\ Train Q4,Q5 low depths\\ - Valid  Q4,Q5 high depths\\ Train Q4,Q5 low \& high depths\\ - Valid Q4,Q5 middle depths\end{tabular} & \begin{tabular}[c]{@{}c@{}}(0.60,0.25)\\ (0.50,0.00)\\ \\ (0.65,0.33) \\ \\ \end{tabular}  \\ \hline
\begin{tabular}[c]{@{}l@{}}NB classification on enhanced\\ datasets with rank vectors\end{tabular} & \begin{tabular}[c]{@{}c@{}}Q4 - Q5 - Q6\\ Q4 - Q5 - Q6 - Q7\\ Q4 - Q5 - Q6 - Q7 - Q8\end{tabular} & \begin{tabular}[c]{@{}c@{}}(0.91,0.83)\\ (0.86,0.72)\\ (0.84,0.67)\end{tabular}  \\ \hline
\begin{tabular}[c]{@{}l@{}}NB classification on enhanced\\ datasets with Diophantine variables\end{tabular} & \begin{tabular}[c]{@{}c@{}}Q9 - Q10 - Q11\\ Q4 - Q5 - Q6\\ Q4 - Q5 - Q6 - Q7\\ Q4 - Q5 - Q6 - Q7 - Q8\end{tabular} & \begin{tabular}[c]{@{}c@{}}(0.91,0.84)\\ (0.91,0.83)\\ (0.86,0.72)\\ (0.84,0.69)\end{tabular}  \\ \hline
\begin{tabular}[c]{@{}l@{}}NN classification between \\ finite-type classes\end{tabular}  & Q12 - Q13 - Q15  & (0.33,$\sim$)  \\ \hline
\begin{tabular}[c]{@{}l@{}}NN classification on mixed \\ mutation type (finite and infinite)\end{tabular}  & Q12 - Q13- Q14    & (0.55,$\sim$)  \\ \hline
\begin{tabular}[c]{@{}l@{}}NN classification against random \\ antisymmetric matrices\end{tabular}   & Q9 - Antisymm    & (0.97,$\sim$) \\ \hline
\begin{tabular}[c]{@{}l@{}}NN classification on enhanced\\ datasets with rank vectors\end{tabular}  & \begin{tabular}[c]{@{}c@{}}Q12 - Q13 - Q14\\ Q12 - Q13 - Q15\end{tabular}    & \begin{tabular}[c]{@{}c@{}}(1.00,$\sim$)\\ (0.71,$\sim$)\end{tabular}     \\ \hline
\begin{tabular}[c]{@{}l@{}}NN extrapolation predictions:\\ validating on different mutation\\ depths to training\\ (with rank vector \\ data enhancement)\end{tabular} & \begin{tabular}[c]{@{}c@{}}Train Q12,Q13,Q14 low depths\\ - Valid Q12,Q13,Q14 high depths\\ Train Q12,Q13,Q14 low \& high depths\\ - Valid Q12,Q13,Q14 middle depths\\ Train Q12,Q13,Q14,Q15 low \& high depths\\ - Valid Q12,Q13,Q14,Q15 middle depths\end{tabular} & \begin{tabular}[c]{@{}c@{}}(0.74,$\sim$)\\ \\ (1.00,1.00)\\ \\ (0.98,$\sim$)\end{tabular}  \\ \hline
\begin{tabular}[c]{@{}l@{}}NN classification against random\\ antisymmetric matrices \\ (with rank vector \\ data enhancement)\end{tabular}                            & \begin{tabular}[c]{@{}c@{}}Q9 - Antisymm\\ Q9 - Q10 - Antisymm\end{tabular}     & \begin{tabular}[c]{@{}c@{}}(1.00,$\sim$)\\ (0.85,$\sim$)\end{tabular}   \\ \hline
\end{tabular}
}
\end{table}

\section{Dramatis Personae}
\subsection{Seiberg Duality}\label{seiberg}
In this section we review Seiberg duality, which is an IR equivalence between $4d$ $\mathcal{N}=1$ gauge theories \cite{Seiberg:1994pq}. We will phrase our discussion in the language of quivers, since all the theories considered in this paper are of this type.

Let us consider dualizing a node $j$ in the quiver, which does not have adjoint chiral fields.\footnote{Generalizations of Seiberg duality to gauge groups with adjoints are known, under certain conditions (see e.g. \cite{Kutasov:1995np,Kutasov:1995ve,Kapustin:1996nb}).} 
The transformation of the gauge theory can be summarized in terms of the following rules:

\paragraph{1. Flavors.} 

\begin{figure}[ht]
	\centering
	\includegraphics[width=12cm]{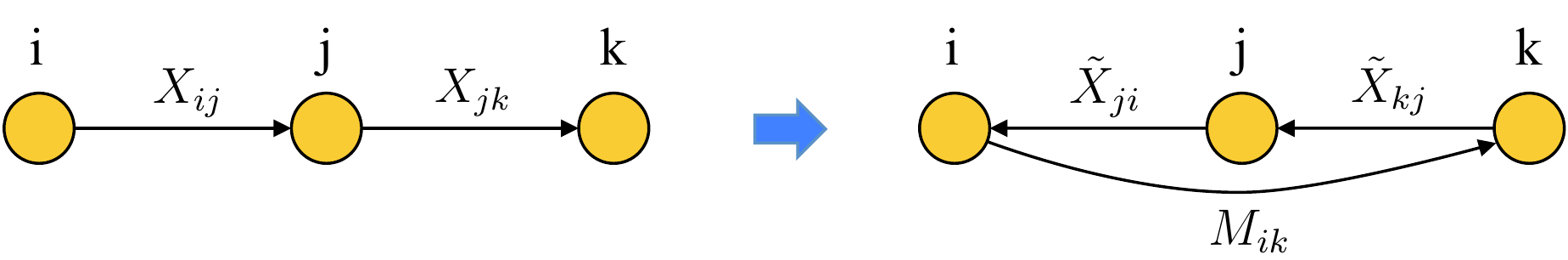}
\caption{Schematic representation of Seiberg duality. The dualized node $j$ can actually be connected to multiple nodes by incoming and outgoing arrows.}
	\label{SD_quiver}
\end{figure}

In physics, the arrows connected to the mutated node are usually referred to as {\it flavors}. The flavors transform by simply reversing their orientation, namely: 
\begin{itemize}
\item[{\bf 1.a)}] Replace every incoming arrow  $i \to j$ with the outgoing arrow $j \to i$. Calling $X_{ij}$ the incoming arrow, we replace it by the dual flavor $\tilde{X}_{ji}$.

\item[{\bf 1.b)}] Replace every outgoing arrow $j \to k$ with the incoming arrow $k \to j$. Calling $X_{jk}$ the outgoing arrow, we replace it by the dual flavor $\tilde{X}_{kj}$.
\end{itemize}
This is the quiver implementation of the fact that the magnetic flavors are in the complex conjugate representations, of both the dualized gauge group and the spectator nodes, of the original flavors.\footnote{In our discussion, including the points that follow, we allow for the possibility of chiral fields connecting a given pair of nodes in both directions.} This transformation is shown in \fref{SD_quiver}.

\paragraph{2. Mesons.} 

Next we add {\it mesons}, i.e. composite arrows, to the quiver as follows. For every $2$-path $i\to j \to k$ we add a new arrow $i \to k$. This meson $M_{ik}$ can be regarded as the composition of the flavors $i\to j$ and $j \to k$ of the original theory, namely $M_{ik}=X_{ij}X_{jk}$. In other words, we generate all possible composite arrows consisting of incoming and outgoing chiral fields. \fref{SD_quiver} also illustrates the addition of a meson.

\paragraph{3. Ranks.} 

The rank of the dualized node transforms as
\beq
N_j'=N_{f_j}-N_j ~,
\eeq
where $N_{f_j}$ is the number of flavors at the dualized node $j$. Later we will consider generic quivers, which are not necessarily anomaly free. These quivers are interesting from a mathematical point of view and, in such cases, we will not consider the ranks of the nodes. Ranks will only be taken into account for anomaly free quivers, i.e. theories for which the gauge (and hence dualizable) nodes have an equal number of incoming and outgoing arrows. In these cases,
\beq
N_{f_j}=N_{\text{in}_j}=N_{\text{out}_j} ~,
\eeq
which, more explicitly, is given by
\beq
N_{f_j}=N_{in_j}=\sum_{i\to j} a_{ij} N_i ~,
\eeq
with $a_{ij}$ the (positive) number of bifundamental arrows going from node $i$ into node $j$.

\paragraph{4. Superpotential.} 

The superpotential transforms as follows:
\begin{itemize}
\item[{\bf 4.a)}] In the original superpotential, we replace instances of $X_{ij} X_{jk}$ with the meson $M_{ik}$ obtained by composing the two arrows. 

\item[{\bf 4.b)}] {\bf Cubic dual flavors-meson couplings.} For every meson, we add a new cubic term in the superpotential, coupling it to the corresponding magnetic flavors. Namely, we add the term $M_{ik} \tilde{X}_{kj} \tilde{X}_{ji}$.

\end{itemize}
If there are fields that acquire mass in this process, we can integrate them out using their equations of motion.

\medskip

All the rules discussed above, with the exception of the one for the ranks, are the same ones that are used for cluster algebras.  Cluster algebras also come equipped with a set of generators known as cluster variables. 

\subsection{Mutation of Cluster Algebras}\label{clusters}

Mathematically speaking, an algebra is a structure that functions like a vector space with the additional feature that elements can be multiplied together.  An algebra can be presented by generators, think of basis vectors, and relations, i.e. algebraic dependencies generalizing linear dependencies of a vector space.  A rank $n$ cluster algebra is a subalgebra of the field of rational functions in $n$ variables where its generators can be grouped together into algebraically independent sets known as clusters, also all of size $n$, such that certain exchange relations allow one to transition from one cluster to another \cite{MR1887642}.  These exchange relations, known as cluster mutation, can be described using the language of quivers, echoing the description of Seiberg duality in physics.  

\paragraph{ 5. Cluster Variables.}
Given an initial cluster $\{x_1,x_2,\dots, x_n\}$, we allow cluster mutations in $n$ directions, each of the form 
$$x_j x_j' = \prod_{i \to j \mathrm{~in~Q}} x_i + \prod_{j \to k \mathrm{~in~Q}} x_k$$ for each $1 \leq j \leq n$, and where the products are over all incoming arrows and outgoing arrows, respectively.  We thus get a new generator, cluster variable, $x_j'$, and yielding the cluster $\{x_1,x_2,\dots, x_{j-1}, x_j', x_{j+1}, \dots, x_n\}$.  The process of cluster mutation may be continued but to mutate while using this new cluster as a reference, we use the quiver $\mu_j Q$ in place of $Q$, where $\mu_j Q$ is the quiver obtained by applying the rules of Seiberg duality at node $j$.

Given a quiver $Q$, we construct the associated cluster algebra $\mathcal{A}_Q$ by applying cluster mutation in all directions and iterating to obtain the full list of cluster variables, i.e. generators of $\mathcal{A}_Q$.  Generically, this process yields an infinite number of generators for the cluster algebra, as well as an infinite number of different quivers along the way.  However, in special cases, a cluster algebra, and its defining quiver, have a specified mutation type.

We refer to a cluster algebra, or its associated quiver, as being of finite type if it has a finite number of generators, i.e. cluster variables, constructed by the cluster mutation process\footnote{This is a different statement than saying the cluster algebra is finitely generated, or Noetherian, as an algebra.  There are examples of Noetherian algebras that admit an infinite number of cluster variables as generators.  The simplest such example corresponds to the quiver associated to $SU(2)$ theories consisting of two nodes and two arrows between them.  There are an infinite number of cluster variables for the associated cluster algebra even though as an algebra, it is generated by four elements \cite{BFZ3,SherZel}.}.  As proven by Fomin and Zelevinsky \cite{MR2004457}, the list of cluster algebras of finite type exactly agree with Gabriel's ADE classification\footnote{Or if we allow cluster algebras associated to skew-symmetrizable matrices rather than only quivers, which must be skew-symmetric, we get the Cartan-Killing or Dynkin classification including types B, C, $F_4$, and $G_2$ as well.} of quivers admitting only finitely many indecomposable representations \cite{Gabriel}, or those equivalent to them  via quiver mutation, i.e. Seiberg duality.

Another important family of cluster algebras are those of finite mutation type. Such cluster algebras are those with only a finite number of quivers reachable via mutation, i.e. Seiberg duality.  This class of cluster algebras completely encompasses the subclass of cluster algebras of finite type.  In totality, this class contains all rank 2 cluster algebras, like the aforementioned cluster algebra associated to $SU(2)$, cluster algebras of surface type, and eleven exceptional types ($E_6, E_7, E_8$, affine $E_6, E_7, E_8$, elliptic $E_6, E_7, E_8$, and two additional quivers known as $X_6$ and $X_7$) \cite{2008arXiv0811.1703F, DerkOwen}.  Such finite mutation type quivers have also been studied previously in the physics literature where they were referred to as complete quantum field theories \cite{Complete}.

Cluster algebras of surface types, i.e. associated to orientable Riemann surfaces, were first described by Fomin, Shapiro, and Thurston \cite{2006math......8367F}.  Generically, the quiver associated to a triangulation of a Riemann surface is obtained by taking the medial graph where nodes of the quiver correspond to non-boundary arcs of the triangulation and we draw an arrow of the quiver between nodes $i$ and $j$ for every triangular face where arcs associated to $i$ and $j$ meet at a vertex and $j$ follows $i$ in clockwise order.  Mutating at a node corresponds to flipping between the two possible diagonals for triangulating a quadrilateral.  Since such triangulations live on an orientable Riemann surface, any associated quiver has at most two arrows between any given pair of nodes, thus demonstrating that such cluster algebras admit only finitely many quivers and are hence of finite mutation type.  The eleven exceptional cases of Felikson, Shapiro and Tumarkin do not have a surface model but at least the finite and affine type $E$ quivers are well-known from previous representation theory, e.g. Gabriel's ADE classification, and Kac's extension to affine quivers \cite{Kac}.
 
In this paper we will focus on the transformation of the quiver (rules 1 and 2) and in some cases include information on the ranks (rule 3), so we will not deal with rule 4 nor with rule 5. Even with this restriction, we will manage to obtain non-trivial results. Having said that, the superpotential is a crucial element of the duality, as is the mutation of cluster variables in the context of cluster algebras. We plan to incorporate both of these in future studies.

\section{Recognizing Mutations}\label{recogmut}
There are various ways to construct the dataset. We can directly assign each mutation class a different label. Then the machine will be asked to do a multiclass classification. We can also have datasets that consists of matrix pairs so that every \{input$\rightarrow$output\} has the form
\begin{equation}
\{(M_1,M_2)\rightarrow1/0\},
\end{equation}
where 1 indicates that $M_1$ and $M_2$ are in the same class while 0 indicates that they are not. Let us first start with the latter using the $\mathtt{Mathematica}$ built-in function $\mathtt{Classify}$.

\subsection{Classifying Two Mutation Classes}\label{twoclasses}
As the simplest example, let us machine learn only two different classes, [`A',4] and [`D',4]\footnote{Henceforth, we will use the same notation as in $\mathtt{Sage}$ \cite{sagemath, 2011arXiv1102.4844M} for known quiver mutation types, and we will not specify the matrices and quivers.}, shown with their adjacency matrices as \textbf{Q1} and \textbf{Q2}, for the cases $n=4$, in the Quivers list of \S\ref{summary}.

Notice that these matrices/quivers are of finite mutation types, i.e., the duality trees are closed. Many (but not all) quivers in finite mutation types\footnote{To be clear, we should point out that finite \emph{mutation types} and finite \emph{types} refer to different concepts. In the sense of \cite{MR2004457, 2008arXiv0811.1703F}, a finite mutation type indicate that there are finitely many dual quivers generated from our starting quiver while a finite type is the namesake of a Dynkin type. Sometimes we will use the term ``finite classes''. This is the same as ``finite mutation types''. However, note the word ``class'' is slightly different from ``mutation type'' in our context. Each class refers to one duality tree. For instance, [`A',4] and [`A',5] are not in the same class as they are certainly not duals, but they are both of finite mutation types.} contain sources and sinks, and are hence anomalous. Albeit not physically meaningful, we are still interested in these quivers from pure mathematics and machine learning viewpoints. Furthermore, we can compare these results with those from infinite mutation types.

The result\footnote{Note the metrics used to evaluate the machine's performance (accuracy, F-score, and MCC $\phi$) are defined in appendix \ref{measures}.} of 5-fold cross validation is tabulated in Table \ref{A4D4table}.  
\begin{table}[h!]
	\centering
	\begin{tabular}{|c|c|c|}
		\hline
		Accuracy	& F-Score & $\phi$ \\ \hline
		1$\pm$0	& 1$\pm$0 & 1$\pm$0 \\ \hline
	\end{tabular}
	\caption{Training and validating two classes: [`A',4] and [`D',4]. We generate (144+50) matrices. There are 9026 1's and 7193 0's. The method is chosen by the machine. The results are accurate to the floating point precision, but decimal points are not shown.}\label{A4D4table}
\end{table}
We also plot the learning curves at different training percentages in Fig. \ref{A4D4curve}.
We can see that the machine gives 100\% accuracy most of the time, which is very inspiring.

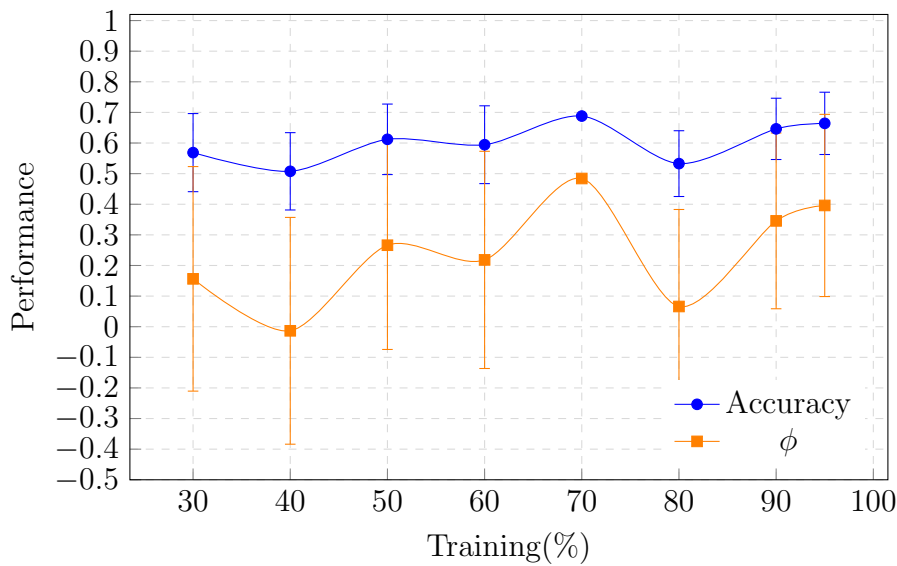
\begin{figure}[h]
	\centering
	\begin{tikzpicture}
	\begin{axis}[ymin=-0.5, ymax=1.02,
	width=0.75\textwidth,
	height=0.5\textwidth,
	ytick={-0.5,-0.4,-0.3,-0.2,-0.1,0,...,1.1}, ytick align=inside, ytick pos=left,
	xtick={20,30,...,100}, xtick align=inside, xtick pos=left,
	xlabel=Training(\%),
	ylabel=Performance,
	grid=major,
	grid style={dashed, gray!30},
	legend pos=south east,
	legend style={draw=none}]
	\addplot+[
	blue, mark options={blue, scale=1},
	smooth, 
	error bars/.cd, 
	y fixed,
	y dir=both, 
	y explicit
	] table [x=x, y=y,y error=error, col sep=comma] {
		x,  y,        error
		30, 0.568684, 0.127578
		40, 0.507619, 0.126295
		50, 0.612129, 0.115163
		60, 0.594532, 0.127098
		70, 0.688095, 0.00538318
		80, 0.532768, 0.107442
		90, 0.64627,  0.100102
		95, 0.664286, 0.101737
	};
	\addlegendentry{Accuracy}
	\addplot+[
	orange, mark options={orange, scale=1},
	smooth, 
	error bars/.cd, 
	y fixed,
	y dir=both, 
	y explicit
	] table [x=x, y=y,y error=error, col sep=comma] {
		x,  y,        error
		30, 0.156344, 0.366748
		40, -0.0133713,0.370416
		50, 0.266207, 0.340513
		60, 0.218254, 0.35476
		70, 0.484128, 0.00663401
		80, 0.0660389,0.316887
		90, 0.345691, 0.287058
		95, 0.396108, 0.297599
	};
	\addlegendentry{$\phi$}
	\end{axis}
	\end{tikzpicture}
	\caption{Training two classes: [`A',6] and [`D',6], and validating [`E',6] (0's from [`E',6]-[`A',6] and [`E',6]-[`D',6] pairs). We generate 517, 572 and 600 matrices respectively. We choose training data out of 14182 pairs and validation data out of 13897 pairs. Data with indeterminate $\phi$'s, which appeared several times, is not plotted. These indeterminate $\phi$'s appeared 7 times in all (training and validation ten times at each training percentage). The method is chosen by the machine.}\label{ADE6curves}
\end{figure}

Before we add more mutation classes to our data, we are also curious about how the machine would behave when it is asked to predict unseen classes. In the above two-class example, the validation set $V$ is the complement of the training set $T$. Therefore, what the machine validates are in the same classes as those being trained. Now let us train the [`A',6] and [`D',6] classes and validate [`E',6] (shown as \textbf{Q3} in the Quivers considered list). In the validation dataset, 1's are always from pairs in [`E',6] while 0's are from [`E',6]/[`A',6] or [`E',6]/[`D',6] pairs\footnote{Unlike 1's, the 0's always have a matrix from trained classes. However, as we will further study in \S\ref{fixmethod} and Appendix \ref{NB} when finding the optimal method of the classifier, assigning 1 or 0 to a given pair is solely determined by the two matrices in this pair. Any other matrices, no matter whether they are related by mutations to the matrices in this pair, are irrelevant. In this sense, the [`E',6]/[`A',6] and the [`E',6]/[`D',6] pairs are always unseen classes.}. The learning result with various training percentages is given in Fig. \ref{ADE6curves}.
We find that the overall result is not very satisfying, and the Matthews $\phi$ could be indeterminate occasionally. From the confusion matrices, we can know that there is still always a zero entry. This zero always appears at FP or TN, i.e., only 1's or only 0's are predicted when the actual values are 0 in each single training. It is reasonable to see that such result as the machine has met some unseen mutation classes. This also shows that the machine is certainly not learning mutations (at least not the whole knowledge thereof) when the dataset only contains two different mutation classes\footnote{One may wonder whether the dimensions of matrices would affect our result, but in fact it is not a main influence. We will further study this when we include more different mutation classes in our training.}.

\subsection{Fixing the Method}\label{fixmethod}
\begin{figure}[h]
	\centering
	\begin{tikzpicture}
	\begin{axis}[ymin=-0.5, ymax=1.02,
	width=0.75\textwidth,
	height=0.5\textwidth,
	ytick={-0.5,-0.4,-0.3,-0.2,-0.1,0,...,1.1}, ytick align=inside, ytick pos=left,
	xtick={20,30,...,100}, xtick align=inside, xtick pos=left,
	xlabel=Training(\%),
	ylabel=Performance,
	grid=major,
	grid style={dashed, gray!30},
	legend pos=south east,
	legend style={draw=none}]
	\addplot+[
	blue, mark options={blue, scale=1},
	smooth, 
	error bars/.cd, 
	y fixed,
	y dir=both, 
	y explicit
	] table [x=x, y=y,y error=error, col sep=comma] {
		x,  y,        error
		20, 0.520154, 0.149151
		40, 0.617333, 0.11617
		60, 0.661071, 0.0950441
		80, 0.699857, 0.00948854
		90, 0.703072, 0.0102189
		95, 0.711143, 0.0152352
	};
	\addlegendentry{Accuracy}
	\addplot+[
	orange, mark options={orange, scale=1},
	smooth, 
	error bars/.cd, 
	y fixed,
	y dir=both, 
	y explicit
	] table [x=x, y=y,y error=error, col sep=comma] {
		x,  y,        error
		20, 0.0555401,0.428822
		40, 0.282095, 0.33579
		60, 0.485422, 0.00762256
		80, 0.499099, 0.0128635
		90, 0.503326, 0.00969516
		95, 0.509562, 0.0206741
	};
	\addlegendentry{$\phi$}
	\end{axis}
	\end{tikzpicture}
	\caption{Training two classes: [`A',6] and [`D',6], and validating [`E',6] (0's from [`E',6]-[`A',6] and [`E',6]-[`D',6] pairs). We generate 517, 572 and 600 matrices respectively. We choose training data out of 14182 pairs and validation data out of 13897 pairs. There are \emph{no} indeterminate $\phi$'s. The method is NB.}\label{ADE6curveNB}
\end{figure}
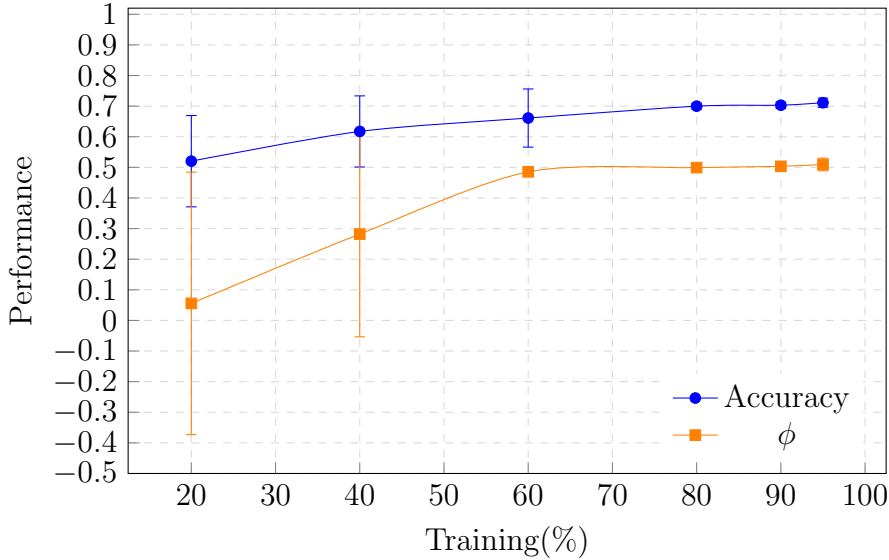

The $\mathtt{Classify}$ function in $\mathtt{Mathematica}$ has an option where one can specify the method used in the classifier. So far, this value is default in our experiments, and the method is chosen automatically by the machine. However, it is worth finding what method can give better predictions. It turns out that the Naive Bayes (NB) is the method we should choose. When studying the ADE Dynkin type quivers with 6 nodes above, we find that at each training percentage, the relatively higher accuracy is obtained only when the machine chooses NB. Hence, we perform this experiment with the same dataset again, but this time, we fix our method to NB. The learning curves are reported in Fig. \ref{ADE6curveNB}.
We find that the standard deviations are indeed reduced. The trends of the curves behave like those of usual learning curves. Moreover, although there is still always a zero entry in the confusion matrix, the Matthews $\phi$ is \emph{never} indeterminate anymore. In contrast, we can try what happens if we fixate on other methods. As an example, the result of only using Random Forest with the same dataset at 80\% training percentage is reported in Table \ref{randomforest}. It is obviously inferior to the result using NB. Henceforth, unless specified, we will always apply NB in the $\mathtt{Classify}$ function for future experiments\footnote{We also tried different methods when machine learning the example in \S\ref{moreclasses} which has four different classes. It turns out that in the built-in $\mathtt{Classify}$ function, NB gives nearly 85\% accuracy at 80\% training percentage while NN gives $\sim$60\% accuracy and SVM gives $\sim$50\% accuracy. Moreover, NN and SVM would take 1-2 minutes while NB would only take 1-2 seconds.}.
\begin{table}[h!]
	\centering
	\begin{tabular}{|c|c|c|c|c|c|}
		\hline
		Accuracy(\%)	& 60.3214 & 46.1786 & 50.2143 & 51.2857 & 53.2500 \\ \hline
		$\phi$	& 0.3468220 & -0.2088830 & Indeterminate & -0.0184085 & 0.1678070 \\ \hline\hline
		Accuracy(\%)	& 54.8571 & 49.8929 & 47.5714 & 49.6071 & 50.6071 \\ \hline
		$\phi$	& 0.2026570 & -0.0377965 & -0.1103220 & 0.0925057 & Indeterminate \\ \hline
	\end{tabular}
	\caption{Learning on ADE quivers with Random Forest method. Compared to Figure \ref{ADE6curveNB}, Naive Bayes performs superiorly and never gives indeterminate.}\label{randomforest}
\end{table}

Now, we would like to understand why NB always yields such good results. In Appendix \ref{NB}, we give a mathematical background of NB. The main reason is that the mutual independence of matrix pairs coincides with the basic assumption of NB.

\subsection{Two Classes Revisit}\label{twoinf}
To some extent, machine learning finite mutation classes would not be that necessary in application simply because we can traverse all the matrices. Let us try another example which contains two infinite mutation classes. The first one is the theory living on D3s probing $F_0$, the $0^\text{th}$ Hirzebruch surface, which is isomorphic to $\mathbb{P}^1\times\mathbb{P}^1$, as depicted in \textbf{Q4} \cite{Feng:2002kk,Franco:2003ja}.
The second one is generated by the quiver and adjacency matrix given in \textbf{Q5}, which is also anomaly free.

The learning result of 5-fold cross validation is tabulated in Table \ref{F0inf15fold}.
\begin{table}[h!]
	\centering
	\begin{tabular}{|c|c|c|}
		\hline
		Accuracy	& F-Score & $\phi$ \\ \hline
		1$\pm$0	& 1$\pm$0 & 1$\pm$0 \\ \hline
	\end{tabular}
	\caption{Training and validating two classes: \textbf{Q4} and \textbf{Q5}. We generate (102+138) matrices. There are 6344 1's and 6268 0's. The method is NB.}\label{F0inf15fold}
\end{table}
We also plot the learning curve at different training percentages in Fig. \ref{F0inf1curve}, showing results as perfect as the example of [`A',4] and [`D',4]. It is also worth noting that comparing Figure \ref{F0inf1curve} with Figure \ref{A4D4curve}, we see that the learning curve now looks smoother and more beautiful when we use NB.

Now that infinite mutation types generate infinitely many quivers under the Seiberg duality mutation, we can do something that is not done in finite mutations. In the training dataset $T$, we include the matrices generated to some depth (equal to the number of mutations from the original quiver) in the duality tree. However, the validation dataset $V$ consists of matrices generated at depths that are far away from those in $T$. We still start with the above two matrices, and generate (102+138) matrices. From these matrices, we create 6933 1's and 6358 0's. Then the 1's and 0's of $T$ will be evenly chosen out of the 13291 pairs. For $V$, we start with the following matrices:
\begin{equation}
\begin{pmatrix}
0&211&-16644&765262\\
-211&0&-1658&76232\\
16644&1658&0&-46\\
-765262&-76232&46&0\\
\end{pmatrix},
\begin{pmatrix}
0&-2586&39&55\\
2586&0&39603&-47\\
-39&-39603&0&843\\
-55&47&-843&0\\
\end{pmatrix},
\end{equation}
and generate (161+161) matrices. From these matrices, we create 5689 1's and 5663 0's. Then the 1's and 0's of $V$ will be evenly chosen out of the 11352 pairs. We make a dataset with 12000 pairs in all. At 90\% training percentage, the result is tabulated in Table \ref{depthfaraway}.
\begin{table}[h]
	\centering
	\begin{tabular}{|c|c|c|c|c|c|}
		\hline
		Accuracy(\%)	& 50.0833 & 51.1667 & 49.8333 & 47.8333 & 47.9167 \\ \hline
		$\phi$	& -0.0807034 & -0.0922570 & -0.0811080 & -0.0520199 & -0.1067660 \\ \hline
	\end{tabular}
	\caption{Training infinite type quiver matrices at low depths from the originals, and validating on matrices at depths far away. The method, NB, performs poorly.}\label{depthfaraway}
\end{table}
This shows that the machine is just guessing. Since it is predicting those of unseen depths, the result is not very surprising. As a matter of fact, the confusion matrices always have a vanishing TP (actual=predicted=1) and an extremely small FP (actual=0, predicted=1). This shows that the machine tends to regard the pairs from unseen depths as unrelated theories.

We now have seen that the machine does a good job for validation, but does not perform well when meeting unseen depths far away. It would be natural to ask, given both matrices of depth 0 to depth $n_1$ and of depth $n_2$ to depth $n_3$ ($n_3>n_2>n_1>0$), whether the machine can extrapolate the matrices of depths between $n_1$ and $n_2$. We still contemplate the above case with two different mutation classes (\textbf{Q4} and \textbf{Q5}), but this time, we have $n_1=3$, $n_2=6$ and $n_3=8$ for both of the two classes (and hence, we are validating matrices of depths 4 and 5)\footnote{In our training set, we also include pairs of 1's from depths 0-3 and depths 6-8. Likewise, in our validation set, we also include pairs of 1's from depths 4-5 and depths 0-3/6-8. Same is for 0's as well.}. The learning result at 90\% training percentage is listed in Table \ref{depthbetween}.
\begin{table}[h]
	\centering
	\begin{tabular}{|c|c|c|c|c|c|}
		\hline
		Accuracy(\%)	& 65.3891 & 65.5761 & 66.6277 & 66.2772 & 65.9500 \\ \hline
		$\phi$	& 0.333734 & 0.329995 & 0.347612 & 0.333243 & 0.336945 \\ \hline
	\end{tabular}
	\caption{For training set, we have 19267 1's and 19243 0's. For validation set, we generate 13020 1's and 13227 0's. Then we choose correspondingly many pairs used for validation, viz, 3946 pairs.}\label{depthbetween}
\end{table}
We can see that the result is better than the one in Table \ref{depthfaraway}. It is very natural to expect this since we are having much more matrices trained (or more precisely, the ratio of seen against unseen matrices is much larger). On the other hand, we should also expect that the result would still have much room to be improved regarding the fearture of NB.

We now make a proposal using the assumption of NB. As discussed in \S\ref{fixmethod} and Appendix \ref{NB}, whether a pair matrices are related to each other by mutations is independent of other matrices. This condition certainly applies here.

\begin{figure}[h]
	\centering
	\tikzset{every picture/.style={line width=0.75pt}}     
	\begin{tikzpicture}[x=0.75pt,y=0.75pt,yscale=-1,xscale=1]
	\draw  [fill={rgb, 255:red, 157; green, 232; blue, 255 }  ,fill opacity=1 ] (315.95,143.25) .. controls (315.95,100.67) and (350.47,66.15) .. (393.05,66.15) .. controls (435.63,66.15) and (470.15,100.67) .. (470.15,143.25) .. controls (470.15,185.83) and (435.63,220.35) .. (393.05,220.35) .. controls (350.47,220.35) and (315.95,185.83) .. (315.95,143.25) -- cycle ;
	\draw  [fill={rgb, 255:red, 255; green, 255; blue, 255 }  ,fill opacity=1 ] (74.5,143.25) .. controls (74.5,112.02) and (99.82,86.7) .. (131.05,86.7) .. controls (162.28,86.7) and (187.6,112.02) .. (187.6,143.25) .. controls (187.6,174.48) and (162.28,199.8) .. (131.05,199.8) .. controls (99.82,199.8) and (74.5,174.48) .. (74.5,143.25) -- cycle ;
	\draw  [fill={rgb, 255:red, 157; green, 232; blue, 255 }  ,fill opacity=1 ] (107.4,143.25) .. controls (107.4,130.19) and (117.99,119.6) .. (131.05,119.6) .. controls (144.11,119.6) and (154.7,130.19) .. (154.7,143.25) .. controls (154.7,156.31) and (144.11,166.9) .. (131.05,166.9) .. controls (117.99,166.9) and (107.4,156.31) .. (107.4,143.25) -- cycle ; 
	\draw    (131.05,143.25) -- (144.91,128.84) ;
	\draw [shift={(146.3,127.4)}, rotate = 493.89] [color={rgb, 255:red, 0; green, 0; blue, 0 }  ][line width=0.75]    (10.93,-3.29) .. controls (6.95,-1.4) and (3.31,-0.3) .. (0,0) .. controls (3.31,0.3) and (6.95,1.4) .. (10.93,3.29)   ;
	\draw    (131.05,143.25) -- (84.07,118.34) ;
	\draw [shift={(82.3,117.4)}, rotate = 387.94] [color={rgb, 255:red, 0; green, 0; blue, 0 }  ][line width=0.75]    (10.93,-3.29) .. controls (6.95,-1.4) and (3.31,-0.3) .. (0,0) .. controls (3.31,0.3) and (6.95,1.4) .. (10.93,3.29)   ;
	\draw  [fill={rgb, 255:red, 255; green, 255; blue, 255 }  ,fill opacity=1 ] (336.5,143.25) .. controls (336.5,112.02) and (361.82,86.7) .. (393.05,86.7) .. controls (424.28,86.7) and (449.6,112.02) .. (449.6,143.25) .. controls (449.6,174.48) and (424.28,199.8) .. (393.05,199.8) .. controls (361.82,199.8) and (336.5,174.48) .. (336.5,143.25) -- cycle ;
	\draw  [fill={rgb, 255:red, 157; green, 232; blue, 255 }  ,fill opacity=1 ] (369.4,143.25) .. controls (369.4,130.19) and (379.99,119.6) .. (393.05,119.6) .. controls (406.11,119.6) and (416.7,130.19) .. (416.7,143.25) .. controls (416.7,156.31) and (406.11,166.9) .. (393.05,166.9) .. controls (379.99,166.9) and (369.4,156.31) .. (369.4,143.25) -- cycle ; 
	\draw    (393.05,143.25) -- (406.91,128.84) ;
	\draw [shift={(408.3,127.4)}, rotate = 493.89] [color={rgb, 255:red, 0; green, 0; blue, 0 }  ][line width=0.75]    (10.93,-3.29) .. controls (6.95,-1.4) and (3.31,-0.3) .. (0,0) .. controls (3.31,0.3) and (6.95,1.4) .. (10.93,3.29)   ;
	\draw    (393.05,143.25) -- (346.07,118.34) ;
	\draw [shift={(344.3,117.4)}, rotate = 387.94] [color={rgb, 255:red, 0; green, 0; blue, 0 }  ][line width=0.75]    (10.93,-3.29) .. controls (6.95,-1.4) and (3.31,-0.3) .. (0,0) .. controls (3.31,0.3) and (6.95,1.4) .. (10.93,3.29)   ;
	\draw    (393.05,143.25) -- (404.01,217.42) ;
	\draw [shift={(404.3,219.4)}, rotate = 261.6] [color={rgb, 255:red, 0; green, 0; blue, 0 }  ][line width=0.75]    (10.93,-3.29) .. controls (6.95,-1.4) and (3.31,-0.3) .. (0,0) .. controls (3.31,0.3) and (6.95,1.4) .. (10.93,3.29)   ;
	\draw (142,146) node   [align=left] {$n_1$};
	\draw (97,141) node   [align=left] {$n_2$};
	\draw (404,146) node   [align=left] {$n_1$};
	\draw (359,141) node   [align=left] {$n_2$};
	\draw (415,204) node   [align=left] {$n_3$};
	\draw (135,262) node   [align=left] {(a)};
	\draw (395,261) node   [align=left] {(b)};
	\end{tikzpicture}
	\caption{(a) The blue disk of radius $n_1$ indicates that the matrices up to depth $n_1$ are trained. The annulus between circles of radii $n_1$ and $n_2$ is the data used in prediction. The behavior of the machine is poor when predicting the matrices in this white annulus. (b) The blue disk and blue annulus indicate the seen matrices in the duality tree. The middle white annulus is used in prediction.}\label{annuli1}
\end{figure}
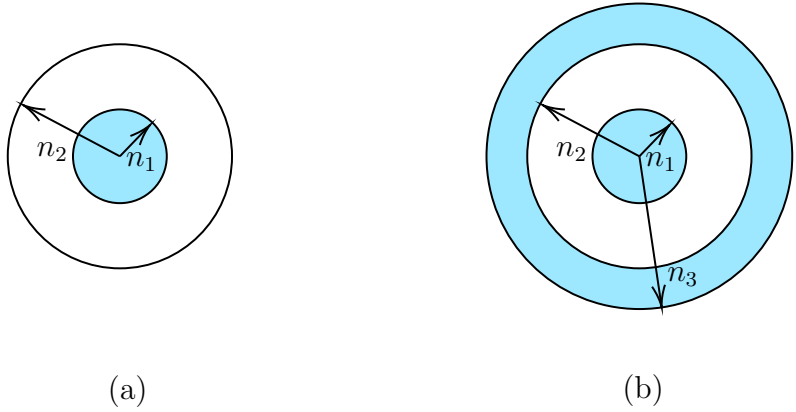

We can actually visualize the duality trees of quivers. Examples can be found in Fig. 2 and 7 in \cite{Franco:2003ja}. Since a mutation can act on every single node of a quiver, an $n$-node quiver is \emph{directly} connected to $n$ other dual quivers. This is true for any quiver in any mutation class. Furthermore, the duality tree of an infinite mutation class is apparently infinite. Thus, it does not matter which quiver we choose to start with due to the symmetry of the duality tree\footnote{For a finite mutation, this is also true as the duality tree will finally close and be symmetric.}. Now, from the example of Table \ref{depthfaraway}, we know that the machine is poor at predicting matrices of depths from ($n_1$+1) to $n_2$ when only matrices within depths $n_1$ are trained. This can be illustrated as in Fig. \ref{annuli1}(a).

\begin{figure}[h]
	\centering
	\tikzset{every picture/.style={line width=0.75pt}}       
	\begin{tikzpicture}[x=0.75pt,y=0.75pt,yscale=-1,xscale=1]
	\draw  [fill={rgb, 255:red, 157; green, 232; blue, 255 }  ,fill opacity=1 ] (313.95,142.25) .. controls (313.95,99.67) and (348.47,65.15) .. (391.05,65.15) .. controls (433.63,65.15) and (468.15,99.67) .. (468.15,142.25) .. controls (468.15,184.83) and (433.63,219.35) .. (391.05,219.35) .. controls (348.47,219.35) and (313.95,184.83) .. (313.95,142.25) -- cycle ;
	\draw  [fill={rgb, 255:red, 255; green, 255; blue, 255 }  ,fill opacity=1 ] (334.5,142.25) .. controls (334.5,111.02) and (359.82,85.7) .. (391.05,85.7) .. controls (422.28,85.7) and (447.6,111.02) .. (447.6,142.25) .. controls (447.6,173.48) and (422.28,198.8) .. (391.05,198.8) .. controls (359.82,198.8) and (334.5,173.48) .. (334.5,142.25) -- cycle ;
	\draw  [fill={rgb, 255:red, 255; green, 255; blue, 255 }  ,fill opacity=1 ] (305.05,191.7) .. controls (305.05,168.37) and (323.97,149.45) .. (347.3,149.45) .. controls (370.63,149.45) and (389.55,168.37) .. (389.55,191.7) .. controls (389.55,215.03) and (370.63,233.95) .. (347.3,233.95) .. controls (323.97,233.95) and (305.05,215.03) .. (305.05,191.7) -- cycle ;
	\draw  [fill={rgb, 255:red, 255; green, 255; blue, 255 }  ,fill opacity=1 ] (397.05,188.7) .. controls (397.05,165.37) and (415.97,146.45) .. (439.3,146.45) .. controls (462.63,146.45) and (481.55,165.37) .. (481.55,188.7) .. controls (481.55,212.03) and (462.63,230.95) .. (439.3,230.95) .. controls (415.97,230.95) and (397.05,212.03) .. (397.05,188.7) -- cycle ;
	\draw  [fill={rgb, 255:red, 157; green, 232; blue, 255 }  ,fill opacity=1 ] (53.95,143.25) .. controls (53.95,100.67) and (88.47,66.15) .. (131.05,66.15) .. controls (173.63,66.15) and (208.15,100.67) .. (208.15,143.25) .. controls (208.15,185.83) and (173.63,220.35) .. (131.05,220.35) .. controls (88.47,220.35) and (53.95,185.83) .. (53.95,143.25) -- cycle ;
	\draw  [fill={rgb, 255:red, 255; green, 255; blue, 255 }  ,fill opacity=1 ] (74.5,143.25) .. controls (74.5,112.02) and (99.82,86.7) .. (131.05,86.7) .. controls (162.28,86.7) and (187.6,112.02) .. (187.6,143.25) .. controls (187.6,174.48) and (162.28,199.8) .. (131.05,199.8) .. controls (99.82,199.8) and (74.5,174.48) .. (74.5,143.25) -- cycle ;
	\draw  [fill={rgb, 255:red, 255; green, 255; blue, 255 }  ,fill opacity=1 ] (137.8,97.7) .. controls (137.8,74.78) and (156.38,56.2) .. (179.3,56.2) .. controls (202.22,56.2) and (220.8,74.78) .. (220.8,97.7) .. controls (220.8,120.62) and (202.22,139.2) .. (179.3,139.2) .. controls (156.38,139.2) and (137.8,120.62) .. (137.8,97.7) -- cycle ;
	\draw  [fill={rgb, 255:red, 157; green, 232; blue, 255 }  ,fill opacity=1 ] (107.4,143.25) .. controls (107.4,130.19) and (117.99,119.6) .. (131.05,119.6) .. controls (144.11,119.6) and (154.7,130.19) .. (154.7,143.25) .. controls (154.7,156.31) and (144.11,166.9) .. (131.05,166.9) .. controls (117.99,166.9) and (107.4,156.31) .. (107.4,143.25) -- cycle ;
	\draw    (131.05,143.25) -- (84.07,118.34) ;
	\draw [shift={(82.3,117.4)}, rotate = 387.94] [color={rgb, 255:red, 0; green, 0; blue, 0 }  ][line width=0.75]    (10.93,-3.29) .. controls (6.95,-1.4) and (3.31,-0.3) .. (0,0) .. controls (3.31,0.3) and (6.95,1.4) .. (10.93,3.29)   ;
	\draw  [fill={rgb, 255:red, 26; green, 229; blue, 91 }  ,fill opacity=1 ] (169.6,97.7) .. controls (169.6,92.34) and (173.94,88) .. (179.3,88) .. controls (184.66,88) and (189,92.34) .. (189,97.7) .. controls (189,103.06) and (184.66,107.4) .. (179.3,107.4) .. controls (173.94,107.4) and (169.6,103.06) .. (169.6,97.7) -- cycle ;
	\draw  [dash pattern={on 4.5pt off 4.5pt}] (74.5,143.25) .. controls (74.5,112.02) and (99.82,86.7) .. (131.05,86.7) .. controls (162.28,86.7) and (187.6,112.02) .. (187.6,143.25) .. controls (187.6,174.48) and (162.28,199.8) .. (131.05,199.8) .. controls (99.82,199.8) and (74.5,174.48) .. (74.5,143.25) -- cycle ;
	\draw  [dash pattern={on 4.5pt off 4.5pt}] (53.95,143.25) .. controls (53.95,100.67) and (88.47,66.15) .. (131.05,66.15) .. controls (173.63,66.15) and (208.15,100.67) .. (208.15,143.25) .. controls (208.15,185.83) and (173.63,220.35) .. (131.05,220.35) .. controls (88.47,220.35) and (53.95,185.83) .. (53.95,143.25) -- cycle ; 
	\draw    (179.3,97.7) -- (150.72,125.99) ;
	\draw [shift={(149.3,127.4)}, rotate = 315.28999999999996] [color={rgb, 255:red, 0; green, 0; blue, 0 }  ][line width=0.75]    (10.93,-3.29) .. controls (6.95,-1.4) and (3.31,-0.3) .. (0,0) .. controls (3.31,0.3) and (6.95,1.4) .. (10.93,3.29)   ;
	\draw  [fill={rgb, 255:red, 255; green, 255; blue, 255 }  ,fill opacity=1 ] (397.8,96.7) .. controls (397.8,73.78) and (416.38,55.2) .. (439.3,55.2) .. controls (462.22,55.2) and (480.8,73.78) .. (480.8,96.7) .. controls (480.8,119.62) and (462.22,138.2) .. (439.3,138.2) .. controls (416.38,138.2) and (397.8,119.62) .. (397.8,96.7) -- cycle ; 
	\draw  [fill={rgb, 255:red, 157; green, 232; blue, 255 }  ,fill opacity=1 ] (367.4,142.25) .. controls (367.4,129.19) and (377.99,118.6) .. (391.05,118.6) .. controls (404.11,118.6) and (414.7,129.19) .. (414.7,142.25) .. controls (414.7,155.31) and (404.11,165.9) .. (391.05,165.9) .. controls (377.99,165.9) and (367.4,155.31) .. (367.4,142.25) -- cycle ; 
	\draw    (391.05,142.25) -- (344.07,117.34) ;
	\draw [shift={(342.3,116.4)}, rotate = 387.94] [color={rgb, 255:red, 0; green, 0; blue, 0 }  ][line width=0.75]    (10.93,-3.29) .. controls (6.95,-1.4) and (3.31,-0.3) .. (0,0) .. controls (3.31,0.3) and (6.95,1.4) .. (10.93,3.29)   ;
	\draw  [fill={rgb, 255:red, 26; green, 229; blue, 91 }  ,fill opacity=1 ] (429.6,96.7) .. controls (429.6,91.34) and (433.94,87) .. (439.3,87) .. controls (444.66,87) and (449,91.34) .. (449,96.7) .. controls (449,102.06) and (444.66,106.4) .. (439.3,106.4) .. controls (433.94,106.4) and (429.6,102.06) .. (429.6,96.7) -- cycle ;
	\draw    (439.3,96.7) -- (410.72,124.99) ;
	\draw [shift={(409.3,126.4)}, rotate = 315.28999999999996] [color={rgb, 255:red, 0; green, 0; blue, 0 }  ][line width=0.75]    (10.93,-3.29) .. controls (6.95,-1.4) and (3.31,-0.3) .. (0,0) .. controls (3.31,0.3) and (6.95,1.4) .. (10.93,3.29)   ;
	\draw  [fill={rgb, 255:red, 26; green, 229; blue, 91 }  ,fill opacity=1 ] (429.6,188.7) .. controls (429.6,183.34) and (433.94,179) .. (439.3,179) .. controls (444.66,179) and (449,183.34) .. (449,188.7) .. controls (449,194.06) and (444.66,198.4) .. (439.3,198.4) .. controls (433.94,198.4) and (429.6,194.06) .. (429.6,188.7) -- cycle ;
	\draw    (439.3,188.7) -- (400.25,179.84) ;
	\draw [shift={(398.3,179.4)}, rotate = 372.78] [color={rgb, 255:red, 0; green, 0; blue, 0 }  ][line width=0.75]    (10.93,-3.29) .. controls (6.95,-1.4) and (3.31,-0.3) .. (0,0) .. controls (3.31,0.3) and (6.95,1.4) .. (10.93,3.29)   ; 
	\draw  [fill={rgb, 255:red, 26; green, 229; blue, 91 }  ,fill opacity=1 ] (337.6,191.7) .. controls (337.6,186.34) and (341.94,182) .. (347.3,182) .. controls (352.66,182) and (357,186.34) .. (357,191.7) .. controls (357,197.06) and (352.66,201.4) .. (347.3,201.4) .. controls (341.94,201.4) and (337.6,197.06) .. (337.6,191.7) -- cycle ;
	\draw  [dash pattern={on 4.5pt off 4.5pt}] (334.5,142.25) .. controls (334.5,111.02) and (359.82,85.7) .. (391.05,85.7) .. controls (422.28,85.7) and (447.6,111.02) .. (447.6,142.25) .. controls (447.6,173.48) and (422.28,198.8) .. (391.05,198.8) .. controls (359.82,198.8) and (334.5,173.48) .. (334.5,142.25) -- cycle ;
	\draw  [dash pattern={on 4.5pt off 4.5pt}] (313.95,142.25) .. controls (313.95,99.67) and (348.47,65.15) .. (391.05,65.15) .. controls (433.63,65.15) and (468.15,99.67) .. (468.15,142.25) .. controls (468.15,184.83) and (433.63,219.35) .. (391.05,219.35) .. controls (348.47,219.35) and (313.95,184.83) .. (313.95,142.25) -- cycle ;
	\draw    (347.3,191.7) -- (382.51,174.29) ;
	\draw [shift={(384.3,173.4)}, rotate = 513.6800000000001] [color={rgb, 255:red, 0; green, 0; blue, 0 }  ][line width=0.75]    (10.93,-3.29) .. controls (6.95,-1.4) and (3.31,-0.3) .. (0,0) .. controls (3.31,0.3) and (6.95,1.4) .. (10.93,3.29)   ;
	\draw (97,141) node   [align=left] {$n_2$};
	\draw (135,262) node   [align=left] {(a)};
	\draw (395,261) node   [align=left] {(b)};
	\draw (171,121) node   [align=left] {$n_4$};
	\draw (357,140) node   [align=left] {$n_2$};
	\draw (431,120) node   [align=left] {$n_4$};
	\draw (420,175) node   [align=left] {$n_4$};
	\draw (363,173) node   [align=left] {$n_4$};
	\end{tikzpicture}
	\caption{(a) We can choose a matrix in the blue annulus and generate the green disks. Then the overlap of disks with radii $n_2$ and $n_4$ form a leaf shape, whose interior consists of unseen matrices. (b) We can draw all the green disks along the blue annulus. Each disk is contained in a white disk of radius $n_4$. The green disks then form the blue annulus, and the leaf-shaped overlaps form the big white annulus in the middle.}\label{annuli2}
\end{figure}
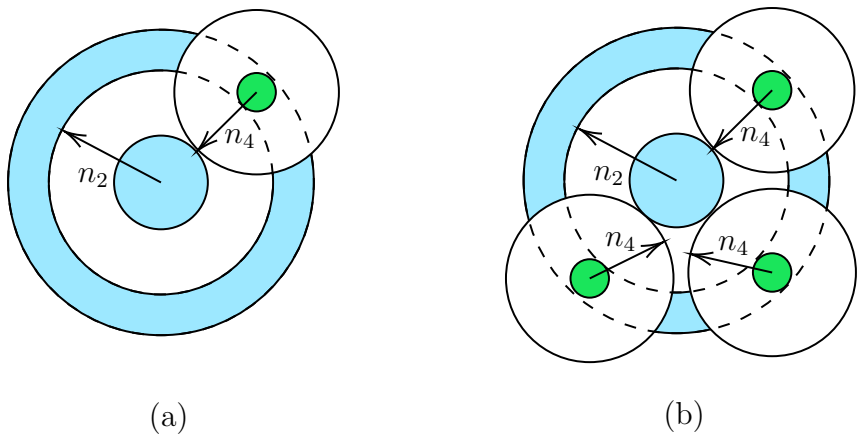

Likewise, for the example of Table \ref{depthbetween}, we have Fig. \ref{annuli1}(b). Then we can have a green disk of trained matrices, centered at each point (up to the azimuth) in the blue annulus, tangent to the two boundaries of the blue annulus as shown in Fig. \ref{annuli2}(a).
We can use such trained green disk/dataset to predict the matrices inside the white annulus bounded by the green disk and the disk of radius $n_4$. By the same reasoning, the machine would give poor predictions to those matrices. Notice that the disks of radii $n_2$ and $n_4$ have a leaf-shaped overlap, which means that given the small blue disk and the green disk as the training set, this leaf would not enjoy a good prediction. If we draw the green disk along the blue annulus, then those green disks, along with the blue disk in the middle, will become the same training set as in Fig. \ref{annuli1}(b). The leaf-shaped overlaps will form the white annulus in the middle bounded by blue disk and the blue annulus, which is the unseen dataset as in Fig. \ref{annuli1}(b). Since the machine cannot learn well in the leaf shapes, although the training set is larger (compared to Fig. \ref{annuli1}(a)) which may improve the result, as a consequence of mutual independence assumption, the performance of the machine would still not be greatly improved. Nevertheless, we should emphasize that this is mainly due to the particular feature of NB. As we will see in \S\ref{multiclass}, this illustration for NB here would be quite different for Neural Networks (NN).

\subsection{Classifying More Mutation Classes} \label{moreclasses}
We now contemplate the datasets containing more mutation classes. It is natural to first consider the case with three mutation classes. We again use [`A',6], [`D',6] and [`E',6] as an example. Of course, unlike the aforementioned case, all the three classes have to appear in the training dataset this time. The learning result of 5-fold cross validation is reported in Table \ref{ADE6seen5fold}.
\begin{table}[H]
	\centering
	\begin{tabular}{|c|c|c|}
		\hline
		Accuracy	& F-Score & $\phi$ \\ \hline
		0.90291800$\pm$0.00920160 & 0.90936100$\pm$0.00886124 & 0.81580000$\pm$0.01625320 \\ \hline
	\end{tabular}
	\caption{Training and validating three classes: [`A',6], [`D',6], [`E',6]. We generate (76+77+77) matrices. There are 6116 1's and 6049 0's. The method is NB.}\label{ADE6seen5fold}
\end{table}
\noindent The learning curve at different training percentage is given in Fig. \ref{ADE6seencurve}.
We can see that the performance, albeit not as perfect as the cases with two classes, is still very satisfying, with $\sim$90\% accuracies and $\sim$0.8 Matthews correlation coefficients when only $\sim$60\% of the data is trained.

We can also add one more class into the two-class example for \textbf{Q4} and \textbf{Q5}. The new one is generated by \textbf{Q6}.
The learning results are reported in Table \ref{F0inf1inf25fold} for 5-fold cross validation and Fig. \ref{F0inf1inf2curve} for learning curves.
\begin{table}[h]
	\centering
	\begin{tabular}{|c|c|c|}
		\hline
		Accuracy	& F-Score & $\phi$ \\ \hline
		0.90553300$\pm$0.00970378	& 0.91187400$\pm$0.00831757 & 0.82051800$\pm$0.01696320 \\ \hline
	\end{tabular}
	\caption{Training and validating three classes: \textbf{Q4}, \textbf{Q5}, and \textbf{Q6}. We generate (102+138+161) matrices. There are 11563 1's and 11482 0's. The method is NB.}\label{F0inf1inf25fold}
\end{table}
The performance is still very nice, though it is not as perfect as the two-example class.

Let us now contemplate examples with four and five mutation classes. To compare this with the three-class example above, we first choose \textbf{Q4}, \textbf{Q5}, and \textbf{Q6} for our data. For the four-class example, the remaining quiver is depicted in \textbf{Q7}.

The learning results are reported in Table \ref{fourclass5fold} for 5-fold cross validation and Fig. \ref{fourclasscurve} for learning curves.
\begin{table}[h]
	\centering
	\begin{tabular}{|c|c|c|}
		\hline
		Accuracy	& F-Score & $\phi$ \\ \hline
		0.85739200$\pm$0.00750116	& 0.86872800$\pm$0.00563417 & 0.72165300$\pm$0.01548070 \\ \hline
	\end{tabular}
	\caption{Training and validating four classes: \textbf{Q4}, \textbf{Q5}, \textbf{Q6}, and \textbf{Q7}. We generate (102+138+161+102) matrices. There are 16059 1's and 16250 0's. The method is NB.}\label{fourclass5fold}
\end{table}

For the five-class example, we further include \textbf{Q8}.
The learning results are reported in Table \ref{fiveclass5fold} for 5-fold cross validation and Fig. \ref{fiveclasscurve} for learning curves.
\begin{table}[h!]
	\centering
	\begin{tabular}{|c|c|c|}
		\hline
		Accuracy	& F-Score & $\phi$ \\ \hline
		0.83572900$\pm$0.00292061	& 0.84193700$\pm$0.00320654 & 0.67346000$\pm$0.00515131 \\ \hline
	\end{tabular}
	\caption{Training and validating five classes: \textbf{Q4}, \textbf{Q5}, \textbf{Q6}, \textbf{Q7}, and \textbf{Q8}. We generate (102+138+161+102+161) matrices. There are 23645 1's and 23698 0's. The method is NB.}\label{fiveclass5fold}
\end{table}
Indeed, we see that the numbers of different classes can affect the performance of the machine.

Nevertheless, a better learning result is always wanted. When we are having more classes, a combinatorial problem arises. If there are more mutation classes in the data, there will be more and more distinct pairs of 0's than pairs of 1's. If we want adequate combinations of 0's, then to keep the dataset well-balanced, correspondingly many 1's are required as well. However, all the distinct pairs of 1's will be included while 0's may still not be enough. On the other hand, if we keep adding pairs to our dataset, although we will have more combinations of 0's, there will be duplicated pairs of 1's. These repeated pairs will not be helpful and hence the dataset will be biased. Thus, how the number of mutation classes is (quantitatively) related to the number of matrices generated and the number of pairs assigned is a newly raised question. Roughly speaking, the best way is perhaps to include all the 1's and correspondingly many 0's. Then the number of distinct pairs is maximized while keeping the dataset balanced. Another possible way to resolve this is to use multiclassification with one single matrix as a data point instead of matrix pairs so that the combinatorial problem could be avoided. Let us now contemplate such multiclassifications.

\subsection{Multiclass Classifications}\label{multiclass}
For datasets consisting of matrix pairs, we have already seen that NB is the best method for learning mutations. To make this more convincing and more clear, we also plot the learning curves with different methods in Fig. \ref{compare} as an example\footnote{At first, we would like to try much more matrices and much larger datasets. However, a normal laptop is not capable of giving the whole learning curve of SVM. Nevertheless, this example with a smaller size can still tell the difference between various methods. Here, although random forest is still inferior to NB, the discrepancy is small. However, one can check that if we include more matrices and more data, the advantage of NB over other methods will be greater.}. We also tabulate the 5-fold cross validation for NN in Table \ref{NN5fold}. We should emphasize that the NN here used in $\mathtt{Mathematica}$ is different from the (C)NN we will use below for multiclassifications. The NN in $\mathtt{Mathematica}$ $\mathtt{Classify}$ is used for matrix pairs while NN in $\mathtt{Python}$ deals with single matrix as one datapoint in the dataset\footnote{We should mention that $\mathtt{Mathematica}$ now also incorporates complicated neural network, though we are using $\mathtt{Python}$ here for CNN to make a more clear distinction between binary and multi classifications in our discussions.}. Unless specified, we will always refer to multiclassifications in $\mathtt{Python}$ when saying NN below.
\begin{figure}[h]
	\centering
	\begin{tikzpicture}
	\begin{axis}[ymin=0, ymax=1.02,
	width=0.75\textwidth,
	height=0.5\textwidth,
	ytick={0,0.1,...,1.1}, ytick align=inside, ytick pos=left,
	xtick={0,10,...,100}, xtick align=inside, xtick pos=left,
	xlabel=Training(\%),
	ylabel=Accuracy,
	grid=major,
	grid style={dashed, gray!30},
	legend pos=south east,
	legend style={draw=none}]
	\addplot+[
	blue, mark options={blue, scale=1},
	smooth, 
	error bars/.cd, 
	y fixed,
	y dir=both, 
	y explicit
	] table [x=x, y=y,y error=error, col sep=comma] {
		x,  y,        error
		10, 0.788203, 0.00976359
		20, 0.849317, 0.0038057
		30, 0.871109, 0.00689812
		40, 0.88607 , 0.00892426
		50, 0.892907, 0.00659157
		60, 0.897624, 0.00508019
		70, 0.901397, 0.00514787
		80, 0.905616, 0.00821445
		90, 0.905572, 0.00306361
	};
	\addlegendentry{NB}
	\addplot+[
	green, mark options={green, scale=1},
	smooth, 
	error bars/.cd, 
	y fixed,
	y dir=both, 
	y explicit
	] table [x=x, y=y,y error=error, col sep=comma] {
		x,  y,        error
		10, 0.654036, 0.0115314
		20, 0.729081, 0.0101981
		30, 0.751455, 0.00930238
		40, 0.80224 , 0.0167716
		50, 0.813608, 0.00917948
		60, 0.832021, 0.0180664
		70, 0.858083, 0.0175952
		80, 0.868067, 0.0177113
		90, 0.876134, 0.0194484
	};
	\addlegendentry{Random Forest}
	\addplot+[
	red, mark options={red, scale=1},mark=triangle*,
	smooth, 
	error bars/.cd, 
	y fixed,
	y dir=both, 
	y explicit
	] table [x=x, y=y,y error=error, col sep=comma] {
		x,  y,        error
		10, 0.568315, 0.0248532
		20, 0.559711, 0.0184656
		30, 0.59712 , 0.0267956
		40, 0.657683, 0.0222722
		50, 0.715498, 0.0527283
		60, 0.698373, 0.0325451
		70, 0.720364, 0.059091
		80, 0.801891, 0.0486697
		90, 0.736134, 0.103354
	};
	\addlegendentry{NN}
	\addplot+[
	brown, mark options={brown, scale=1},mark=diamond*,
	smooth, 
	error bars/.cd, 
	y fixed,
	y dir=both, 
	y explicit
	] table [x=x, y=y,y error=error, col sep=comma] {
		x,  y,        error
		10, 0.487634, 0.0182175
		20, 0.60357 , 0.0502131
		30, 0.544134, 0.0370762
		40, 0.551278, 0.0558001
		50, 0.521126, 0.0376653
		60, 0.607559, 0.0466463
		70, 0.585864, 0.0912071
		80, 0.666807, 0.0110768
		90, 0.62521 , 0.0661448
	};
	\addlegendentry{SVM}
	\end{axis}
	\end{tikzpicture}
	\caption{Training and validating three classes: \textbf{Q4}, \textbf{Q5}, and \textbf{Q6}. We generate (38+48+53) matrices. There are 2365 1's and 2397 0's.}\label{compare}
\end{figure}
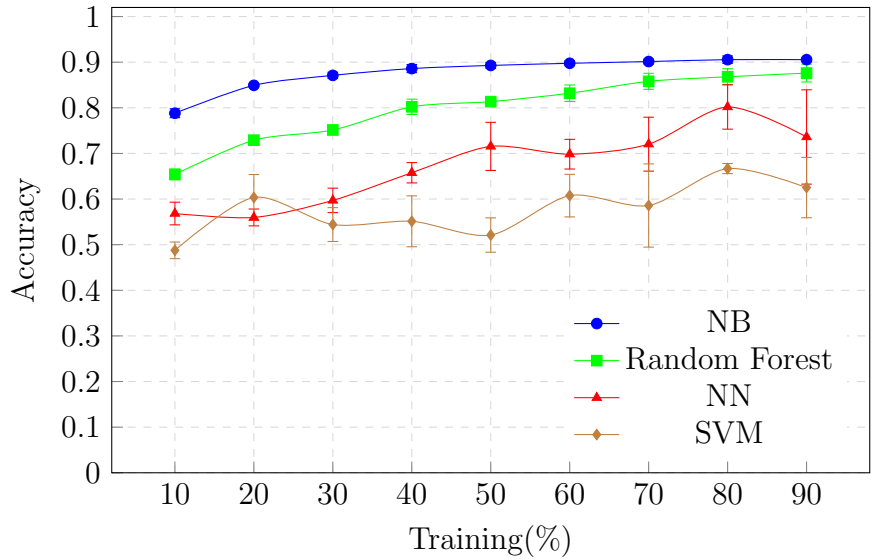
\begin{table}[H]
	\centering
	\begin{tabular}{|c|c|c|}
		\hline
		Accuracy	& F-Score & $\phi$ \\ \hline
		0.76590900$\pm$0.05281270	& 0.77165100$\pm$0.04618850 & 0.53412100$\pm$0.10287100 \\ \hline
	\end{tabular}
	\caption{Training on 3 infinite type quivers using the Neural network method within Mathematica's classify function.}\label{NN5fold}
\end{table}

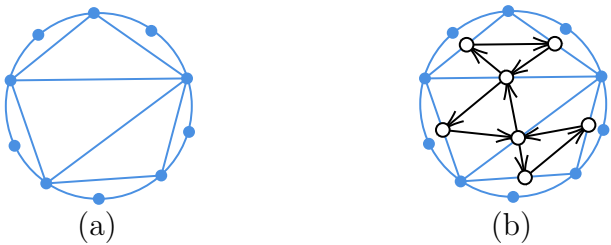
\begin{figure}[h]
	\centering
	\tikzset{every picture/.style={line width=0.75pt}}      
	\begin{tikzpicture}[x=0.75pt,y=0.75pt,yscale=-1,xscale=1]
	\draw    (361.85,195) -- (392.36,169.77) ;
	\draw [shift={(393.9,168.5)}, rotate = 500.41] [color={rgb, 255:red, 0; green, 0; blue, 0 }  ][line width=0.75]    (10.93,-3.29) .. controls (6.95,-1.4) and (3.31,-0.3) .. (0,0) .. controls (3.31,0.3) and (6.95,1.4) .. (10.93,3.29)   ;
	\draw    (358.4,175) -- (361.51,193.03) ;
	\draw [shift={(361.85,195)}, rotate = 260.21] [color={rgb, 255:red, 0; green, 0; blue, 0 }  ][line width=0.75]    (10.93,-3.29) .. controls (6.95,-1.4) and (3.31,-0.3) .. (0,0) .. controls (3.31,0.3) and (6.95,1.4) .. (10.93,3.29)   ;
	\draw    (393.9,168.5) -- (360.37,174.64) ;
	\draw [shift={(358.4,175)}, rotate = 349.62] [color={rgb, 255:red, 0; green, 0; blue, 0 }  ][line width=0.75]    (10.93,-3.29) .. controls (6.95,-1.4) and (3.31,-0.3) .. (0,0) .. controls (3.31,0.3) and (6.95,1.4) .. (10.93,3.29)   ; 
	\draw    (358.4,175) -- (352.83,146.46) ;
	\draw [shift={(352.45,144.5)}, rotate = 438.96] [color={rgb, 255:red, 0; green, 0; blue, 0 }  ][line width=0.75]    (10.93,-3.29) .. controls (6.95,-1.4) and (3.31,-0.3) .. (0,0) .. controls (3.31,0.3) and (6.95,1.4) .. (10.93,3.29)   ;
	\draw    (320.4,171) -- (356.41,174.79) ;
	\draw [shift={(358.4,175)}, rotate = 186.01] [color={rgb, 255:red, 0; green, 0; blue, 0 }  ][line width=0.75]    (10.93,-3.29) .. controls (6.95,-1.4) and (3.31,-0.3) .. (0,0) .. controls (3.31,0.3) and (6.95,1.4) .. (10.93,3.29)   ; 
	\draw    (352.45,144.5) -- (321.94,169.73) ;
	\draw [shift={(320.4,171)}, rotate = 320.40999999999997] [color={rgb, 255:red, 0; green, 0; blue, 0 }  ][line width=0.75]    (10.93,-3.29) .. controls (6.95,-1.4) and (3.31,-0.3) .. (0,0) .. controls (3.31,0.3) and (6.95,1.4) .. (10.93,3.29)   ;
	\draw    (352.45,144.5) -- (333.94,129.27) ;
	\draw [shift={(332.4,128)}, rotate = 399.45] [color={rgb, 255:red, 0; green, 0; blue, 0 }  ][line width=0.75]    (10.93,-3.29) .. controls (6.95,-1.4) and (3.31,-0.3) .. (0,0) .. controls (3.31,0.3) and (6.95,1.4) .. (10.93,3.29)   ;
	\draw    (376.9,127.5) -- (354.09,143.36) ;
	\draw [shift={(352.45,144.5)}, rotate = 325.19] [color={rgb, 255:red, 0; green, 0; blue, 0 }  ][line width=0.75]    (10.93,-3.29) .. controls (6.95,-1.4) and (3.31,-0.3) .. (0,0) .. controls (3.31,0.3) and (6.95,1.4) .. (10.93,3.29)   ;
	\draw    (332.4,128) -- (374.9,127.52) ;
	\draw [shift={(376.9,127.5)}, rotate = 539.36] [color={rgb, 255:red, 0; green, 0; blue, 0 }  ][line width=0.75]    (10.93,-3.29) .. controls (6.95,-1.4) and (3.31,-0.3) .. (0,0) .. controls (3.31,0.3) and (6.95,1.4) .. (10.93,3.29)   ;
	\draw [color={rgb, 255:red, 74; green, 144; blue, 226 }  ,draw opacity=1 ]   (102.4,146) -- (144.4,112) ;
	\draw  [color={rgb, 255:red, 74; green, 144; blue, 226 }  ,draw opacity=1 ] (100,158.9) .. controls (100,133) and (121,112) .. (146.9,112) .. controls (172.8,112) and (193.8,133) .. (193.8,158.9) .. controls (193.8,184.8) and (172.8,205.8) .. (146.9,205.8) .. controls (121,205.8) and (100,184.8) .. (100,158.9) -- cycle ;
	\draw  [color={rgb, 255:red, 74; green, 144; blue, 226 }  ,draw opacity=1 ][fill={rgb, 255:red, 74; green, 144; blue, 226 }  ,fill opacity=1 ] (146.9,112) .. controls (146.9,110.62) and (145.78,109.5) .. (144.4,109.5) .. controls (143.02,109.5) and (141.9,110.62) .. (141.9,112) .. controls (141.9,113.38) and (143.02,114.5) .. (144.4,114.5) .. controls (145.78,114.5) and (146.9,113.38) .. (146.9,112) -- cycle ;
	\draw  [color={rgb, 255:red, 74; green, 144; blue, 226 }  ,draw opacity=1 ][fill={rgb, 255:red, 74; green, 144; blue, 226 }  ,fill opacity=1 ] (104.9,146) .. controls (104.9,144.62) and (103.78,143.5) .. (102.4,143.5) .. controls (101.02,143.5) and (99.9,144.62) .. (99.9,146) .. controls (99.9,147.38) and (101.02,148.5) .. (102.4,148.5) .. controls (103.78,148.5) and (104.9,147.38) .. (104.9,146) -- cycle ;
	\draw  [color={rgb, 255:red, 74; green, 144; blue, 226 }  ,draw opacity=1 ][fill={rgb, 255:red, 74; green, 144; blue, 226 }  ,fill opacity=1 ] (122.9,198) .. controls (122.9,196.62) and (121.78,195.5) .. (120.4,195.5) .. controls (119.02,195.5) and (117.9,196.62) .. (117.9,198) .. controls (117.9,199.38) and (119.02,200.5) .. (120.4,200.5) .. controls (121.78,200.5) and (122.9,199.38) .. (122.9,198) -- cycle ;
	\draw  [color={rgb, 255:red, 74; green, 144; blue, 226 }  ,draw opacity=1 ][fill={rgb, 255:red, 74; green, 144; blue, 226 }  ,fill opacity=1 ] (180.9,194) .. controls (180.9,192.62) and (179.78,191.5) .. (178.4,191.5) .. controls (177.02,191.5) and (175.9,192.62) .. (175.9,194) .. controls (175.9,195.38) and (177.02,196.5) .. (178.4,196.5) .. controls (179.78,196.5) and (180.9,195.38) .. (180.9,194) -- cycle ;
	\draw  [color={rgb, 255:red, 74; green, 144; blue, 226 }  ,draw opacity=1 ][fill={rgb, 255:red, 74; green, 144; blue, 226 }  ,fill opacity=1 ] (193.9,145) .. controls (193.9,143.62) and (192.78,142.5) .. (191.4,142.5) .. controls (190.02,142.5) and (188.9,143.62) .. (188.9,145) .. controls (188.9,146.38) and (190.02,147.5) .. (191.4,147.5) .. controls (192.78,147.5) and (193.9,146.38) .. (193.9,145) -- cycle ;
	\draw  [color={rgb, 255:red, 74; green, 144; blue, 226 }  ,draw opacity=1 ][fill={rgb, 255:red, 74; green, 144; blue, 226 }  ,fill opacity=1 ] (118.9,123) .. controls (118.9,121.62) and (117.78,120.5) .. (116.4,120.5) .. controls (115.02,120.5) and (113.9,121.62) .. (113.9,123) .. controls (113.9,124.38) and (115.02,125.5) .. (116.4,125.5) .. controls (117.78,125.5) and (118.9,124.38) .. (118.9,123) -- cycle ;
	\draw  [color={rgb, 255:red, 74; green, 144; blue, 226 }  ,draw opacity=1 ][fill={rgb, 255:red, 74; green, 144; blue, 226 }  ,fill opacity=1 ] (106.9,179) .. controls (106.9,177.62) and (105.78,176.5) .. (104.4,176.5) .. controls (103.02,176.5) and (101.9,177.62) .. (101.9,179) .. controls (101.9,180.38) and (103.02,181.5) .. (104.4,181.5) .. controls (105.78,181.5) and (106.9,180.38) .. (106.9,179) -- cycle ; 
	\draw  [color={rgb, 255:red, 74; green, 144; blue, 226 }  ,draw opacity=1 ][fill={rgb, 255:red, 74; green, 144; blue, 226 }  ,fill opacity=1 ] (149.4,205.8) .. controls (149.4,204.42) and (148.28,203.3) .. (146.9,203.3) .. controls (145.52,203.3) and (144.4,204.42) .. (144.4,205.8) .. controls (144.4,207.18) and (145.52,208.3) .. (146.9,208.3) .. controls (148.28,208.3) and (149.4,207.18) .. (149.4,205.8) -- cycle ;
	\draw  [color={rgb, 255:red, 74; green, 144; blue, 226 }  ,draw opacity=1 ][fill={rgb, 255:red, 74; green, 144; blue, 226 }  ,fill opacity=1 ] (194.9,172) .. controls (194.9,170.62) and (193.78,169.5) .. (192.4,169.5) .. controls (191.02,169.5) and (189.9,170.62) .. (189.9,172) .. controls (189.9,173.38) and (191.02,174.5) .. (192.4,174.5) .. controls (193.78,174.5) and (194.9,173.38) .. (194.9,172) -- cycle ;
	\draw  [color={rgb, 255:red, 74; green, 144; blue, 226 }  ,draw opacity=1 ][fill={rgb, 255:red, 74; green, 144; blue, 226 }  ,fill opacity=1 ] (175.9,121) .. controls (175.9,119.62) and (174.78,118.5) .. (173.4,118.5) .. controls (172.02,118.5) and (170.9,119.62) .. (170.9,121) .. controls (170.9,122.38) and (172.02,123.5) .. (173.4,123.5) .. controls (174.78,123.5) and (175.9,122.38) .. (175.9,121) -- cycle ;
	\draw [color={rgb, 255:red, 74; green, 144; blue, 226 }  ,draw opacity=1 ]   (120.4,198) -- (102.4,146) ;
	\draw [color={rgb, 255:red, 74; green, 144; blue, 226 }  ,draw opacity=1 ]   (120.4,198) -- (178.4,194) ;
	\draw [color={rgb, 255:red, 74; green, 144; blue, 226 }  ,draw opacity=1 ]   (178.4,194) -- (191.4,145) ;
	\draw [color={rgb, 255:red, 74; green, 144; blue, 226 }  ,draw opacity=1 ]   (191.4,145) -- (144.4,112) ;
	\draw [color={rgb, 255:red, 74; green, 144; blue, 226 }  ,draw opacity=1 ]   (102.4,146) -- (191.4,145) ;
	\draw [color={rgb, 255:red, 74; green, 144; blue, 226 }  ,draw opacity=1 ]   (120.4,198) -- (191.4,145) ;
	\draw [color={rgb, 255:red, 74; green, 144; blue, 226 }  ,draw opacity=1 ]   (311.4,145) -- (353.4,111) ;
	\draw  [color={rgb, 255:red, 74; green, 144; blue, 226 }  ,draw opacity=1 ] (309,157.9) .. controls (309,132) and (330,111) .. (355.9,111) .. controls (381.8,111) and (402.8,132) .. (402.8,157.9) .. controls (402.8,183.8) and (381.8,204.8) .. (355.9,204.8) .. controls (330,204.8) and (309,183.8) .. (309,157.9) -- cycle ;
	\draw  [color={rgb, 255:red, 74; green, 144; blue, 226 }  ,draw opacity=1 ][fill={rgb, 255:red, 74; green, 144; blue, 226 }  ,fill opacity=1 ] (355.9,111) .. controls (355.9,109.62) and (354.78,108.5) .. (353.4,108.5) .. controls (352.02,108.5) and (350.9,109.62) .. (350.9,111) .. controls (350.9,112.38) and (352.02,113.5) .. (353.4,113.5) .. controls (354.78,113.5) and (355.9,112.38) .. (355.9,111) -- cycle ;
	\draw  [color={rgb, 255:red, 74; green, 144; blue, 226 }  ,draw opacity=1 ][fill={rgb, 255:red, 74; green, 144; blue, 226 }  ,fill opacity=1 ] (313.9,145) .. controls (313.9,143.62) and (312.78,142.5) .. (311.4,142.5) .. controls (310.02,142.5) and (308.9,143.62) .. (308.9,145) .. controls (308.9,146.38) and (310.02,147.5) .. (311.4,147.5) .. controls (312.78,147.5) and (313.9,146.38) .. (313.9,145) -- cycle ;
	\draw  [color={rgb, 255:red, 74; green, 144; blue, 226 }  ,draw opacity=1 ][fill={rgb, 255:red, 74; green, 144; blue, 226 }  ,fill opacity=1 ] (331.9,197) .. controls (331.9,195.62) and (330.78,194.5) .. (329.4,194.5) .. controls (328.02,194.5) and (326.9,195.62) .. (326.9,197) .. controls (326.9,198.38) and (328.02,199.5) .. (329.4,199.5) .. controls (330.78,199.5) and (331.9,198.38) .. (331.9,197) -- cycle ;
	\draw  [color={rgb, 255:red, 74; green, 144; blue, 226 }  ,draw opacity=1 ][fill={rgb, 255:red, 74; green, 144; blue, 226 }  ,fill opacity=1 ] (389.9,193) .. controls (389.9,191.62) and (388.78,190.5) .. (387.4,190.5) .. controls (386.02,190.5) and (384.9,191.62) .. (384.9,193) .. controls (384.9,194.38) and (386.02,195.5) .. (387.4,195.5) .. controls (388.78,195.5) and (389.9,194.38) .. (389.9,193) -- cycle ;
	\draw  [color={rgb, 255:red, 74; green, 144; blue, 226 }  ,draw opacity=1 ][fill={rgb, 255:red, 74; green, 144; blue, 226 }  ,fill opacity=1 ] (402.9,144) .. controls (402.9,142.62) and (401.78,141.5) .. (400.4,141.5) .. controls (399.02,141.5) and (397.9,142.62) .. (397.9,144) .. controls (397.9,145.38) and (399.02,146.5) .. (400.4,146.5) .. controls (401.78,146.5) and (402.9,145.38) .. (402.9,144) -- cycle ; 
	\draw  [color={rgb, 255:red, 74; green, 144; blue, 226 }  ,draw opacity=1 ][fill={rgb, 255:red, 74; green, 144; blue, 226 }  ,fill opacity=1 ] (327.9,122) .. controls (327.9,120.62) and (326.78,119.5) .. (325.4,119.5) .. controls (324.02,119.5) and (322.9,120.62) .. (322.9,122) .. controls (322.9,123.38) and (324.02,124.5) .. (325.4,124.5) .. controls (326.78,124.5) and (327.9,123.38) .. (327.9,122) -- cycle ;
	\draw  [color={rgb, 255:red, 74; green, 144; blue, 226 }  ,draw opacity=1 ][fill={rgb, 255:red, 74; green, 144; blue, 226 }  ,fill opacity=1 ] (315.9,178) .. controls (315.9,176.62) and (314.78,175.5) .. (313.4,175.5) .. controls (312.02,175.5) and (310.9,176.62) .. (310.9,178) .. controls (310.9,179.38) and (312.02,180.5) .. (313.4,180.5) .. controls (314.78,180.5) and (315.9,179.38) .. (315.9,178) -- cycle ;
	\draw  [color={rgb, 255:red, 74; green, 144; blue, 226 }  ,draw opacity=1 ][fill={rgb, 255:red, 74; green, 144; blue, 226 }  ,fill opacity=1 ] (358.4,204.8) .. controls (358.4,203.42) and (357.28,202.3) .. (355.9,202.3) .. controls (354.52,202.3) and (353.4,203.42) .. (353.4,204.8) .. controls (353.4,206.18) and (354.52,207.3) .. (355.9,207.3) .. controls (357.28,207.3) and (358.4,206.18) .. (358.4,204.8) -- cycle ;
	\draw  [color={rgb, 255:red, 74; green, 144; blue, 226 }  ,draw opacity=1 ][fill={rgb, 255:red, 74; green, 144; blue, 226 }  ,fill opacity=1 ] (403.9,171) .. controls (403.9,169.62) and (402.78,168.5) .. (401.4,168.5) .. controls (400.02,168.5) and (398.9,169.62) .. (398.9,171) .. controls (398.9,172.38) and (400.02,173.5) .. (401.4,173.5) .. controls (402.78,173.5) and (403.9,172.38) .. (403.9,171) -- cycle ;
	\draw  [color={rgb, 255:red, 74; green, 144; blue, 226 }  ,draw opacity=1 ][fill={rgb, 255:red, 74; green, 144; blue, 226 }  ,fill opacity=1 ] (384.9,120) .. controls (384.9,118.62) and (383.78,117.5) .. (382.4,117.5) .. controls (381.02,117.5) and (379.9,118.62) .. (379.9,120) .. controls (379.9,121.38) and (381.02,122.5) .. (382.4,122.5) .. controls (383.78,122.5) and (384.9,121.38) .. (384.9,120) -- cycle ;
	\draw [color={rgb, 255:red, 74; green, 144; blue, 226 }  ,draw opacity=1 ]   (329.4,197) -- (311.4,145) ;
	\draw [color={rgb, 255:red, 74; green, 144; blue, 226 }  ,draw opacity=1 ]   (329.4,197) -- (387.4,193) ;
	\draw [color={rgb, 255:red, 74; green, 144; blue, 226 }  ,draw opacity=1 ]   (387.4,193) -- (400.4,144) ;
	\draw [color={rgb, 255:red, 74; green, 144; blue, 226 }  ,draw opacity=1 ]   (400.4,144) -- (353.4,111) ;
	\draw [color={rgb, 255:red, 74; green, 144; blue, 226 }  ,draw opacity=1 ]   (311.4,145) -- (400.4,144) ;
	\draw [color={rgb, 255:red, 74; green, 144; blue, 226 }  ,draw opacity=1 ]   (329.4,197) -- (400.4,144) ;
	\draw  [fill={rgb, 255:red, 255; green, 255; blue, 255 }  ,fill opacity=1 ] (328.95,128) .. controls (328.95,126.09) and (330.49,124.55) .. (332.4,124.55) .. controls (334.31,124.55) and (335.85,126.09) .. (335.85,128) .. controls (335.85,129.91) and (334.31,131.45) .. (332.4,131.45) .. controls (330.49,131.45) and (328.95,129.91) .. (328.95,128) -- cycle ;
	\draw  [fill={rgb, 255:red, 255; green, 255; blue, 255 }  ,fill opacity=1 ] (349,144.5) .. controls (349,142.59) and (350.54,141.05) .. (352.45,141.05) .. controls (354.36,141.05) and (355.9,142.59) .. (355.9,144.5) .. controls (355.9,146.41) and (354.36,147.95) .. (352.45,147.95) .. controls (350.54,147.95) and (349,146.41) .. (349,144.5) -- cycle ;
	\draw  [fill={rgb, 255:red, 255; green, 255; blue, 255 }  ,fill opacity=1 ] (373.45,127.5) .. controls (373.45,125.59) and (374.99,124.05) .. (376.9,124.05) .. controls (378.81,124.05) and (380.35,125.59) .. (380.35,127.5) .. controls (380.35,129.41) and (378.81,130.95) .. (376.9,130.95) .. controls (374.99,130.95) and (373.45,129.41) .. (373.45,127.5) -- cycle ;
	\draw  [fill={rgb, 255:red, 255; green, 255; blue, 255 }  ,fill opacity=1 ] (316.95,171) .. controls (316.95,169.09) and (318.49,167.55) .. (320.4,167.55) .. controls (322.31,167.55) and (323.85,169.09) .. (323.85,171) .. controls (323.85,172.91) and (322.31,174.45) .. (320.4,174.45) .. controls (318.49,174.45) and (316.95,172.91) .. (316.95,171) -- cycle ;
	\draw  [fill={rgb, 255:red, 255; green, 255; blue, 255 }  ,fill opacity=1 ] (354.95,175) .. controls (354.95,173.09) and (356.49,171.55) .. (358.4,171.55) .. controls (360.31,171.55) and (361.85,173.09) .. (361.85,175) .. controls (361.85,176.91) and (360.31,178.45) .. (358.4,178.45) .. controls (356.49,178.45) and (354.95,176.91) .. (354.95,175) -- cycle ;
	\draw  [fill={rgb, 255:red, 255; green, 255; blue, 255 }  ,fill opacity=1 ] (390.45,168.5) .. controls (390.45,166.59) and (391.99,165.05) .. (393.9,165.05) .. controls (395.81,165.05) and (397.35,166.59) .. (397.35,168.5) .. controls (397.35,170.41) and (395.81,171.95) .. (393.9,171.95) .. controls (391.99,171.95) and (390.45,170.41) .. (390.45,168.5) -- cycle ;
	\draw  [fill={rgb, 255:red, 255; green, 255; blue, 255 }  ,fill opacity=1 ] (358.4,195) .. controls (358.4,193.09) and (359.94,191.55) .. (361.85,191.55) .. controls (363.76,191.55) and (365.3,193.09) .. (365.3,195) .. controls (365.3,196.91) and (363.76,198.45) .. (361.85,198.45) .. controls (359.94,198.45) and (358.4,196.91) .. (358.4,195) -- cycle ;
	\node[] at (146,220){(a)};
	\node[] at (355,220){(b)};
	\end{tikzpicture}
	\caption{The quiver obtained from triangulation of a surface \cite{2006math......8367F}. (a) The triangulation of a 10-gon. (b) The quiver from a triangulated 10-gon.  Note that this quiver is in the mutation class of type ['A', 7]. }\label{Q11_traingulation}
\end{figure}

Besides pairing matrices and assigning 1's and 0's, there is a more direct way to classify theories in distinct duality trees as aforementioned. We can simply assign different mutation classes with different labels, and then let the machine tell which classes the given quivers belong to. So far, we have been using $\mathtt{Mathematica}$ and its built-in function to do the machine learning. One can still use $\mathtt{Classify}$ and NB to do the training, but it turns out that NB (and $\mathtt{Mathematica}$ classifier) is only good when the data is a set of pairs. Thus, we turn to $\mathtt{Python}$ to perform machine learning on mutations with the help of $\mathtt{Sage}$ \cite{sagemath} and $\mathtt{TensorFlow}$ \cite{tensorflow2015-whitepaper}. Henceforth, when we say that the method is NB (or NN), we simultaneously mean that the type of the dataset used is the one suitable for this method. This time, we choose three classes generated by \textbf{Q12}, \textbf{Q13}, \textbf{Q14}. Quiver \textbf{Q12} is defined through triangulation of a 10-gon, this process is shown in Fig. \ref{Q11_traingulation}.

According to the theorem by Felikson, Shapiro and Tumarkin \cite{2008arXiv0811.1703F}, the first two classes are finite while the third one is infinite. Now we label the three classes with [1,0,0], [0,1,0] and [0,0,1] respectively. Thus, when the machine predicts [$a_1$,$a_2$,$a_3$], it is giving probabilities of which class the matrix being predicted should belong to, where $a_i$'s are the probabilities of the three classes respectively. For instance, if the output is [0.9,0.06,0.04], then the machine classifies the matrix into the first class.

We use Convolutional Neural Networks (CNNs) to deal with the dataset which contains (1547+1956+1828) matrices. We find that there is only $\sim$55\% of accuracy when 80\% of data is trained. However, it is quite remarkable that for the last class, which is the only infinite one, the machine has a 100\% accuracy, i.e., it always correctly recognizes the matrices in this class and never misclassify other matrices to this class. Hence, the machine seems to have learnt something related to finite and infinite mutations. We will explore this in \S\ref{fininfmut}.

\subsection{Classifying against Random Antisymmetric Matrices}\label{randommats}
There is also another possible way to have a machine learning model on quiver mutations. If we are given some quiver and a class of dual theories, we may wonder whether this quiver also belongs to the duals. Therefore, we can train the machine using a specific class of matrices, along with some randomly generated antisymmetric matrices.

So as not to just learn anomalies, when we are dealing with anomaly-free quivers, we should mainly have random matrices that are anomaly free as well. For simplicity, let us contemplate the 3$\times$3 matrices. As the nullity of a non-zero 3$\times$3 matrix is at most 1, it should be easier to generate matrices that are anomaly free\footnote{For matrices of higher dimensions, anomalous matrices might be more easily generated randomly. What one could do is to use other different known classes of (anomaly-free) quivers to form a randomly generated set.}. We first test the dP$_0$ theory, viz, the class generated by \textbf{Q9}, with correspondingly many random antisymmetric matrices. We generate matrices up to depth 7, and we have (382+388) matrices for training and validation. The learning curves are plotted in Fig. \ref{random3x3}.

As we can see, the result is pretty good with $\sim$90\% accuracy when only $\sim$60\% of data is trained. If we use this model to predict unseen matrices, i.e. the 384 matrices at depth 8 plus 377 random matrices, the prediction can still reach $\sim$97\% accuracy. The accuracy for the matrices in the dP$_0$ duals is $\sim$93\% while the accuracy for random matrices is 100\%.

\section{Examples with Different Types}\label{differenttypes}
Let us go back to NB with matrix pairs and contemplate a heuristic example with four different classes. We use [`A',4], [`D',4], [`A',(3,1),1] and [`A',(2,2),1] here, where the latter two are called affine types. The learning results are again reported using 5-fold cross validation and learning curves as in Table \ref{AD4Aaffine5fold} and Fig. \ref{AD4Aaffinecurve} respectively.
\begin{table}[h]
	\centering
	\begin{tabular}{|c|c|c|}
		\hline
		Accuracy	& F-Score & $\phi$ \\ \hline
		0.84648200$\pm$0.00502814 & 0.85653500$\pm$0.00533456 & 0.69999000$\pm$0.0073538 \\ \hline
	\end{tabular}
	\caption{Training and validating four classes: [`A',4], [`D',4], [`A',(3,1),1] and [`A',(2,2),1]. We generate (52+50+70+54) matrices. There are 5503 1's and 5512 0's. The method is NB.}\label{AD4Aaffine5fold}
\end{table}
Even at 95\% training percentage, the accuracy is $\sim$85\%, and the Matthews $\phi$ is only $\sim$0.7. This is certainly not that satisfying\footnote{We already know that the numbers of mutation classes in the training data can affect our result. Nevertheless, it is reasonable to speculate that other factors such as the quiver types may also have influence.}.

We can simply put all the finite and affine types we meet so far ([`A',4], [`D',4], [`A',(3,1),1], [`A',(2,2),1], [`A',6], [`D',6], [`E',6]) together to create a dataset containing seven different mutation classes. We try the following three experiments:
\begin{enumerate}
	\item We generate 52, 50, 70, 54, 76, 77 and 77 matrices respectively. We have 14821 pairs in our dataset with 7360 1's and 7461 0's.
	\item We generate 144, 50, 120, 54, 76, 77 and 77 matrices respectively. We have 46332 pairs in our dataset with 22387 1's and 23945 0's.
	\item We generate 144, 50, 120, 54, 200, 213 and 213 matrices respectively. We have 43588 pairs in our dataset with 21229 1's and 22359 0's.
\end{enumerate}
Notice that in these three experiments, we also have matrix pairs \{($M_{4\times4}$,$M_{6\times6}$)$\rightarrow$0\} in our data, that is, we also include the trivial zeros from pairs of two quivers with different numbers of nodes. In all of the experiments, when we train 95\% of the dataset and validate the remaining 5\%, the accuracy is about 70\%-80\%, and $\phi$ is about 0.4-0.6. As expected, when we have more mutation classes, the performance of the machine becomes worse.

As a sanity check, we remove \{($M_{4\times4}$,$M_{6\times6}$)$\rightarrow$0\} in our data. For instance, we generate 52, 50, 70, 54, 76, 77 and 77 matrices respectively, and create 14254 pairs with 7375 1's and 7529 0's. We find that the accuracy becomes 65\%-75\%, and $\phi$ becomes 0.4-0.5. Getting a lower accuracy and a lower $\phi$ completely makes sense. Quivers with different nodes are apparently not dual to each other. Henceforth, we will not include pairs of matrices with different dimensions for 0's in our datasets which are easily learnt to classify as 0's.

\subsection{Dynkin and Affine Types}\label{finaffine}
So far in this section, we have discussed two different (finite) mutation types. We mainly deal with ADE types and include affine types as well. In light of the above learning results, we wonder whether different types would affect our result. A simple check would involve only two mutation classes with one Dynkin and one affine. For instance, we test [`D',4] and [`A',(3,1),1] here. We pick out two points in the whole learning curve as in Table \ref{dynkinaffine}.
\begin{table}[h]
	\centering
	\begin{tabular}{|c|c|c|c|c|c|}
		\hline
		\multirow{2}{*}{\begin{tabular}[c]{@{}l@{}}Training\\ Percentage\end{tabular}} & \multicolumn{5}{c|}{Accuracy(\%)} \\ \cline{2-6} 
		& \multicolumn{5}{c|}{$\phi$}          \\ \hline\hline
		\multirow{2}{*}{90\%}  &    100   &   100   &   100   &   100   &   100   \\ \cline{2-6} 
		&   1.00    &   1.00   &   1.00   &   1.00   &   1.00   \\ \hline
		\multirow{2}{*}{55\%}  &    100   &   100   &   100   &   100   &   100   \\ \cline{2-6} 
		&   1.00    &   1.00   &   1.00   &   1.00   &   1.00   \\ \hline
	\end{tabular}
	\caption{We machine learn [`D',4] and [`A',(3,1),1] mutation classes. We generate 92 and 104 matrices respectively. There are 6347 1's and 6320 0's. The method is NB.}\label{dynkinaffine}
\end{table}
The learning result is as perfect as the result in the example of [`A',4] and [`D',4]. From the viewpoint of machine learning, this is definitely a successful and exciting result. More importantly, our point here is to seek out the influence of different types. We find that learning mutation classes of the same type (e.g. only Dynkin) and learning those of different types (e.g. Dynkin+affine) have the same performance.

Let us further try an example with one finite mutation type ([`D',4]) and one infinite mutation type. For the infinite one, we choose the quiver \textbf{Q4}. We pick out two points on the learning curve as tabulated in Table \ref{fininf}.
\begin{table}[h]
	\centering
	\begin{tabular}{|c|c|c|c|c|c|}
		\hline
		\multirow{2}{*}{\begin{tabular}[c]{@{}l@{}}Training\\ Percentage\end{tabular}} & \multicolumn{5}{c|}{Accuracy(\%)} \\ \cline{2-6} 
		& \multicolumn{5}{c|}{$\phi$}          \\ \hline\hline
		\multirow{2}{*}{90\%}  &    100   &   100   &   100   &   100   &   100   \\ \cline{2-6} 
		&   1.00    &   1.00   &   1.00   &   1.00   &   1.00   \\ \hline
		\multirow{2}{*}{55\%}  &    100   &   100   &   100   &   100   &   100   \\ \cline{2-6} 
		&   1.00    &   1.00   &   1.00   &   1.00   &   1.00   \\ \hline
	\end{tabular}
	\caption{We machine learn [`D',4]'s and $F_0$ theory's quiver mutation classes. We generate 92 and 102 matrices respectively. There are 6313 1's and 6316 0's. The method is NB.}\label{fininf}
\end{table}
We see that it is still as perfect as the case with two Dynkin types ([`A',4],[`D',4]). To summarize, the mutation types would not really affect our learning performance for NB.

We return to our example with seven classes ([`A',4], [`D',4], [`A',(3,1),1], [`A',(2,2),1], [`A',6], [`D',6], [`E',6]). This time let us remove the two affine types and study the learning performance of the data with 5 classes. The results are reported in Table \ref{dynkinfiveclass5fold} for 5-fold cross validation and \ref{dynkinfiveclasscurve} for learning curves. We find that the result is improved. It is even better than the result of 4 classes ([`A',4], [`D',4], [`A',(3,1),1], [`A',(2,2),1]). Unlike the above tests, this seems to tell us that the influence from different types outcompetes the influence from the number of mutation classes. However, as we will see next, this is not the real reason.
\begin{table}[h]
	\centering
	\begin{tabular}{|c|c|c|}
		\hline
		Accuracy	& F-Score & $\phi$ \\ \hline
		0.88710300$\pm$0.00751058 & 0.89368900$\pm$0.00711335 & 0.78158800$\pm$0.01641340 \\ \hline
	\end{tabular}
	\caption{Training and validating five classes: [`A',4], [`D',4], [`A',6], [`D',6] and [`E',6]. We generate (52+50+76+77+77) matrices. There are 6791 1's and 6726 0's. The method is NB.}\label{dynkinfiveclass5fold}
\end{table}

\subsection{T Type}\label{ttype}
Now, we perform a test on 3 infinite classes, all of which are T types \cite{2011arXiv1102.4844M}: [`T',(4,4,4)], [`T',(4,5,3)] and [`T',(4,6,2)]. A quiver of T type is an orientation of a tree containing a unique trivalent vertex, three leaves of degree one, and with the remaining vertices in the branches being of degree two. When we say a quiver is of type $['T', (a,b,c)]$, we mean there are a total of $(a-2) + (b-2)+(c-2)$ vertices of degree two, summing up the contributions from the three branches. They are all 10$\times$10 matrices\footnote{We already know that the sizes of matrices will not have a big influence on our results, so we are free to choose matrices of any dimension.}. The learning results are given in Table \ref{ttype5fold} for 5-fold cross validation and Figure \ref{ttypecurve} for learning curves.
\begin{table}[h]
	\centering
	\begin{tabular}{|c|c|c|}
		\hline
		Accuracy	& F-Score & $\phi$ \\ \hline
		0.88569500$\pm$0.00987409	& 0.89199500$\pm$0.00793421 & 0.77648300$\pm$0.01925770 \\ \hline
	\end{tabular}
	\caption{Training and validating three classes: [`T',(4,4,4)], [`T',(4,5,3)] and [`T',(4,6,2)]. We generate (65+65+66) matrices. There are 2553 1's and 2565 0's. The method is NB.}\label{ttype5fold}
\end{table}
\noindent We see that the performance is basically the same as the three-infinite-class example in \S\ref{moreclasses}. Therefore we do not see the influence of mutation types here. Again, the influence of numbers of classes should dominate the performance of the $\mathtt{Classify}$ function in $\mathtt{Mathematica}$.

\subsection{Splitting the Dataset}\label{splitting}
Let us now try to solve the puzzle left at the end of \S\ref{finaffine}. Consider the quivers and matrices in \textbf{Q9} and \textbf{Q10}.
We can machine learn the dataset with these two classes. This yields 100\% accuracy and $\phi=1$ most of the time, which is good as expected. However, we can put these two quivers and the two quivers in \S\ref{twoinf} (\textbf{Q4} and \textbf{Q5}) together and machine learn the four classes generated from these four quivers. The 5-fold cross validation is given in Table \ref{split5fold}.
\begin{table}[h]
	\centering
	\begin{tabular}{|c|c|c|}
		\hline
		Accuracy	& F-Score & $\phi$ \\ \hline
		1$\pm$0	& 1$\pm$0 & 1$\pm$0 \\ \hline
	\end{tabular}
	\caption{Training and validating four classes: \textbf{Q4}, \textbf{Q5}, \textbf{Q9}, and \textbf{Q10}. We generate (102+138+94+138) matrices. There are 12276 1's and 11915 0's. The method is NB.}\label{split5fold}
\end{table}
We also pick out three points on the learning curve, which is tabulated in Table \ref{split4}.
\begin{table}[h]
	\centering
	\begin{tabular}{|c|c|c|c|c|c|}
		\hline
		\multirow{2}{*}{\begin{tabular}[c]{@{}l@{}}Training\\ Percentage\end{tabular}} & \multicolumn{5}{c|}{Accuracy(\%)} \\ \cline{2-6} 
		& \multicolumn{5}{c|}{$\phi$}          \\ \hline\hline
		\multirow{2}{*}{90\%}   &   100.0000    &   100.0000   &   100.0000   &   100.0000   &   100.0000   \\ \cline{2-6} 
		&   1.000000    &   1.000000   &   1.000000   &   1.000000   &   1.000000   \\ \hline
		\multirow{2}{*}{80\%}   &    100.0000   &   100.0000  &   100.0000   &   100.0000   &   100.0000   \\ \cline{2-6} 
		&    1.000000   &   1.000000   &   1.000000   &   1.000000   &   1.000000   \\ \hline
		\multirow{2}{*}{50\%}   &    99.9532   &   99.9844   &   99.9922   &   99.9844   &   99.9922   \\ \cline{2-6} 
		&    0.999065   &   0.999688   &   0.999844   &   0.999688   &   0.999844   \\ \hline
	\end{tabular}
	\caption{We generate 102, 138, 94 and 138 matrices respectively. There are 13199 1's and 12469 0's.}\label{split4}
\end{table}

Unlike the usual result one should expect from a four-class case, this learning result is almost as good as two-class cases. In fact, this is the key. Since we have two classes of 3$\times$3 matrices and two classes of 4$\times$4 matrices, the machine actually splits the dataset into two pieces, viz, it treats 3$\times$3 and 4$\times$4 matrices separately. Just like including zeros from pairs of matrices of different sizes, although machine learning is not affected by dimensions of matrices \emph{longitudinally}\footnote{For the sake of brevity, by this, we mean that if we have two datasets with, say, $k$ different mutation classes of $m\times m$ matrices and $k$ different mutation classes of $n\times n$ matrices ($m\neq n$), the performance should roughly be the same. On the other hand, if we have matrices of different sizes in one dataset, we shall say that we are studying how the matrix dimensions affect the results transversally.}, there is a \emph{transversal} influence of the matrix dimensions. Now we are able to explain why in \S\ref{finaffine}, the example with five classes ([`A',4], [`D',4], [`A',6], [`D',6], [`E',6]) has a better result than the one with four classes ([`A',4], [`D',4], [`A',(3,1),1], [`A',(2,2),1]). Effectively, the machine is dealing with (2+3) classes and 4 classes respectively.

\section{Enhancing the Dataset}\label{rankinfo}
\subsection{Adding Ranks of Nodes for NB}\label{rank}
Since physically interesting quivers have (round) nodes as gauge groups, each node carries the rank information of the gauge group. Thus, we can further add the rank information to ``help'' the machine learn Seiberg duality. Above all, these quivers should be anomaly free, which is encoded by the kernel of the adjacency matrix $M$ with certain rules under Seiberg duality as discussed in \S\ref{seiberg} \cite{Benvenuti:2004dw,Hanany:2012mb}. We simply add the ranks of nodes as a column vector $\bm{v}$ to our dataset by
\begin{equation}
\{(M_1,\bm{v}_1),(M_2,\bm{v}_2)\rightarrow1/0\}.
\end{equation}

We first test this on three classes as in \textbf{Q4}, \textbf{Q5}, and \textbf{Q6}. The results are given in Table \ref{threeclasskernel5fold} for 5-fold cross validation and Fig. \ref{threeclasskernelcurve} for learning curves.
\begin{table}[h]
	\centering
	\begin{tabular}{|c|c|c|}
		\hline
		Accuracy	& F-Score & $\phi$ \\ \hline
		0.91041400$\pm$0.00306970	& 0.91662600$\pm$0.00340356 & 0.82855000$\pm$0.00626524 \\ \hline
	\end{tabular}
	\caption{Training and validating three classes: \textbf{Q4}, \textbf{Q5}, and \textbf{Q6}. We generate (102+138+161) matrices. There are 11506 1's and 11645 0's. The method is NB. The rank information is included.}\label{threeclasskernel5fold}
\end{table}
We find that the learning result is the same compared to the former example with bare matrix input.

Now we add the class generated by \textbf{Q7} to our data. The four-class result is reported in Table \ref{fourclasskernel5fold} for 5-fold cross validation and Fig. \ref{fourclasskernelcurve} for learning curves.
\begin{table}[h]
	\centering
	\begin{tabular}{|c|c|c|}
		\hline
		Accuracy	& F-Score & $\phi$ \\ \hline
		0.85520000$\pm$0.00674474	& 0.86390500$\pm$0.00619574 & 0.71583900$\pm$0.01142870 \\ \hline
	\end{tabular}
	\caption{Training and validating four classes: \textbf{Q4}, \textbf{Q5}, \textbf{Q6}, and \textbf{Q7}. We generate (102+138+161+102) matrices. There are 13930 1's and 14005 0's. The method is NB. The rank information is included.}\label{fourclasskernel5fold}
\end{table}

We also further include \textbf{Q8} to construct the five-class example with extra rank information. The result can again be found in Table \ref{fiveclasskernel5fold} for 5-fold cross validation and Fig. \ref{fiveclasskernelcurve} for learning curves.
\begin{table}[h]
	\centering
	\begin{tabular}{|c|c|c|}
		\hline
		Accuracy	& F-Score & $\phi$ \\ \hline
		0.84267400$\pm$0.00915047	& 0.84683800$\pm$0.00846313 & 0.68633100$\pm$0.01791090 \\ \hline
	\end{tabular}
	\caption{Training and validating five classes: \textbf{Q4}, \textbf{Q5}, \textbf{Q6}, \textbf{Q7}, and \textbf{Q8}. We generate (102+138+161+102+161) matrices. There are 22770 1's and 22823 0's. The method is NB. The rank information is included.}\label{fiveclasskernel5fold}
\end{table}

Again, we learn that the learning results are not improved with the extra vectors. Based on the above results, it is possible that the machine already sees the rank information when we only feed it with bare matrix input (since it is related to the adjacency matrix kernels), therefore it does not require us to give the rank vector explicitly.

Moreover, we can try predicting totally unseen matrices as well. Let us use the three-class example (\textbf{Q4}, \textbf{Q5}, and \textbf{Q6}). We still train (102+138+161) matrices, viz, generate to (and include) depths 4. Then our validation contains matrices of depths 5 and 6, which has (688+978+1258) matrices. The training set has 12938 1's and 12961 0's while the validation set has 8987 1's and 8974 0's. After picking out correspondingly many pairs from each set, at 90\% training, we find that the accuracy is 0.50632400$\pm$0.00932148, and $\phi$ is 0.01286830$\pm$0.01174640. As a result, the performance is the same as before. Therefore, we would say for NB, the machine already sees the rank information to some extent even if we only have bare matrix input\footnote{However, as we will see shortly, rank information would make improvements when we have neural network and use multiclassification.}.

\subsection{Adding Diophantine Variables}\label{diophantine}
It is also natural to ask what would happen if we use some other ways of dataset enhancement. For superconformal chiral quivers, physical constraints should be imposed to those block quivers. The following conditions: chiral anomaly cancellation for the gauge groups, vanishing NSVZ $\beta$-function for each coupling as well as their weighted sum, and marginality of chiral operators in the superpotential at interacting fixed point, leads to a Diophantine equation \cite{Feng:2002kk,Franco:2002mu,Hanany:2012mb}.\footnote{More generally, monodromies give rise to mutation invariants, which in turn can be formulated as a set of Diophantine equations characterizing the space of dual theories (see e.g. \cite{Cachazo:2001sg,Franco:2020ijt}).} For three-block quivers, the Diophantine equation reads
\begin{equation}
\frac{a_{23}^2}{\alpha_1}+\frac{a_{31}^2}{\alpha_2}+\frac{a_{12}^2}{\alpha_3}=a_{12}a_{23}a_{31},
\end{equation}
where $a_{ij}$'s are the numbers of arrows among blocks (i.e., entries of the matrix) and $\alpha_i$'s the numbers of nodes in the blocks. Motivated by this intrinsic  structure of the mutation classes rooted in these physical constraints, we simply arrange $a_{ij}^2$'s and $a_{12}a_{23}a_{31}$ (which we shall call Diophantine variables for simplicity) into a vector and add it to the data. Now each pair looks like
\begin{equation}
\Big\{\left(M,(a_{12}^2,a_{23}^2,a_{31}^2,a_{12}a_{23}a_{31})^\text{T}\right),\left(N,(b_{12}^2,b_{23}^2,b_{31}^2,b_{12}b_{23}b_{31})^\text{T}\right)\rightarrow1/0\Big\}.
\end{equation}
However, we should emphasize that we are not actually telling the machine that the quivers/matrices should obey the Diophantine equation. Otherwise, for instance, for superconformal three-\emph{block} quivers, we would only have 16 of them \cite{Benvenuti:2004dw}. We are just using some specific combinations of $a_{ij}$'s (inspired by Diophantine equations), and putting this extra explicit vector in the data to see if this would give any improvement.

We first try an example with three mutation classes of 3$\times$3 matrices\footnote{Since we have already seen that the machine almost always gives correct predictions for two classes, we will start from three classes.}. We use the quivers \textbf{Q9}, \textbf{Q10} and \textbf{Q11}.
We list the 5-fold cross validation result in Table \ref{threeclass3x3diophantine}.
\begin{table}[h]
	\centering
	\begin{tabular}{|c|c|c|}
		\hline
		Accuracy	& F-Score & $\phi$ \\ \hline
		0.91148800$\pm$0.00091432	& 0.91759000$\pm$0.00115878 & 0.83179900$\pm$0.00135928 \\ \hline
	\end{tabular}
	\caption{Training and validating three classes: \textbf{Q9}, \textbf{Q10} and \textbf{Q11}. We generate (94+138+123) matrices. There are 11271 1's and 11301 0's. The method is NB. The Diophantine variables are included.}\label{threeclass3x3diophantine}
\end{table}
For reference, the learning result without including any extra information/vectors is also given in Table \ref{threeclass3x3bare}.
\begin{table}[h]
	\centering
	\begin{tabular}{|c|c|c|}
		\hline
		Accuracy	& F-Score & $\phi$ \\ \hline
		0.91431900$\pm$0.00644304	& 0.91987000$\pm$0.00657059 & 0.83621300$\pm$0.01123010 \\ \hline
	\end{tabular}
	\caption{Training and validating three classes: Fig. \textbf{Q9}, \textbf{Q10} and \textbf{Q11}. We generate (94+138+123) matrices without the augmented Diophantine variable information. There are 11239 1's and 11298 0's. The method is NB. The dataset is composed of bare matrix pairs only.}\label{threeclass3x3bare}
\end{table}
We can see that there is no improvement.

Let us now try 4$\times$4 matrices. Again we have three classes as in \textbf{Q4}, \textbf{Q5} and \textbf{Q6}. The Diophantine equation for four-block quivers reads \cite{Hanany:2012mb}
\begin{eqnarray}
a_{12}a_{23}a_{34}a_{14}&=&\frac{a_{12}^2}{\alpha_3\alpha_4}+\frac{a_{13}^2}{\alpha_2\alpha_4}+\frac{a_{14}^2}{\alpha_2\alpha_3}+\frac{a_{23}^2}{\alpha_1\alpha_4}+\frac{a_{24}^2}{\alpha_1\alpha_3}+\frac{a_{34}^2}{\alpha_1\alpha_2}\nonumber\\
&&+\frac{a_{12}a_{24}a_{14}}{\alpha_3}-\frac{a_{12}a_{23}a_{13}}{\alpha_4}+\frac{a_{13}a_{34}a_{13}}{\alpha_4}-\frac{a_{23}a_{34}a_{24}}{\alpha_1}.
\end{eqnarray}
We therefore add the vector
\begin{equation}
\left(a_{12}^2,a_{13}^2,a_{14}^2,a_{23}^2,a_{24}^2,a_{34}^2,a_{12}a_{24}a_{14},a_{12}a_{23}a_{13},a_{13}a_{34}a_{14},a_{23}a_{34}a_{24},a_{12}a_{23}a_{34}a_{14}\right)^\text{T}
\end{equation}\label{diophantinevec4}
to our data\footnote{Again, we are essentially adding these specific combinations of variables to the dataset, not the equation.}. The learning result are given in Table \ref{threeclassdiophantine5fold} and Fig. \ref{threeclassdiophantinecurve} for 5-fold cross validation. The performance is not really improved.
\begin{table}[h]
	\centering
	\begin{tabular}{|c|c|c|}
		\hline
		Accuracy	& F-Score & $\phi$ \\ \hline
		0.90980400$\pm$0.00358550	& 0.91565100$\pm$0.00323882 & 0.82811200$\pm$0.0057953 \\ \hline
	\end{tabular}
	\caption{Training and validating three classes: \textbf{Q4}, \textbf{Q5} and \textbf{Q6}. We generate (102+138+161) matrices. There are 11490 1's and 11449 0's. The method is NB. The Diophantine variables are included.}\label{threeclassdiophantine5fold}
\end{table}

Let us contemplate an example with four mutation classes. This time, we use the quivers \textbf{Q4}, \textbf{Q5}, \textbf{Q6}, and \textbf{Q7}. We report the results in Table \ref{fourclassdiophantine5fold} for 5-fold cross validation and Fig. \ref{fourclassdiophantinecurve} for learning curves. Again, the performance is the same.
\begin{table}[h]
	\centering
	\begin{tabular}{|c|c|c|}
		\hline
		Accuracy	& F-Score & $\phi$ \\ \hline
		0.858965$\pm$0.00349098	& 0.868007$\pm$0.0032153 & 0.72425500$\pm$0.00712124 \\ \hline
	\end{tabular}
	\caption{Training and validating four classes: \textbf{Q4}, \textbf{Q5}, \textbf{Q6}, and \textbf{Q7}. We generate (102+138+161+102) matrices. There are 14040 1's and 14109 0's. The method is NB. The Diophantine variables are included.}\label{fourclassdiophantine5fold}
\end{table}

Now move on to the case with five mutation classes. Besides the above four matrices, we further include the quiver \textbf{Q8}. The experiment without adding the Diophantine variables is done in \S\ref{moreclasses}. The new learning results are given in Table \ref{fiveclassdiophantine5fold} for 5-fold cross validation and Fig. \ref{fiveclassdiophantinecurve} for learning curves. We find that this is still not improved.
\begin{table}[h]
	\centering
	\begin{tabular}{|c|c|c|}
		\hline
		Accuracy	& F-Score & $\phi$ \\ \hline
		0.84443400$\pm$0.00325140	& 0.84887800$\pm$0.00285711 & 0.68986700$\pm$0.00652878 \\ \hline
	\end{tabular}
	\caption{Training and validating five classes: \textbf{Q4}, \textbf{Q5}, \textbf{Q6}, \textbf{Q7}, and \textbf{Q8}. We generate (102+138+161+102+161) matrices. There are 23211 1's and 23316 0's. The method is NB. The Diophantine variables are included.}\label{fiveclassdiophantine5fold}
\end{table}

Moreover, we can try predicting totally unseen matrices as well. Let us use the three-class example (\textbf{Q4}, \textbf{Q5}, and \textbf{Q6}). We still train (102+138+161) matrices, viz, generate to (and include) depths 4. Then our validation contains matrices of depths 5 and 6, which has (688+978+1258) matrices. The training set has 12886 1's and 13029 0's while the validation set has 8979 1's and 8981 0's. After picking out correspondingly many pairs from each set, at 90\% training, we find that the accuracy is 0.50191000$\pm$0.01061240, and $\phi$ is 0.00206997$\pm$0.025543800. We also have the similar experiment for NN, where this extra Diophantine-inspired structure does not improve learning as well. This suggests that such information does not help encode the structure of the quivers, which may be reasonable as we are also considering more general quivers and classes.

\subsection{Adding Ranks of Nodes for NN}\label{nnrk}
\begin{figure}[h]
	\centering
	\begin{tikzpicture}
	\begin{axis}[ymin=0, ymax=1.02,
	width=0.75\textwidth,
	height=0.5\textwidth,
	ytick={0,0.1,...,1.1}, ytick align=inside, ytick pos=left,
	xtick={0,10,...,100}, xtick align=inside, xtick pos=left,
	xlabel=Training(\%),
	ylabel=Performance,
	grid=major,
	grid style={dashed, gray!30},
	legend pos=south east,
	legend style={draw=none}]
	\addplot+[
	blue, mark options={blue, scale=1},
	smooth, 
	error bars/.cd, 
	y fixed,
	y dir=both, 
	y explicit
	] table [x=x, y=y,y error=error, col sep=comma] {
		x,  y,        error
		10, 0.514, 0.0181659
		20, 0.682, 0.144291
		30, 0.842, 0.0676018
		40, 0.932, 0.0704982
		50, 0.930, 0.0578792
		60, 0.924, 0.0371484
		70, 0.950, 0.0374166
		80, 0.980, 0.03937
		90, 0.986, 0.0260768
	};
	\addlegendentry{Total}
	\addplot+[
	green, mark options={green, scale=1},
	smooth, 
	error bars/.cd, 
	y fixed,
	y dir=both, 
	y explicit
	] table [x=x, y=y,y error=error, col sep=comma] {
		x,  y,        error
		10, 0.000, 0
		20, 0.412, 0.42228
		30, 0.720, 0.131719
		40, 0.882, 0.143597
		50, 0.882, 0.108028
		60, 0.844, 0.0838451
		70, 0.904, 0.0901665
		80, 0.962, 0.0849706
		90, 0.970, 0.0565685
	};
	\addlegendentry{\textbf{Q12}}
	\addplot+[
	red, mark options={red, scale=1},mark=triangle*,
	smooth, 
	error bars/.cd, 
	y fixed,
	y dir=both, 
	y explicit
	] table [x=x, y=y,y error=error, col sep=comma] {
		x,  y,        error
		10, 0.546, 0.0602495
		20, 0.654, 0.0288097
		30, 0.832, 0.0712039
		40, 0.920, 0.0608276
		50, 0.908, 0.0687023
		60, 0.942, 0.0248998
		70, 0.960, 0.0212132
		80, 0.980, 0.0291548
		90, 0.988, 0.0178885
	};
	\addlegendentry{\textbf{Q13}}
	\addplot+[
	brown, mark options={brown, scale=1},mark=diamond*,
	smooth, 
	error bars/.cd, 
	y fixed,
	y dir=both, 
	y explicit
	] table [x=x, y=y,y error=error, col sep=comma] {
		x,  y,        error
		10, 0.994, 0.0134164
		20, 0.982, 0.0204939
		30, 0.972, 0.0258844
		40, 0.998, 0.00447214
		50, 0.996, 0.00547723
		60, 0.986, 0.0114018
		70, 0.988, 0.0164317
		80, 1.000, 0
		90, 0.996, 0.00894427
	};
	\addlegendentry{\textbf{Q14}}
	\end{axis}
	\end{tikzpicture}
	\caption{Training and validating three classes: \textbf{Q12}, \textbf{Q13}, and \textbf{Q14}. We have (496+898+484) matrices. We use multiclass classification in NN. The rank information is included via imposing the null vector. The learning curves are all accuracies.}\label{nnkernel}
\end{figure}
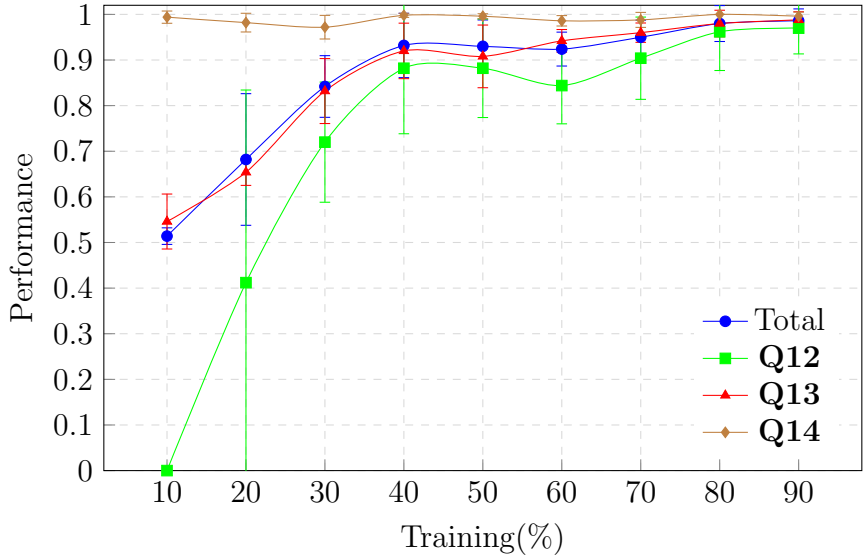

Now back to the example of \textbf{Q12}, \textbf{Q13}, and \textbf{Q14} in the multiclass classification, let us add the rank information to our dataset by augmenting the data input matrices to include the rank vectors as before. We have (496+898+484) matrices for training and validation. The learning curves of accuracies are plotted in Fig. \ref{nnkernel}.
We can see that the result is greatly improved after we include the rank information. With enough data trained, the accuracies approach 1, which is much better than the examples using NB. We also notice that at very low training percentage, the machine again confuses the two finite mutation classes while almost always gives correct results for the infinite one\footnote{Notice that the machine tends to classify the matrices in the first class as in the second class when making mistakes. This is due to the imbalance in the data. In spite of this, we can still get a very good result.}. The test without rank information above looks like the ``limit'' at low training percentage of the test with rank information. To see whether this model is really useful, we use it to predict matrices at unseen depths in these classes. For the predicted (1051+3263+1344) matrices, we get $\sim$74\% accuracy and $\sim$71\% F1 score. Although this has not reached perfectness, in particular for the purpose of application, the result for unseen matrices are still much better than those in NB. It is not just guessing any more, and we are on track to further improve this.

\subsection{Finite and Infinite Mutations}\label{fininfmut}
Recall that in \S\ref{multiclass}, the machines seems to treat finite and infinite mutations separately. Hence, we replace the infinite one (\textbf{Q14}) with another finite class as shown in \textbf{Q15}, which is anomalous.

We have tried CNN, as well as MLP and RNN, and find that all of them predict [$\sim$0.333,$\sim$0.333,$\sim$0.333]. This means that the machine is not able to decide the classes of the matrices. Hence, comparing the two examples (\textbf{Q11-13} and \textbf{Q11,12,14}), whether a mutation class is finite or infinite could affect the learning result. More precisely, the machine is learning something that helps it distinguish between finite and infinite mutation types.

\begin{figure}[h]
	\centering
	\begin{tikzpicture}
	\begin{axis}[ymin=0, ymax=1.02,
	width=0.75\textwidth,
	height=0.5\textwidth,
	ytick={0,0.1,...,1.1}, ytick align=inside, ytick pos=left,
	xtick={0,10,...,100}, xtick align=inside, xtick pos=left,
	xlabel=Training(\%),
	ylabel=Performance,
	grid=major,
	grid style={dashed, gray!30},
	legend pos=south east,
	legend style={draw=none}]
	\addplot+[
	blue, mark options={blue, scale=1},
	smooth, 
	error bars/.cd, 
	y fixed,
	y dir=both, 
	y explicit
	] table [x=x, y=y,y error=error, col sep=comma] {
		x,  y,        error
		10, 0.574, 0.115022
		20, 0.570, 0.0959166
		30, 0.818, 0.0334664
		40, 0.858, 0.0719027
		50, 0.856, 0.0482701
		60, 0.920, 0.0406202
		70, 0.828, 0.0614003
		80, 0.940, 0.0212132
		90, 0.952, 0.0178885
	};
	\addlegendentry{Total}
	\addplot+[
	green, mark options={green, scale=1},
	smooth, 
	error bars/.cd, 
	y fixed,
	y dir=both, 
	y explicit
	] table [x=x, y=y,y error=error, col sep=comma] {
		x,  y,        error
		10, 0.324, 0.299833
		20, 0.406, 0.227332
		30, 0.770, 0.0509902
		40, 0.776, 0.126412
		50, 0.768, 0.0759605
		60, 0.886, 0.0439318
		70, 0.714, 0.1076
		80, 0.912, 0.0396232
		90, 0.912, 0.0319374
	};
	\addlegendentry{\textbf{Q12}}
	\addplot+[
	red, mark options={red, scale=1},mark=triangle*,
	smooth, 
	error bars/.cd, 
	y fixed,
	y dir=both, 
	y explicit
	] table [x=x, y=y,y error=error, col sep=comma] {
		x,  y,        error
		10, 1.000, 0
		20, 1.000, 0
		30, 1.000, 0
		40, 1.000, 0
		50, 1.000, 0
		60, 1.000, 0
		70, 1.000, 0
		80, 1.000, 0
		90, 1.000, 0
	};
	\addlegendentry{\textbf{Q13}}
	\addplot+[
	brown, mark options={brown, scale=1},mark=diamond*,
	smooth, 
	error bars/.cd, 
	y fixed,
	y dir=both, 
	y explicit
	] table [x=x, y=y,y error=error, col sep=comma] {
		x,  y,        error
		10, 0.400, 0.225056
		20, 0.304, 0.277723
		30, 0.676, 0.0581378
		40, 0.794, 0.0884873
		50, 0.800, 0.0845577
		60, 0.880, 0.0886002
		70, 0.764, 0.0792465
		80, 0.910, 0.0324037
		90, 0.948, 0.0228035
	};
	\addlegendentry{\textbf{Q15}}
	\end{axis}
	\end{tikzpicture}
	\caption{Training and validating three classes: \textbf{Q12}, \textbf{Q13}, and \textbf{Q15}. We have (496+499+484) matrices. We use multiclass classification in NN. The rank information is included via imposing the null vector. The learning curves are all accuracies.}\label{nnkernelallfin}
\end{figure}
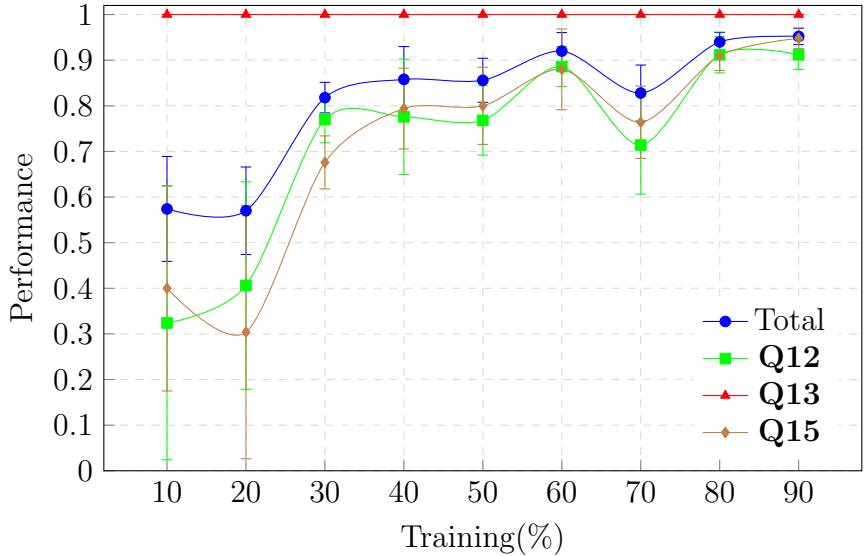

We can also include the rank information for the example of \textbf{Q12}, \textbf{Q13}, and \textbf{Q15}. Although the quiver \textbf{Q15} is anomalous, we can still assign some vector, say (1,1,1,1,1,1,1)$^\text{T}$ to it. Then the anomalies for every node should still add some consistent information on the duality operation among duals\footnote{Incidentally, this is also true for anomaly-free quivers. For example, the rank of Fig. \textbf{Q14} is (1,1,2,1,1,1,1)$^\text{T}$, but we can get the same good result if we assign a different vector, say (1,1,1,1,1,1,1)$^\text{T}$, as long as the following generated quivers and additional vectors are consistent with this choice.}. We have (496+484+499) matrices for training and validation\footnote{This time we do not choose all the 614 matrices in 0-4 depths for the third class so that the data would not be biased.}, and the model will be used to predict (1051+1344+1631) matrices. For training and validation, the learning curves are plotted in Fig. \ref{nnkernelallfin}.
We can see that with enough training, the result is still very good. It is also worth noting that when the machine meets a matrix belonging to the second class (\textbf{Q13}), it never misclassifies the matrix to other classes, viz, the red learning curve is a constant equal 100\%. Now for prediction, the machine again gives $\sim$71\% accuracy and $\sim$0.71 F1 score.

The above two examples show promising results for both physicists and mathematicians. We see that imposing rank information in NN significantly improves the performance of the machine to learn Seiberg duality. From a pure mathematical point of view, in particular the second example with all finite mutation types, this shows that the machine can learn which quivers are from which surfaces (or the 11 sporadic quivers) if we enhance the data as above.

\subsection{Predicting Matrices at Middle Depths}\label{predmid}
Now we would like to know whether the results for unseen data in predictions can be improved. Our strategy is again to train the matrices up to some depths, as well as some matrices at depths far away. Then we can check how NN behaves when predicting the matrices at middle depths. As a toy model, we train the matrices generated from \textbf{Q12}, \textbf{Q13}, and \textbf{Q14} at depths 0-3 and 5. Then we use the trained model to predict the (351+705+350) matrices at depth 4. In order to have a more balanced dataset, we choose 1062 matrices out of 3263 matrices at depth 5 for the class of \textbf{Q13}. Therefore, we have (1196+1255+1478) matrices for training and validation. We train 90\% and validate the remaining 10\% for our model, which gives almost always 100\% accuracy as expected. Impressively, after repeating training/validation and prediction a few times, we find that the machine almost always gives 100\% accuracy on the matrices at unseen depth (with only several errors out of tens of thousands of predictions, and in particular these few errors never happen for the infinite class). Such things do not happen for the NB cases. This is a perfect result, especially in the sense of application of machine learning on quiver mutations. It means that we can have a model to make good predictions on data of a different style to the training data (here at unseen depths).

One may also wonder whether things would change if more mutation classes are involved. Hence, we further include \textbf{Q15} to the above dataset. For just training and validation, we find that the result is still that good. Having more classes does not seem to affect the learning result too much. Now we apply this model to matrices at unseen depth just like the above case. Again, the machine gives $\sim$98\% accuracy and $\sim$0.98 F1 score, which is an impressive result.

\subsection{Classifying Against Random Antisymmetric Matrices}\label{randommatsrk}
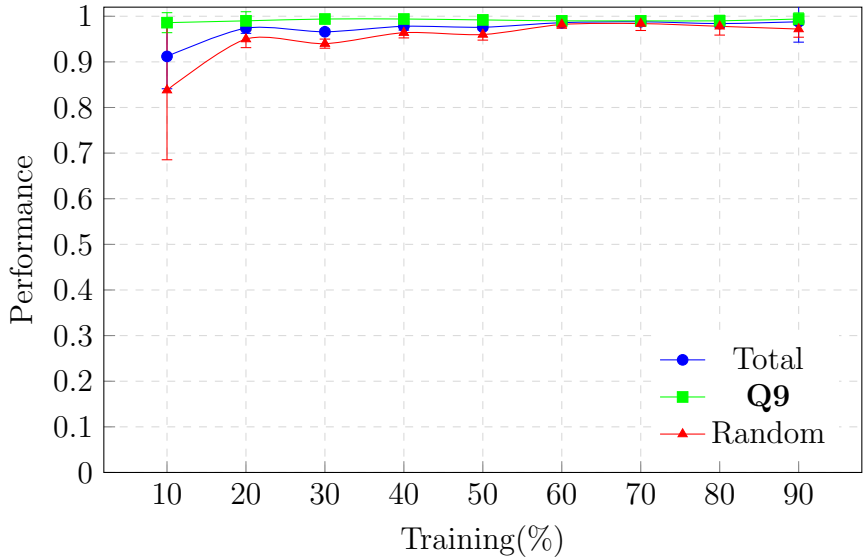
\begin{figure}[h]
	\centering
	\begin{tikzpicture}
	\begin{axis}[ymin=0, ymax=1.02,
	width=0.75\textwidth,
	height=0.5\textwidth,
	ytick={0,0.1,...,1.1}, ytick align=inside, ytick pos=left,
	xtick={0,10,...,100}, xtick align=inside, xtick pos=left,
	xlabel=Training(\%),
	ylabel=Performance,
	grid=major,
	grid style={dashed, gray!30},
	legend pos=south east,
	legend style={draw=none}]
	\addplot+[
	blue, mark options={blue, scale=1},
	smooth, 
	error bars/.cd, 
	y fixed,
	y dir=both, 
	y explicit
	] table [x=x, y=y,y error=error, col sep=comma] {
		x,  y,        error
		10, 0.912, 0.070852
		20, 0.974, 0.0114018
		30, 0.966, 0.00547723
		40, 0.978, 0.0083666
		50, 0.976, 0.00547723
		60, 0.986, 0.00547723
		70, 0.988, 0.0083666
		80, 0.984, 0.00894427
		90, 0.988, 0.0447214
	};
	\addlegendentry{Total}
	\addplot+[
	green, mark options={green, scale=1},
	smooth, 
	error bars/.cd, 
	y fixed,
	y dir=both, 
	y explicit
	] table [x=x, y=y,y error=error, col sep=comma] {
		x,  y,        error
		10, 0.986, 0.0219089
		20, 0.990, 0.02
		30, 0.994, 0.00547723
		40, 0.994, 0.00547723
		50, 0.992, 0.00447214
		60, 0.990, 0.00707107
		70, 0.990, 0.00707107
		80, 0.990, 0.0122474
		90, 0.994, 0.0134164
	};
	\addlegendentry{\textbf{Q9}}
	\addplot+[
	red, mark options={red, scale=1},mark=triangle*,
	smooth, 
	error bars/.cd, 
	y fixed,
	y dir=both, 
	y explicit
	] table [x=x, y=y,y error=error, col sep=comma] {
		x,  y,        error
		10, 0.838, 0.152545
		20, 0.950, 0.0187083
		30, 0.940, 0.01
		40, 0.964, 0.0114018
		50, 0.960, 0.0122474
		60, 0.982, 0.0083666
		70, 0.984, 0.0151658
		80, 0.978, 0.0192354
		90, 0.972, 0.0178885
	};
	\addlegendentry{Random}
	\end{axis}
	\end{tikzpicture}
	\caption{Training and validating one class, \textbf{Q9}, with random matrices. We have (382+384) matrices. We use classification in NN. The rank information is included via imposing the null vectors. The learning curves are all accuracies.}\label{randomrk}
\end{figure}

Let us do the same test involving randomly generated antisymmetric matrices again, but with rank information included. We still generate the matrices to depth 7 so that there are 382 matrices. We train these together with 384 random antisymmetric matrices. The learning curves are plotted in Fig. \ref{randomrk}\footnote{Incidentally, one can still try to use $\mathtt{Classify}$ and NB in $\mathtt{Mathematica}$. However, as aforementioned, NB is only good when the data is a set of pairs. For the example here, even at 90\% training, the accuracy is only 0.4619350$\pm$0.0148527. Even if we try only two classes (without random matrices), but not making pairs, the accuracy is only 0.6835440$\pm$0.2462260.}.
As we can see, this again improves the result significantly. Even at low training percentage, the accuracy still looks perfect. Now we use this to predict the 384 matrices at depth 8, along with 461 unseen random matrices. It turns out the accuracy is almost 100\%, with roughly ten mistakes only. Thus, if we would like to know whether a quiver belongs to some specific class of theories, this kind of model would be very useful. It is also worth noting that here we do not even need to include matrices at depths outside those used for predictions.

We can further try an example with two classes and some random matrices. This time, \textbf{Q10} is involved as well. We now generate to depth 6 and choose 384 out of the 506 matrices for this newly added class. It turns out at 90\% training, the accuracy is only 0.8460000$\pm$0.0336155, with F1 score being 0.8420000$\pm$0.0258844. If we use this model to predict matrices at next unseen depths, along with unseen random matrices, the accuracy is $\sim$80\%, with F1 score being $\sim$0.81. This does not decrease too much compared to the validation result. However, using a NN to identify whether a random quiver belongs to a particular duality class works best when only considering one class at a time.

\section{Conclusions and Outlook}\label{outlook}
Based on all the tests above, we can see that Seiberg duality and quiver mutations are very machine learnable. Several points are summarized as below. We first list the conclusions for NB and $\mathtt{Mathematica}$ classifier:
\begin{itemize}
	
	\item The number of different mutation classes is the \emph{dominant} influence in our machine learning. Fewer classes in the dataset would give better learning results. Other factors (such as mutation types, dimensions of matrices and adding rank information) are outcompeted for influence on the learning when there is a larger number of mutation classes.
	
	\item One reason that numbers of classes greatly affect our result would be the large number of matrices we have. In particular, (\#[combinations of assigning 0] $-$ \#[combinations of assigning 1]) gets larger when we include more mutation classes. We need to find a balance between avoiding duplicated 1's and taking care of various combinations of 0's. Our strategy would be to generate as many distinct 1's as possible, and then generate approximately same number of 0's. Thus, we could maximize the combinations of 0 without duplicated 1's while keeping the dataset unbiased.
	
	\item The dimensions of matrices affect the result ``transversally'' rather than ``longitudinally''. If we have two datasets with, say, $k$ different mutation classes of $m\times m$ matrices and $k'$ different mutation classes of $n\times n$ matrices ($m\neq n$), the performance should roughly be the same. On the other hand, the machine would spontaneously split the data into smaller parts in terms of the dimensions of matrices. For instance, a dataset with 2 classes of 4$\times$4 matrices and 3 classes of 5$\times$5 matrices would lead to a better result than the dataset 4 classes of 4$\times$4 matrices does. The former effectively has (2+3) classes, and hence the machine would have better performance in contrast to those with pure 4 or 5 classes. Of course, the (2+3)-class case would still be a bit worse than a pure 2-class example. Moreover, in light of the above two points, we shall \emph{never} include trivial 0's where each pair consists of matrices with different sizes. Although the transversal influence of dimensions does improve our result, this would bring a larger discrepancy between combinations of 0's and 1's, which can be cumbersome as aforementioned, especially for the dataset with many mutation classes. Now that these 0's represent theories that are obviously not dual to each other, there is no necessity to have them in the dataset.
	
	\item NB is the best method in the $\mathtt{Classify}$ function due to its mutual independence assumption.
	
	\item The NB classifier already sees the hint of rank information when we only have bare matrices as input, and thus imposing rank information would not further improve the machine learning result of the NB classifier.
	
	\item When the machine encounters mutation classes that are not seen in the training data, the performance gets worse. This is a reasonable result.
\end{itemize}
For multiclass classifications (and cases with random antisymmetric matrices), we mainly use CNNs here, and we see they behaves differently compared to NB. What NB is good at does not seem to work for a NN method, and vice versa. NB gives good results when the data is arranged in pairs while NN has great performance in multiclass classifications. It turns out that NN would be more useful in application of machine learning mutations in light of the following points:
\begin{itemize}
	\item We find that NN can distinguish whether a mutation class is finite or infinite, even without adding rank information. If we have a finite (infinite) mutation class among infinite (finite) mutation classes, the machine can almost always give 100\% accuracy to single out that finite/infinite class.
	
	\item We can impose the ranks as additional vectors augmented to the matrices. Then an NN classifier can give extremely good results for validation. This means the ranks of nodes would somehow reveal the structure behind a quiver to some extent. If we include some matrices at depths far away, then the unseen matrices at middle depths can be perfectly classified (as depicted in Fig. \ref{annuli1}(b))\footnote{Notice the argument on unseen matrices when discussing NB does not apply here for NN, as we have already seen from the learning results. This should be due to NB's mutual independence condition, while NN does not have this.}. The machine almost always give nearly 100\% accuracy when making predictions. Furthermore, the number of distinct mutation classes does not seem to strongly affect the performance of NN in this case.
	
	\item We can train one class of matrices with some other randomly generated matrices. Even without rank information, the results are still quite nice (e.g. see the results at the end of \S\ref{randommats}). To improve these results, including rank information can bring great improvements. If we use this model to predict matrices at unseen depths in that class (as depicted in Fig. \ref{annuli1}(a)), as well as unseen random matrices, the results are still almost-perfect (i.e., almost 100\% accuracy). Unlike the above bullet point, this does not even require matrices at depths far away to be involved in training. However, this kind of model only works best for classification with one class (against the random matrices). Having more classes would make it lose efficacy (e.g., two classes plus random matrices would decrease the accuracy of predictions to 80\%).
\end{itemize}
We see that  $\sim$100\% accuracy for \emph{predictions} can be obtained in all the above three points. These are the key results that might be useful in real-world application.

\paragraph{Outlook} It would also be interesting to ask whether the machine can recognize totally unseen \emph{classes} (rather than just matrices at unseen depths in trained classes) after training. For NB and $\mathtt{Mathematica}$, we can use matrix pairs and the predictions on pairs involving unseen classes will still be 0 or 1. However, as we have already seen, such model is poor at prediction on unseen data, hence it may not be that useful here. On the other hand, NN performs well for predictions. However, it is not suitable for dataset with matrix pairs. Therefore, we can only apply these classification networks to multiclass classification problems. Unfortunately, due to the problem structure of multiclassification, NNs can only recognize, and classify into, categories that are trained. When meeting an unseen class, it would treat the matrix as some element from a trained class. The design of supervised learning used with these NNs implies no machine can even tell that such matrix does not belong to any trained class, let alone recognizing a totally unseen class. Perhaps the closest realization so far would be the model containing random matrices. Then the machine would at least know that the unseen classes are different from the class being trained.

Thus, it would be natural to ask whether the advantages of the above two methods can be combined. NB has better behavior when the matrices are paired, and NN can have really good results when dealing with matrices at unseen depths. From the perspective of machine learning, the network structure, such as the choices of layers and loss functions, might be improved. We hope that in future we can develop new techniques for our models, especially for NNs or similar models, to make good predictions for matrix pairs and hence be useful for unseen classes.

More generally, we can imagine training the machine with a large number of pairs consisting of a randomly generated quiver and a dual connected to it by a single Seiberg duality on one of its nodes. We could then investigate if the machine can determine whether a pair of quivers are dual. If successful, this would arguably amount to the machine ``learning Seiberg duality".

There are many other directions for future work as well. For instance, supervised learning is used in this paper. We would also like to see what would happen if we do not label the matrices and let the machine learn without supervision. We are also not taking superpotentials into account here. All the bidirectional arrows get cancelled as we integrate out these fields. It would be intriguing to explore non-trivial superpotential quivers. Such data may be constructed with the help of Kasteleyn matrices \cite{Hanany:2005ve,Franco:2005rj}. Moreover, similarly to what we have done for Seiberg duality in $4d$, we can try applying machine to $2d$ $\mathcal{N}=(0,2)$ triality \cite{Gadde:2013lxa,Franco:2016nwv}, $0d$ $\mathcal{N}=1$ quadrality \cite{Franco:2016tcm}, and to the order $(m+1)$ dualities of $m$-graded quivers that generalize them \cite{Franco:2017lpa}. It is also worth noting that in \cite{Krefl:2017yox}, machine learning is applied to D-branes probing toric CY cones. Therefore, it is possible for us to study volume minimizations with machine learning. Finally, it would be interesting to ask whether the concept of finite types could be machine learnt. Such types are exactly the ADE Dynkin types and their matrices have eigenvalues less than 2 \cite{Smith}. Matrices and their eigenspaces are ubiquitous in mathematics, physics and machine learning. This would lead to a deeper study of matrices in machine learning.

\paragraph{Acknowledgements} The authors wish to thank the hospitality of the Institute for Mathematics and its Applications and their hosting of a workshop ``SageMath and Macaulay2: An Open Source Initiative'' that inspired the genesis of this paper.  The open source software Sage \cite{sagemath}, including its cluster algebra and quiver package \cite{2011arXiv1102.4844M}, was especially fundamental to this project. JB would like to thank Zijing Wu for useful discussions. The research of SF was supported by the U.S. National Science Foundation grants PHY-1820721 and DMS-1854179. YHH would like to thank STFC for grant ST/J00037X/1. EH would like to thank STFC for the PhD studentship. GM would like to thank the NSF for grants DMS-1745638 and 1854162.

\newpage
\appendix

\section{Machine Learning Structure}\label{MLstructure}
\subsection{$\mathtt{Mathematica}$'s $\mathtt{Classify}$}\label{mathematicaclassify}
Within the $\mathtt{Mathematica}$ software, the $\mathtt{Classify}$ function allows analysis of a variety of allowed input data types. These input data types include strings, sounds, and images, as well as the familiar numerical inputs. In our case the input data are tensor structures with integer entries. It may hence be noted that the generality of this function's data inputs may reduce the likelihood of it being optimised for use exclusively with tensors.

The $\mathtt{Classify}$ function takes as input training and validation sets, in our case these were lists of pairs of square matrices (or pairs of matrices along with vectors of their respective rank data). In addition within the calling of the function, the user can specify the classification method used, as well as the classification performance goal, and even allow the option for pseudo-random number seeding for the classification process.

The performance goal used was the standard ``automatic'' option. This selection calculates a weighted tradeoff for the final classifier that is trained such that it has high accuracy of output whilst still running quickly in subsequent classifications, and not requiring excessive memory storage.

More importantly in the creation of the classifier is the classification method used. $\mathtt{Mathematica}$ allows 9 method options, which among them include: Decision Trees, Markov Sequence Classifiers, Support Vector Machines, and Simple Artificial Neural Networks. When running $\mathtt{Classify}$ without specifying a method the program will run all methods and output a learning curve to allow comparison of performance between the methods on the input dataset (using parameters for comparison based on the validation data) \cite{Mathematica}. 

In initial testing of the $\mathtt{Classify}$ function with some of the datasets, the Naive Bayes method was consistently superior in the performance of its classifier. This is linked to the independence of the pair structure of the input data. Therefore, to avoid superfluous classifier training the method was specified to be Naive Bayes for the remainder of the investigation. Further discussion of the design and success of this method is discussed in Appendix \ref{NB}.

\subsection{The Naive Bayes Method}\label{NB}
We have seen that the Naive Bayes method, as a machine learning classifier, always gives us the best result when applying the built-in $\mathtt{Classify}$ to learn the matrix mutations. Essentially, our model is a conditional probability problem: $p(v_i|T)$, where $T$ acts as the condition for the machine to predict each $v_i\in V$ to be 0 or 1. Then Bayes' theorem yields
\begin{equation}
p(v_i|T)=\frac{p(T|v_i)p(v_i)}{p(T)}=\frac{p(T,v_i)}{p(T)}.
\end{equation}
Since $p(T)$ does not affect our result as this is solely determined by the fixed training set $T=\{t_1,t_2,\dots,t_n\}$ in each single experiment, we can fixate on the numerator:
\begin{eqnarray}
p(T,v_i)&=&p(t_1,\dots,t_n,v_i)\nonumber\\
&=&p(t_1|t_2,\dots,t_n,v_i)p(t_2,\dots,t_n,v_i)\nonumber\\
&=&\dots\nonumber\\
&=&p(t_1|t_2,\dots,t_n,v_i)p(t_2|t_3,\dots,t_n,v_i)\dots p(t_n|v_i)p(v_i).
\end{eqnarray}

Naive Bayes is ``naive'' because it assumes that every $t_i$ is \emph{independent} of the other conditions in $T$, which is exactly the property of matrix mutations. Whether a pair of matrices/quivers are related by mutations is always \emph{independent} of other matrices/quivers. This is the reason why the NB method is always the ideal choice.

Therefore, we may omit all the $t_k$'s in the conditional probability of $t_j$,viz,
\begin{equation}
p(t_j|t_{j+1},\dots,t_n,v_i)=p(t_j|v_i).
\end{equation}
As a result, we have
\begin{equation}
p(v_i|T)\propto p(v_i)\prod_jp(t_j|v_i).
\end{equation}
For our binary classification, the output is either 0 or 1. Then the Bayesian classifier $C_\text{B}$ should output $n$ ($n=0,1$) if $p(v_i=n|T)\geq p(v_i=1-n|T)$ \cite{NaiveBayes}. Hence, we require
\begin{equation}
C_\text{B}(v_i=n)=\frac{p(v_i=n|T)}{p(v_i=1-n|T)}\geq 1.
\end{equation}
For the NB classifier, we get
\begin{equation}
C_\text{NB}(v_i=n)=\frac{p(v_i=n)}{p(v_i=1-n)}\prod_j\frac{p(t_j|v_i=n)}{p(t_j|v_i=1-n)}\geq 1.
\end{equation}
As NB is the simplest (Bayes) network, it is often faster than other methods. More importantly, the assumption of conditional independence in NB reflects the special feature of the data.

\subsection{$\mathtt{Python}$'s CNNs}\label{python_appendix}
In investigations requiring multiclass classification, a more technical machine learning structure is needed to allow high-performance classification. To facilitate this the $\mathtt{TensorFlow}$ library, and within this the machine learning specific sub-library $\mathtt{Keras}$, were used \cite{tensorflow2015-whitepaper}.

Artificial Neural Networks (NNs) are code structures for non-linear function fitting. Their design was generally inspired by that of a biological brain, and they have seen significant success in recent years where computation speed can now account for the computational inefficiency of using these networks compared to traditional algorithms. The networks used in this investigation were dense and deep, in that they had all neurons fully connected between layers, and there were multiple hidden layers in the network. 

More specifically the network style used was a Convolutional Neural Network (CNN). The defining feature of these networks is the local action at the neurons in the hidden layers which preserves the multidimensional structure of the tensor input, acting with a simple linear $2d$ function, and then applying non-linear activation. Important to stress is the importance of the non-linearity in the activation functions at each neuron, allowing NNs to well address problems of higher complexity. These networks are traditionally used for image recognition, as the use of convolution is good for identifying local structure in arrays with dimension larger than 1 - this motivated their use for this matrix-based datatype \cite{DBLP:journals/corr/OSheaN15}.

The specific CNN used in this investigation had a sequential structure such that it was a linear stack of layers. The network had 3 convolutional layers, each with LeakyReLU activation, and each followed by a Maxpooling layer. Then 2 generic dense layers, one with LeakyReLU activation, and the other with softmax activation. The Maxpooling layers simply assign to an entry the maximum value of a set of some of the surrounding entries. They are traditionally used in the CNN structure.

LeakyReLU was used as the standard activation function at each layer. This activation is simple to compute, it is monotonic, and inherently non-linear, with the added benefit of fast gradient descent in training due to its proportional derivative form. This function leaves positive inputs to the neuron unchanged, but scales negative inputs down (in our case by a factor of 10). The additional dense layers are needed in CNNs to recreate the vector data structure for classification. Softmax was used as the final activation as it is a sigmoid equivalent, however with traditionally better results and a normalized output essential for classification problems with multiple classes.

When compiling the NN, additional inputs of loss function, optimizer, and metric are required. The loss function is a measure of the performance of the model, it is the function whose optimal value will indicate a well-trained NN, and hence a good model.``Mean squared error'' was used for the loss function in this investigation, this measure is simple, and computationally inexpensive. It is calculated as the sum of squares of the difference between each input and its predicted value by the model, therefore the output values used in training are vector floats bounded by 0 and 1 to reflect the hot encoding of the Boolean output nature in this classification. The optimizer is the method by which the parameters of the network are updated in accordance with the performance of the loss function. Here the ``Adam'' optimizer was used, which is an inexpensive first-order gradient based method \cite{2014arXiv1412.6980K}. Finally, the metric used was ``accuracy'', this gives the final measure of the NNs performance and is simply the proportion of correct classifications the model performs on the validation dataset.

\subsection{Measures of the Machine's Performance}\label{measures}
Measures of the performance of a classification method are essential for justifying the use of machine learning. The most standard measure of a classifier is ``accuracy'', as mentioned in Appendix \ref{python_appendix} this is the proportion of correct classifications performed by the classifier on a validation dataset. To ensure the measure is unbiased, it is important the validation dataset is not used for training whilst still being representative.

To ensure representative validation datasets, as well as providing a means of calculating error for these measures, $k$-fold cross-validation was used. In these investigations $k = 5$, and hence in each investigation the full dataset (all data points with their respective classification labels) were first randomized, then split into 5 equal size sub-datasets. The machine learning process for training and then validating the classifier was then iterated 5 times, where in each case the validation dataset was a different sub-dataset from the split, and the training dataset was the remaining 4 sets combined. For each of the 5 iterations the measures of performance were calculated and recorded, giving a small dataset for each measure from which a mean and standard error could be calculated \cite{Kohavi95astudy}.

More technical measures of performance used include Matthew's correlation coefficient (MCC, $\phi$), and F1 score (also called just F-Score). Both these measures take into account Type I and II errors from misclassification. A Type I error is a ``false positive'' (FP), where for example a random matrix is classified as in the mutation class, and conversely a Type II error is a ``false negative'' (FN), where a quiver matrix is classified as not in the class being trained by the machine.

The F1 score measure gives equal weight to Type I and II errors, whereas the MCC measure uses variable weights based on the occurrence of true positives and negatives (TP/TN). These factors make MCC a more favorable measure in this style of binary classification problem \cite{Chicco2020TheAO}. 

All three measures can be summarized as functions over the ``confusion matrix'', defined:
\begin{equation}
M = \begin{pmatrix} \text{TP} & \text{FN} \\ \text{FP} & \text{TN} \end{pmatrix}\,,
\end{equation}
such that
\begin{equation}
\begin{split}
\text{accuracy} & \vcentcolon = \frac{\text{TP} + \text{TN}}{\text{TP} + \text{TN} + \text{FP} + \text{FN}}\,,\\
\text{F1 score} & \vcentcolon = \frac{2 \cdot \text{TP}}{2 \cdot \text{TP} + \text{FP} + \text{FN}}\,,\\
\text{MCC} \quad & \vcentcolon = \frac{\text{TP}\cdot \text{TN} - \text{FP} \cdot \text{FN}}{\sqrt{(\text{TP} + \text{FP})\cdot (\text{TP} + \text{FN}) \cdot (\text{TN} + \text{FP}) \cdot (\text{TN} + \text{FN})}}\,.
\end{split}
\end{equation}
The first two measures, accuracy and F1 score, evaluate in the range $[0,1]$, whilst the MCC measure takes values in $[-1,1]$. In all cases a value of 1 indicates perfect prediction of the model. All measures can be generalized to the multiclassification cases also, evaluating in the same ranges.

\section{Investigation Learning Curves}\label{learning_curves}
This appendix section presents additional learning curves calculated for the investigations, as discussed in the paper. Each graph shows the performance of the investigation's classification method on the specified dataset for varying proportional splits of the dataset into training and validation data. Measures of classification performance considered were accuracy, and Matthew's correlation coefficient, $\phi$, as discussed in \S\ref{measures}.
\begin{figure}[H]
	\centering
	\begin{tikzpicture}
	\begin{axis}[ymin=0, ymax=1.02,
	width=0.75\textwidth,
	height=0.5\textwidth,
	ytick={0,0.1,...,1.1}, ytick align=inside, ytick pos=left,
	xtick={0,10,...,100}, xtick align=inside, xtick pos=left,
	xlabel=Training(\%),
	ylabel=Performance,
	grid=major,
	grid style={dashed, gray!30},
	legend pos=south east,
	legend style={draw=none}]
	\addplot+[
	blue, mark options={blue, scale=1},
	smooth, 
	error bars/.cd, 
	y fixed,
	y dir=both, 
	y explicit
	] table [x=x, y=y,y error=error, col sep=comma] {
		x,  y,        error
		5,  0.916518, 0.016848
		10, 0.971216, 0.0147798
		15, 0.971666, 0.0344917
		20, 0.977929, 0.116573
		25, 0.961076, 0.0352966
		30, 0.902807, 0.106844
		35, 0.961021, 0.0355866
		40, 0.948811, 0.0286174
		45, 1,        0
		50, 0.961088, 0.0355285
		55, 0.987147, 0.0287393
		60, 1,        0
		65, 1,        0
		70, 1,        0
		75, 1,        0
		80, 1,        0
		85, 0.988202, 0.0263807
		90, 1,        0
	};
	\addlegendentry{Accuracy}
	\addplot+[
	orange, mark options={orange, scale=1},
	smooth, 
	error bars/.cd, 
	y fixed,
	y dir=both, 
	y explicit
	] table [x=x, y=y,y error=error, col sep=comma] {
		x,  y,        error
		5,  0.82468,  0.0214413
		10, 0.939824, 0.0296322
		15, 0.944422, 0.0669723
		20, 0.954746, 0.0238331
		25, 0.924303, 0.0686022
		30, 0.792179, 0.232979
		35, 0.924171, 0.0692285
		40, 0.90145,  0.0551207
		45, 1,        0
		50, 0.924572, 0.0688662
		55, 0.975089, 0.0557022
		60, 1,        0
		65, 1,        0
		70, 1,        0
		75, 1,        0
		80, 1,        0
		85, 0.976878, 0.0517015
		90, 1,        0
	};
	\addlegendentry{$\phi$}
	\end{axis}
	\end{tikzpicture}
	\caption{Training and validating two classes: [`A',4] and [`D',4]. We generate (144+50) matrices. There are 11784 1's and 7200 0's. The method is automatically chosen by the machine within Mathematica's classify function.}\label{A4D4curve}
\end{figure}
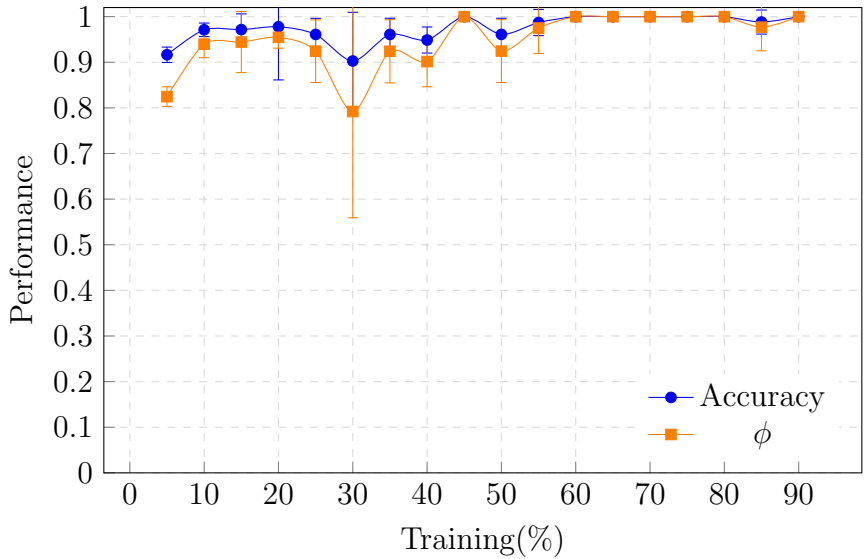

\begin{figure}[h]
	\centering
	\begin{tikzpicture}
	\begin{axis}[ymin=0, ymax=1.02,
	width=0.75\textwidth,
	height=0.5\textwidth,
	ytick={0,0.1,...,1.1}, ytick align=inside, ytick pos=left,
	xtick={0,10,...,100}, xtick align=inside, xtick pos=left,
	xlabel=Training(\%),
	ylabel=Performance,
	grid=major,
	grid style={dashed, gray!30},
	legend pos=south east,
	legend style={draw=none}]
	\addplot+[
	blue, mark options={blue, scale=1},
	smooth, 
	error bars/.cd, 
	y fixed,
	y dir=both, 
	y explicit
	] table [x=x, y=y,y error=error, col sep=comma] {
		x,  y,        error
		10, 0.902112, 0.0124352
		20, 0.982973, 0.00752375
		30, 0.996163, 0.00229954
		40, 0.998368, 0.00119662
		50, 0.999854, 0.000221741
		60, 0.999859, 0.000270448
		70, 0.999973, 0.0000853815
		80, 1       , 0
		90, 1       , 0
	};
	\addlegendentry{Accuracy}
	\addplot+[
	orange, mark options={orange, scale=1},
	smooth, 
	error bars/.cd, 
	y fixed,
	y dir=both, 
	y explicit
	] table [x=x, y=y,y error=error, col sep=comma] {
		x,  y,        error
		10, 0.805395, 0.0254436
		20, 0.965981, 0.0150145
		30, 0.992399, 0.00458646
		40, 0.996741, 0.00238987
		50, 0.999709, 0.000443303
		60, 0.999717, 0.00054073
		70, 0.999946, 0.000170447
		80, 1       , 0
		90, 1       , 0
	};
	\addlegendentry{$\phi$}
	\end{axis}
	\end{tikzpicture}
	\caption{Training and validating two classes: \textbf{Q4} and \textbf{Q5}. We generate (102+138) matrices. There are 6208 1's and 6154 0's. The method is NB.}\label{F0inf1curve}
\end{figure}
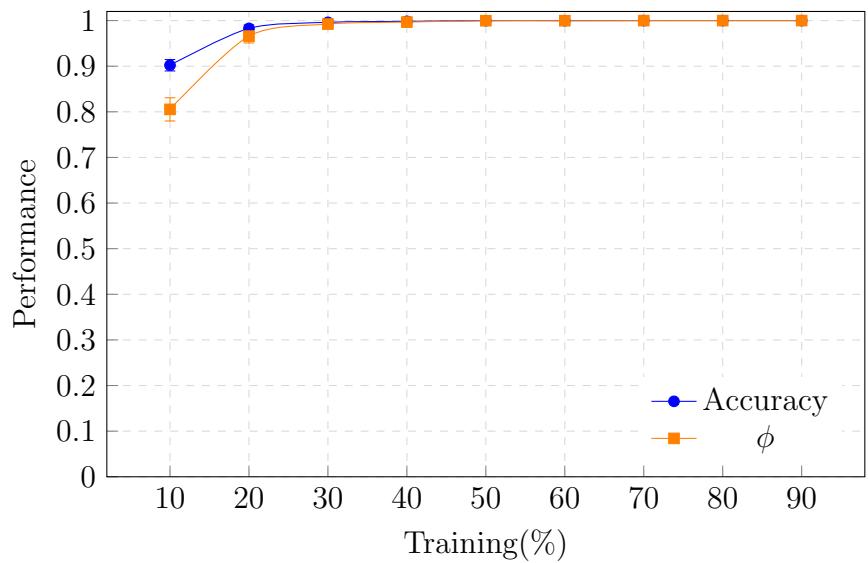

\begin{figure}[H]
	\centering
	\begin{tikzpicture}
	\begin{axis}[ymin=0, ymax=1.02,
	width=0.75\textwidth,
	height=0.5\textwidth,
	ytick={0,0.1,...,1.1}, ytick align=inside, ytick pos=left,
	xtick={0,10,...,100}, xtick align=inside, xtick pos=left,
	xlabel=Training(\%),
	ylabel=Performance,
	grid=major,
	grid style={dashed, gray!30},
	legend pos=south east,
	legend style={draw=none}]
	\addplot+[
	blue, mark options={blue, scale=1},
	smooth, 
	error bars/.cd, 
	y fixed,
	y dir=both, 
	y explicit
	] table [x=x, y=y,y error=error, col sep=comma] {
		x,  y,        error
		10, 0.769812, 0.005552
		20, 0.840716, 0.00746514
		30, 0.8584,   0.00368698
		40, 0.873405, 0.00817822
		50, 0.884933, 0.00552674
		60, 0.893306, 0.00897855
		70, 0.899945, 0.00688427
		80, 0.907535, 0.00225055
		90, 0.907617, 0.0091823
		95, 0.912602, 0.0131242
	};
	\addlegendentry{Accuracy}
	\addplot+[
	orange, mark options={orange, scale=1},
	smooth, 
	error bars/.cd, 
	y fixed,
	y dir=both, 
	y explicit
	] table [x=x, y=y,y error=error, col sep=comma] {
		x,  y,        error
		10, 0.540181, 0.0104313
		20, 0.682437, 0.0145195
		30, 0.717777, 0.00815793
		40, 0.749677, 0.0164298
		50, 0.774031, 0.014636
		60, 0.793536, 0.0178368
		70, 0.808282, 0.0142498
		80, 0.823278, 0.00431318
		90, 0.824874, 0.0169236
		95, 0.835353, 0.0237028
	};
	\addlegendentry{$\phi$}
	\end{axis}
	\end{tikzpicture}
	\caption{Training and validating three classes: [`A',6], [`D',6] and [`E',6]. We generate (76+77+77) matrices. There are 6122 1's and 6090 0's. The method is NB.}\label{ADE6seencurve}
\end{figure}
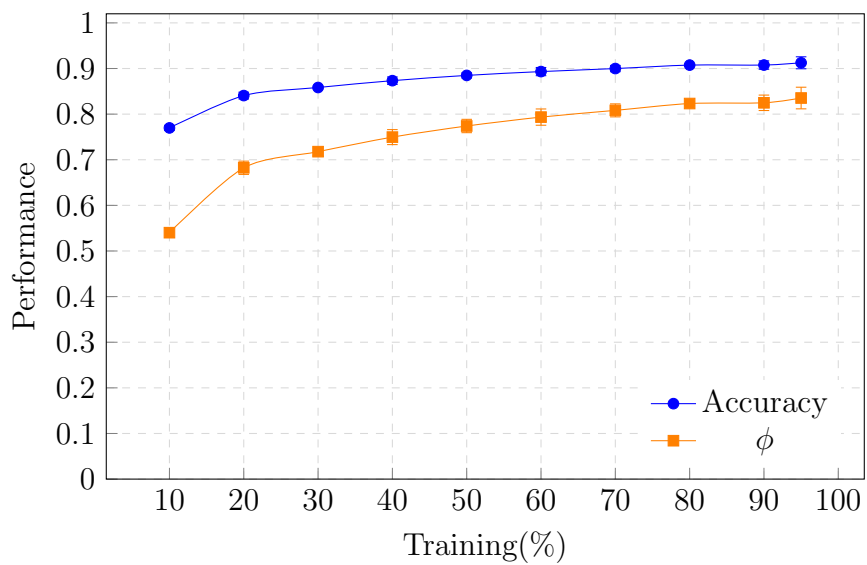

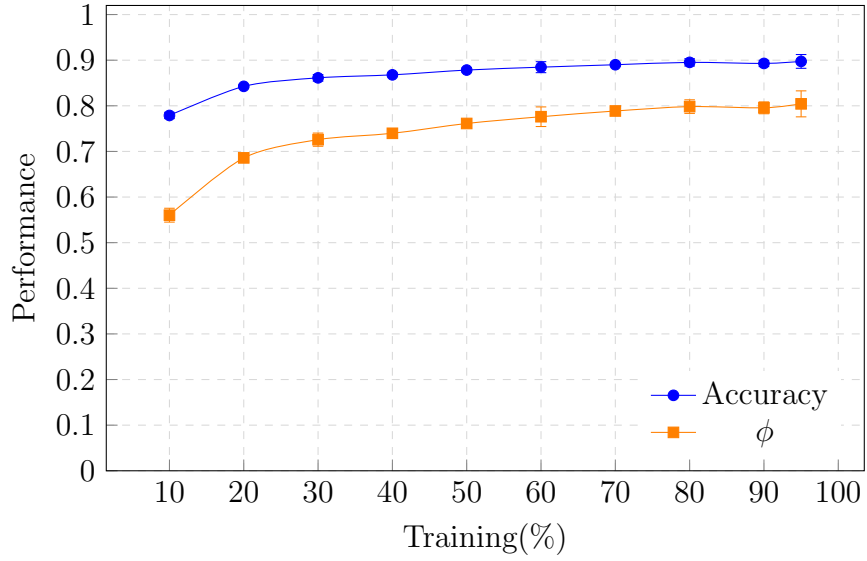
\begin{figure}[H]
	\centering
	\begin{tikzpicture}
	\begin{axis}[ymin=0, ymax=1.02,
	width=0.75\textwidth,
	height=0.5\textwidth,
	ytick={0,0.1,...,1.1}, ytick align=inside, ytick pos=left,
	xtick={0,10,...,100}, xtick align=inside, xtick pos=left,
	xlabel=Training(\%),
	ylabel=Performance,
	grid=major,
	grid style={dashed, gray!30},
	legend pos=south east,
	legend style={draw=none}]
	\addplot+[
	blue, mark options={blue, scale=1},
	smooth, 
	error bars/.cd, 
	y fixed,
	y dir=both, 
	y explicit
	] table [x=x, y=y,y error=error, col sep=comma] {
		x,  y,        error
		10, 0.778905, 0.00690726
		20, 0.8427  , 0.00350171
		30, 0.861332, 0.007208
		40, 0.867903, 0.00427015
		50, 0.878431, 0.0048027
		60, 0.88474 , 0.0120358
		70, 0.890083, 0.00350804
		80, 0.895269, 0.00798103
		90, 0.892924, 0.00688915
		95, 0.897357, 0.0150897
	};
	\addlegendentry{Accuracy}
	\addplot+[
	orange, mark options={orange, scale=1},
	smooth, 
	error bars/.cd, 
	y fixed,
	y dir=both, 
	y explicit
	] table [x=x, y=y,y error=error, col sep=comma] {
		x,  y,        error
		10, 0.560288, 0.0151849
		20, 0.685908, 0.00717791
		30, 0.725847, 0.0144453
		40, 0.739787, 0.00807975
		50, 0.761325, 0.0091569
		60, 0.77614 , 0.0215022
		70, 0.78869 , 0.00744624
		80, 0.798324, 0.0149083
		90, 0.795643, 0.0130037
		95, 0.804341, 0.0284869
	};
	\addlegendentry{$\phi$}
	\end{axis}
	\end{tikzpicture}
	\caption{Training and validating three classes: \textbf{Q4}, \textbf{Q5} and \textbf{Q6}. We generate (102+138+161) matrices. There are 11966 1's and 11494 0's. The method is NB.}\label{F0inf1inf2curve}
\end{figure}

\begin{figure}[H]
	\centering
	\begin{tikzpicture}
	\begin{axis}[ymin=0, ymax=1.02,
	width=0.75\textwidth,
	height=0.5\textwidth,
	ytick={0,0.1,...,1.1}, ytick align=inside, ytick pos=left,
	xtick={0,10,...,100}, xtick align=inside, xtick pos=left,
	xlabel=Training(\%),
	ylabel=Performance,
	grid=major,
	grid style={dashed, gray!30},
	legend pos=south east,
	legend style={draw=none}]
	\addplot+[
	blue, mark options={blue, scale=1},
	smooth, 
	error bars/.cd, 
	y fixed,
	y dir=both, 
	y explicit
	] table [x=x, y=y,y error=error, col sep=comma] {
		x,  y,        error
		10, 0.736092, 0.00716878
		20, 0.790714, 0.00458154
		30, 0.812868, 0.00608461
		40, 0.829361, 0.00319356
		50, 0.84019 , 0.00252728
		60, 0.843009, 0.00315505
		70, 0.858906, 0.00515449
		80, 0.857772, 0.00754941
		90, 0.862172, 0.00506446
	};
	\addlegendentry{Accuracy}
	\addplot+[
	orange, mark options={orange, scale=1},
	smooth, 
	error bars/.cd, 
	y fixed,
	y dir=both, 
	y explicit
	] table [x=x, y=y,y error=error, col sep=comma] {
		x,  y,        error
		10, 0.472087, 0.0143349
		20, 0.581696, 0.00913534
		30, 0.627368, 0.0124301
		40, 0.661816, 0.00533468
		50, 0.685162, 0.00421121
		60, 0.690953, 0.00561845
		70, 0.722629, 0.0102765
		80, 0.722008, 0.014387
		90, 0.732164, 0.0087208
	};
	\addlegendentry{$\phi$}
	\end{axis}
	\end{tikzpicture}
	\caption{Training and validating four classes: \textbf{Q4}, \textbf{Q5}, \textbf{Q6}, and \textbf{Q7}. We generate (102+138+161+102) matrices. There are 16059 1's and 16250 0's. The method is NB.}\label{fourclasscurve}
\end{figure}
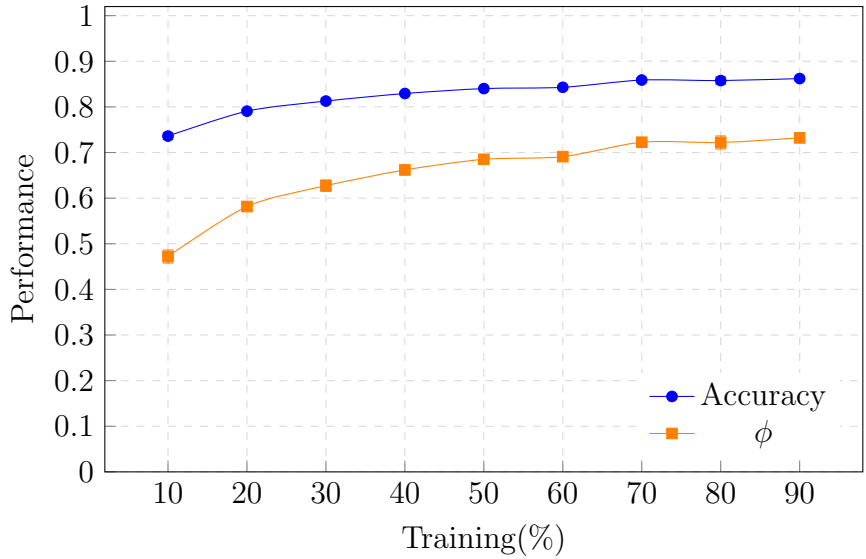

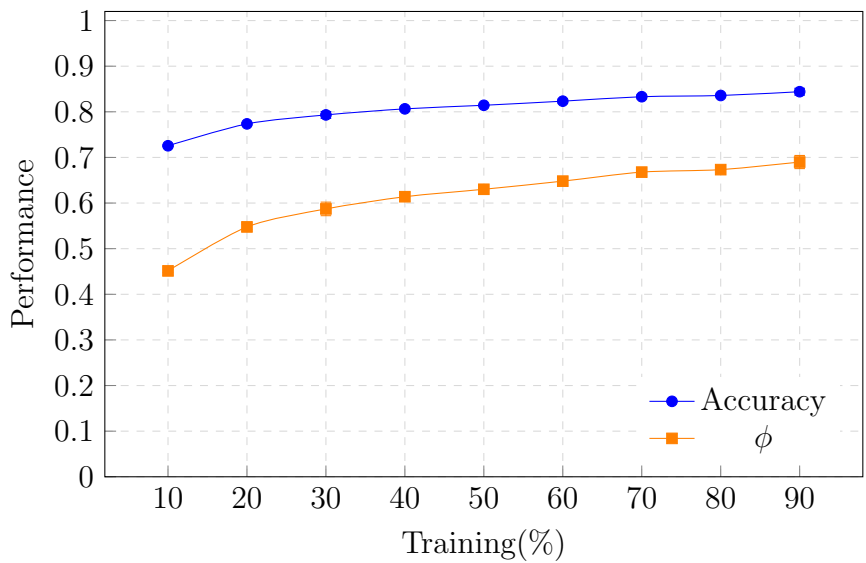
\begin{figure}[H]
	\centering
	\begin{tikzpicture}
	\begin{axis}[ymin=0, ymax=1.02,
	width=0.75\textwidth,
	height=0.5\textwidth,
	ytick={0,0.1,...,1.1}, ytick align=inside, ytick pos=left,
	xtick={0,10,...,100}, xtick align=inside, xtick pos=left,
	xlabel=Training(\%),
	ylabel=Performance,
	grid=major,
	grid style={dashed, gray!30},
	legend pos=south east,
	legend style={draw=none}]
	\addplot+[
	blue, mark options={blue, scale=1},
	smooth, 
	error bars/.cd, 
	y fixed,
	y dir=both, 
	y explicit
	] table [x=x, y=y,y error=error, col sep=comma] {
		x,  y,        error
		10, 0.725586, 0.00174999
		20, 0.773579, 0.00286421
		30, 0.793232, 0.00755719
		40, 0.806529, 0.00491321
		50, 0.814353, 0.000768579
		60, 0.823366, 0.0045804
		70, 0.833134, 0.0046124
		80, 0.835797, 0.00236828
		90, 0.844126, 0.00752319
	};
	\addlegendentry{Accuracy}
	\addplot+[
	orange, mark options={orange, scale=1},
	smooth, 
	error bars/.cd, 
	y fixed,
	y dir=both, 
	y explicit
	] table [x=x, y=y,y error=error, col sep=comma] {
		x,  y,        error
		10, 0.451296, 0.00354002
		20, 0.547708, 0.00570722
		30, 0.587178, 0.0148583
		40, 0.613908, 0.00963629
		50, 0.63025 , 0.0016394
		60, 0.648075, 0.00886968
		70, 0.667863, 0.00933279
		80, 0.673292, 0.00509406
		90, 0.689965, 0.0144048
	};
	\addlegendentry{$\phi$}
	\end{axis}
	\end{tikzpicture}
	\caption{Training and validating five classes: \textbf{Q4}, \textbf{Q5}, \textbf{Q6}, \textbf{Q7}, and \textbf{Q8}. We generate (102+138+161+102+161) matrices. There are 23377 1's and 23442 0's. The method is NB.}\label{fiveclasscurve}
\end{figure}

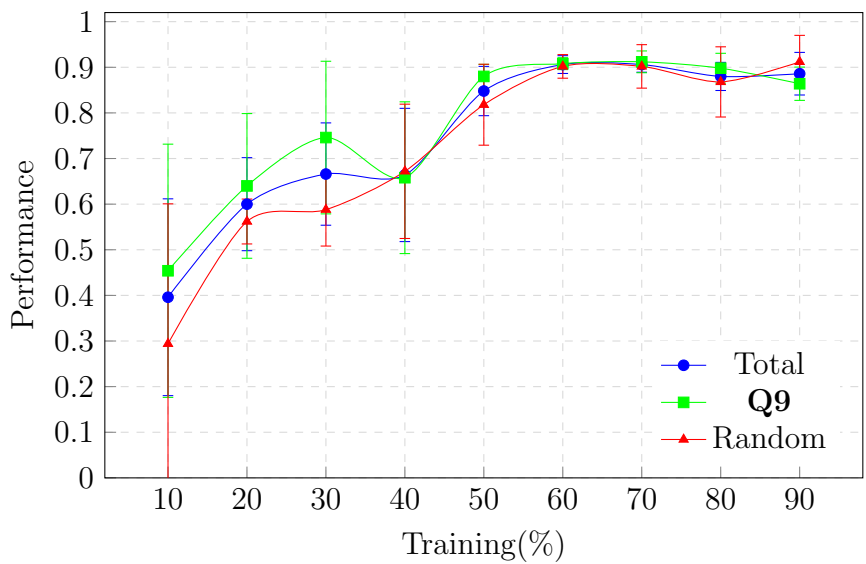
\begin{figure}[H]
	\centering
	\begin{tikzpicture}
	\begin{axis}[ymin=0, ymax=1.02,
	width=0.75\textwidth,
	height=0.5\textwidth,
	ytick={0,0.1,...,1.1}, ytick align=inside, ytick pos=left,
	xtick={0,10,...,100}, xtick align=inside, xtick pos=left,
	xlabel=Training(\%),
	ylabel=Performance,
	grid=major,
	grid style={dashed, gray!30},
	legend pos=south east,
	legend style={draw=none}]
	\addplot+[
	blue, mark options={blue, scale=1},
	smooth, 
	error bars/.cd, 
	y fixed,
	y dir=both, 
	y explicit
	] table [x=x, y=y,y error=error, col sep=comma] {
		x,  y,        error
		10, 0.396, 0.215708
		20, 0.600, 0.10198
		30, 0.666, 0.112161
		40, 0.664, 0.146048
		50, 0.848, 0.054037
		60, 0.906, 0.0194936
		70, 0.906, 0.0167332
		80, 0.880, 0.0308221
		90, 0.886, 0.0466905
	};
	\addlegendentry{Total}
	\addplot+[
	green, mark options={green, scale=1},
	smooth, 
	error bars/.cd, 
	y fixed,
	y dir=both, 
	y explicit
	] table [x=x, y=y,y error=error, col sep=comma] {
		x,  y,        error
		10, 0.454, 0.277543
		20, 0.640, 0.158588
		30, 0.746, 0.167272
		40, 0.658, 0.166343
		50, 0.880, 0.0264575
		60, 0.908, 0.0130384
		70, 0.912, 0.0238747
		80, 0.898, 0.0327109
		90, 0.864, 0.0364692
	};
	\addlegendentry{\textbf{Q9}}
	\addplot+[
	red, mark options={red, scale=1},mark=triangle*,
	smooth, 
	error bars/.cd, 
	y fixed,
	y dir=both, 
	y explicit
	] table [x=x, y=y,y error=error, col sep=comma] {
		x,  y,        error
		10, 0.294, 0.306725
		20, 0.562, 0.0491935
		30, 0.588, 0.0798123
		40, 0.672, 0.147207
		50, 0.818, 0.088713
		60, 0.902, 0.0258844
		70, 0.902, 0.0476445
		80, 0.868, 0.0769415
		90, 0.912, 0.0580517
	};
	\addlegendentry{Random}
	\end{axis}
	\end{tikzpicture}
	\caption{Training and validating one class, \textbf{Q9}, with random matrices. We have (382+388) matrices. We use a NN classifier. The learning curves are all accuracies.}\label{random3x3}
\end{figure}

\begin{figure}[H]
	\centering
	\begin{tikzpicture}
	\begin{axis}[ymin=0, ymax=1.02,
	width=0.75\textwidth,
	height=0.5\textwidth,
	ytick={0,0.1,...,1.1}, ytick align=inside, ytick pos=left,
	xtick={0,10,...,100}, xtick align=inside, xtick pos=left,
	xlabel=Training(\%),
	ylabel=Performance,
	grid=major,
	grid style={dashed, gray!30},
	legend pos=south east,
	legend style={draw=none}]
	\addplot+[
	blue, mark options={blue, scale=1},
	smooth, 
	error bars/.cd, 
	y fixed,
	y dir=both, 
	y explicit
	] table [x=x, y=y,y error=error, col sep=comma] {
		x,  y,        error
		10, 0.705821, 0.0100919
		20, 0.779843, 0.0109857
		30, 0.802464, 0.00490042
		40, 0.810682, 0.00539156
		50, 0.824006, 0.00566482
		60, 0.830458, 0.0138392
		70, 0.842131, 0.00993409
		80, 0.84503,  0.00733762
		90, 0.852679, 0.0150262
		95, 0.844646, 0.00582461
	};
	\addlegendentry{Accuracy}
	\addplot+[
	orange, mark options={orange, scale=1},
	smooth, 
	error bars/.cd, 
	y fixed,
	y dir=both, 
	y explicit
	] table [x=x, y=y,y error=error, col sep=comma] {
		x,  y,        error
		10, 0.41208,  0.020322
		20, 0.560011, 0.0218702
		30, 0.606596, 0.0105213
		40, 0.623634, 0.00877204
		50, 0.652859, 0.0126063
		60, 0.668079, 0.0259422
		70, 0.688586, 0.0170927
		80, 0.695668, 0.0139501
		90, 0.712   , 0.0287905
		95, 0.698198, 0.014297
	};
	\addlegendentry{$\phi$}
	\end{axis}
	\end{tikzpicture}
	\caption{Training and validating four classes: [`A',4], [`D',4], [`A',(3,1),1] and [`A',(2,2),1]. We generate (52+50+70+54) matrices. There are 5503 1's and 5512 0's. The method is NB.}\label{AD4Aaffinecurve}
\end{figure}
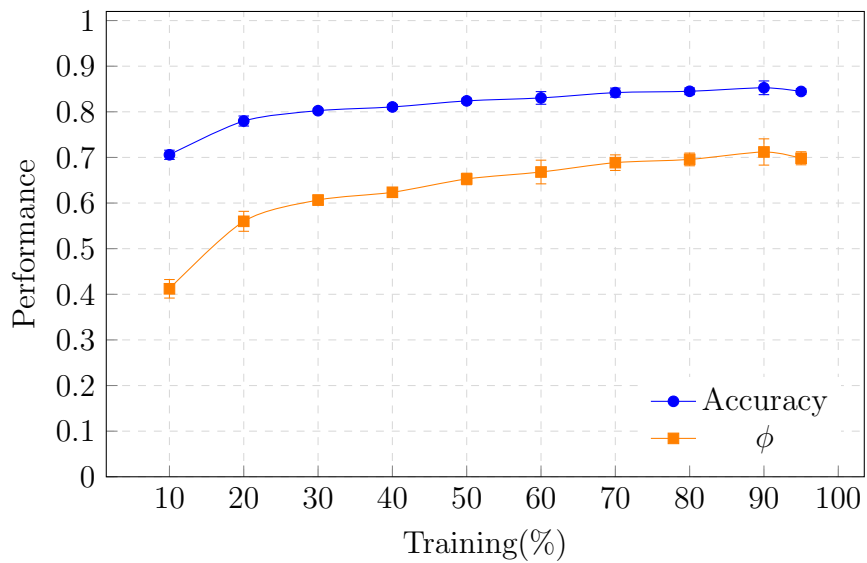

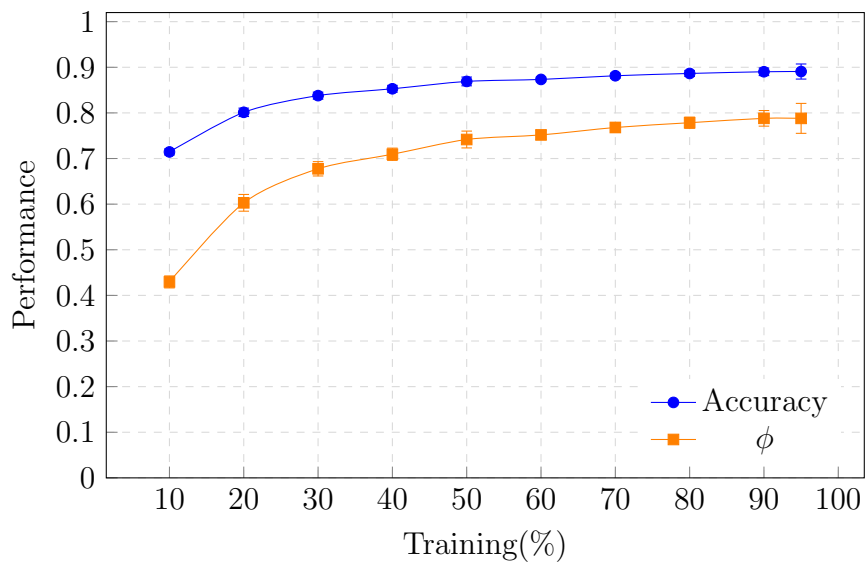
\begin{figure}[H]
	\centering
	\begin{tikzpicture}
	\begin{axis}[ymin=0, ymax=1.02,
	width=0.75\textwidth,
	height=0.5\textwidth,
	ytick={0,0.1,...,1.1}, ytick align=inside, ytick pos=left,
	xtick={0,10,...,100}, xtick align=inside, xtick pos=left,
	xlabel=Training(\%),
	ylabel=Performance,
	grid=major,
	grid style={dashed, gray!30},
	legend pos=south east,
	legend style={draw=none}]
	\addplot+[
	blue, mark options={blue, scale=1},
	smooth, 
	error bars/.cd, 
	y fixed,
	y dir=both, 
	y explicit
	] table [x=x, y=y,y error=error, col sep=comma] {
		x,  y,        error
		10, 0.714492, 0.00657501
		20, 0.801342, 0.00921141
		30, 0.837904, 0.00774891
		40, 0.852921, 0.00756031
		50, 0.869053, 0.00961947
		60, 0.873415, 0.00485987
		70, 0.881481, 0.00551466
		80, 0.886503, 0.00667299
		90, 0.890231, 0.00839356
		95, 0.890746, 0.0165331
	};
	\addlegendentry{Accuracy}
	\addplot+[
	orange, mark options={orange, scale=1},
	smooth, 
	error bars/.cd, 
	y fixed,
	y dir=both, 
	y explicit
	] table [x=x, y=y,y error=error, col sep=comma] {
		x,  y,        error
		10, 0.429548, 0.013145
		20, 0.602972, 0.0183529
		30, 0.67752,  0.015703
		40, 0.709403, 0.0134494
		50, 0.741727, 0.0184571
		60, 0.751777, 0.00990781
		70, 0.768078, 0.0109469
		80, 0.778519, 0.0122396
		90, 0.78814,  0.0171006
		95, 0.788077, 0.032816
	};
	\addlegendentry{$\phi$}
	\end{axis}
	\end{tikzpicture}
	\caption{Training and validating five classes: [`A',4], [`D',4], [`A',6], [`D',6] and [`E',6]. We generate (52+50+76+77+77) matrices. There are 6699 1's and 6711 0's. The method is NB.}\label{dynkinfiveclasscurve}
\end{figure}

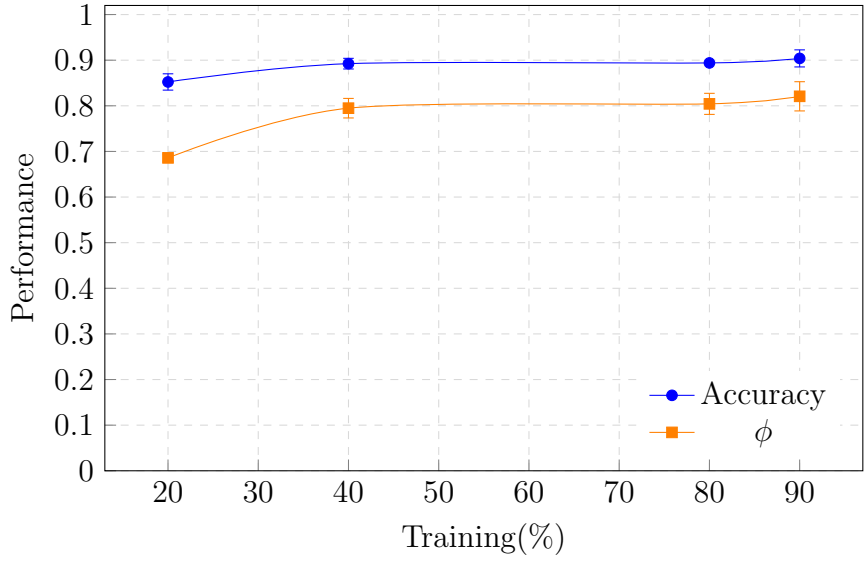
\begin{figure}[H]
	\centering
	\begin{tikzpicture}
	\begin{axis}[ymin=0, ymax=1.02,
	width=0.75\textwidth,
	height=0.5\textwidth,
	ytick={0,0.1,...,1.1}, ytick align=inside, ytick pos=left,
	xtick={0,10,...,100}, xtick align=inside, xtick pos=left,
	xlabel=Training(\%),
	ylabel=Performance,
	grid=major,
	grid style={dashed, gray!30},
	legend pos=south east,
	legend style={draw=none}]
	\addplot+[
	blue, mark options={blue, scale=1},
	smooth, 
	error bars/.cd, 
	y fixed,
	y dir=both, 
	y explicit
	] table [x=x, y=y,y error=error, col sep=comma] {
		x,  y,        error
		20, 0.852329, 0.0178922
		40, 0.892459, 0.0116768
		80, 0.894031, 0.00139587
		90, 0.903983, 0.0187159
	};
	\addlegendentry{Accuracy}
	\addplot+[
	orange, mark options={orange, scale=1},
	smooth, 
	error bars/.cd, 
	y fixed,
	y dir=both, 
	y explicit
	] table [x=x, y=y,y error=error, col sep=comma] {
		x,  y,        error
		20, 0.685908, 0.00717791
		40, 0.79486 , 0.0214059
		80, 0.804325, 0.0231001
		90, 0.820861, 0.0320227
	};
	\addlegendentry{$\phi$}
	\end{axis}
	\end{tikzpicture}
	\caption{Training and validating three classes: [`T',(4,4,4)], [`T',(4,5,3)] and [`T',(4,6,2)]. We generate (65+65+66) matrices. There are 2476 1's and 2301 0's. The method is NB.}\label{ttypecurve}
\end{figure}

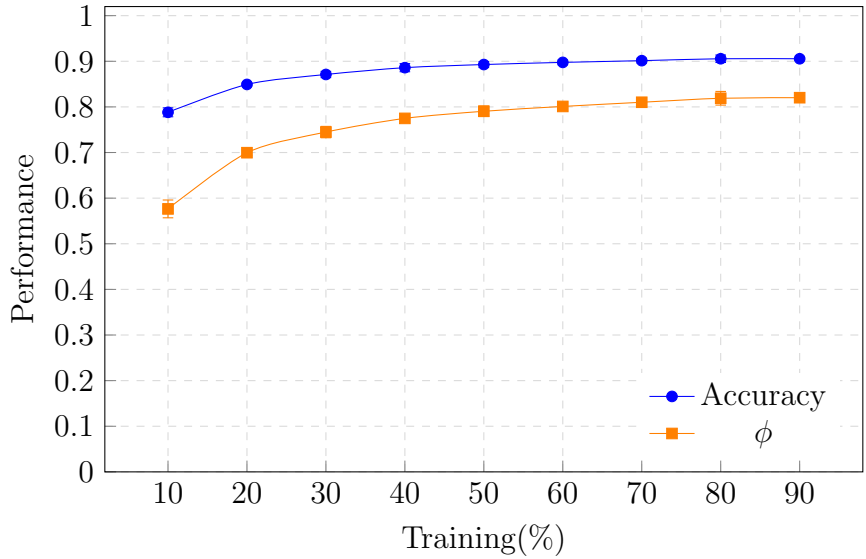
\begin{figure}[H]
	\centering
	\begin{tikzpicture}
	\begin{axis}[ymin=0, ymax=1.02,
	width=0.75\textwidth,
	height=0.5\textwidth,
	ytick={0,0.1,...,1.1}, ytick align=inside, ytick pos=left,
	xtick={0,10,...,100}, xtick align=inside, xtick pos=left,
	xlabel=Training(\%),
	ylabel=Performance,
	grid=major,
	grid style={dashed, gray!30},
	legend pos=south east,
	legend style={draw=none}]
	\addplot+[
	blue, mark options={blue, scale=1},
	smooth, 
	error bars/.cd, 
	y fixed,
	y dir=both, 
	y explicit
	] table [x=x, y=y,y error=error, col sep=comma] {
		x,  y,        error
		10, 0.788203, 0.00976359
		20, 0.849317, 0.0038057
		30, 0.871109, 0.00689812
		40, 0.88607 , 0.00892426
		50, 0.892907, 0.00659157
		60, 0.897624, 0.00508019
		70, 0.901397, 0.00514787
		80, 0.905616, 0.00821445
		90, 0.905572, 0.00306361
	};
	\addlegendentry{Accuracy}
	\addplot+[
	orange, mark options={orange, scale=1},
	smooth, 
	error bars/.cd, 
	y fixed,
	y dir=both, 
	y explicit
	] table [x=x, y=y,y error=error, col sep=comma] {
		x,  y,        error
		10, 0.576453, 0.0195109
		20, 0.699651, 0.00653435
		30, 0.74486 , 0.012488
		40, 0.774889, 0.00171319
		50, 0.790484, 0.011507
		60, 0.801057, 0.00926014
		70, 0.81024 , 0.0100118
		80, 0.818981, 0.0149968
		90, 0.820222, 0.00504527
	};
	\addlegendentry{$\phi$}
	\end{axis}
	\end{tikzpicture}
	\caption{Training and validating three classes: \textbf{Q4}, \textbf{Q5}, and \textbf{Q6}. We generate (102+138+161) matrices. There are 11506 1's and 11645 0's. The method is NB. The rank information is included.}\label{threeclasskernelcurve}
\end{figure}

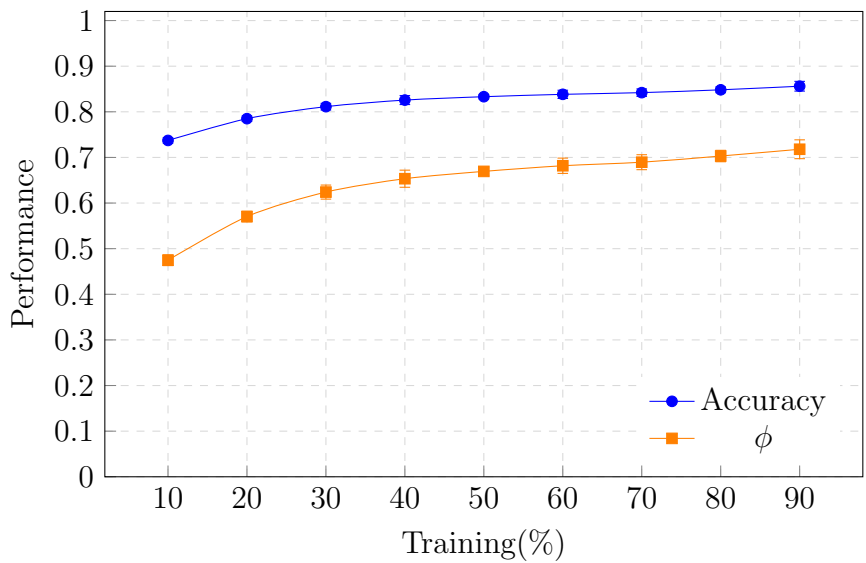
\begin{figure}[H]
	\centering
	\begin{tikzpicture}
	\begin{axis}[ymin=0, ymax=1.02,
	width=0.75\textwidth,
	height=0.5\textwidth,
	ytick={0,0.1,...,1.1}, ytick align=inside, ytick pos=left,
	xtick={0,10,...,100}, xtick align=inside, xtick pos=left,
	xlabel=Training(\%),
	ylabel=Performance,
	grid=major,
	grid style={dashed, gray!30},
	legend pos=south east,
	legend style={draw=none}]
	\addplot+[
	blue, mark options={blue, scale=1},
	smooth, 
	error bars/.cd, 
	y fixed,
	y dir=both, 
	y explicit
	] table [x=x, y=y,y error=error, col sep=comma] {
		x,  y,        error
		10, 0.737274, 0.00592113
		20, 0.785028, 0.00588793
		30, 0.811342, 0.00736965
		40, 0.825762, 0.00984072
		50, 0.83315 , 0.00490497
		60, 0.838339, 0.0090239
		70, 0.842071, 0.00824267
		80, 0.848255, 0.00625482
		90, 0.85614 , 0.0106915
	};
	\addlegendentry{Accuracy}
	\addplot+[
	orange, mark options={orange, scale=1},
	smooth, 
	error bars/.cd, 
	y fixed,
	y dir=both, 
	y explicit
	] table [x=x, y=y,y error=error, col sep=comma] {
		x,  y,        error
		10, 0.474707, 0.0117465
		20, 0.570422, 0.0116023
		30, 0.624074, 0.0154044
		40, 0.653396, 0.0187496
		50, 0.66946 , 0.00908065
		60, 0.68181 , 0.0167427
		70, 0.689579, 0.0163942
		80, 0.703112, 0.0125336
		90, 0.717965, 0.0206565
	};
	\addlegendentry{$\phi$}
	\end{axis}
	\end{tikzpicture}
	\caption{Training and validating four classes: \textbf{Q4}, \textbf{Q5}, \textbf{Q6}, and \textbf{Q7}. We generate (102+138+161+102) matrices. There are 13930 1's and 14005 0's. The method is NB. The rank information is included.}\label{fourclasskernelcurve}
\end{figure}

\begin{figure}[H]
	\centering
	\begin{tikzpicture}
	\begin{axis}[ymin=0, ymax=1.02,
	width=0.75\textwidth,
	height=0.5\textwidth,
	ytick={0,0.1,...,1.1}, ytick align=inside, ytick pos=left,
	xtick={0,10,...,100}, xtick align=inside, xtick pos=left,
	xlabel=Training(\%),
	ylabel=Performance,
	grid=major,
	grid style={dashed, gray!30},
	legend pos=south east,
	legend style={draw=none}]
	\addplot+[
	blue, mark options={blue, scale=1},
	smooth, 
	error bars/.cd, 
	y fixed,
	y dir=both, 
	y explicit
	] table [x=x, y=y,y error=error, col sep=comma] {
		x,  y,        error
		10, 0.728532, 0.0102497
		20, 0.770083, 0.00433987
		30, 0.795024, 0.00451021
		40, 0.80533 , 0.0101619
		50, 0.814879, 0.00655523
		60, 0.824719, 0.00298241
		70, 0.832563, 0.00345224
		80, 0.838886, 0.00428058
		90, 0.848826, 0.00900309
	};
	\addlegendentry{Accuracy}
	\addplot+[
	orange, mark options={orange, scale=1},
	smooth, 
	error bars/.cd, 
	y fixed,
	y dir=both, 
	y explicit
	] table [x=x, y=y,y error=error, col sep=comma] {
		x,  y,        error
		10, 0.457102, 0.0205287
		20, 0.540266, 0.00857363
		30, 0.590414, 0.00884585
		40, 0.611156, 0.0199922
		50, 0.630763, 0.0129997
		60, 0.650132, 0.00608921
		70, 0.666037, 0.00702776
		80, 0.678467, 0.00785069
		90, 0.698067, 0.017765
	};
	\addlegendentry{$\phi$}
	\end{axis}
	\end{tikzpicture}
	\caption{Training and validating five classes: \textbf{Q4}, \textbf{Q5}, \textbf{Q6}, \textbf{Q7}, and \textbf{Q8}. We generate (102+138+161+102+161) matrices. There are 22770 1's and 22823 0's. The method is NB. The rank information is included via imposing the null vectors.}\label{fiveclasskernelcurve}
\end{figure}
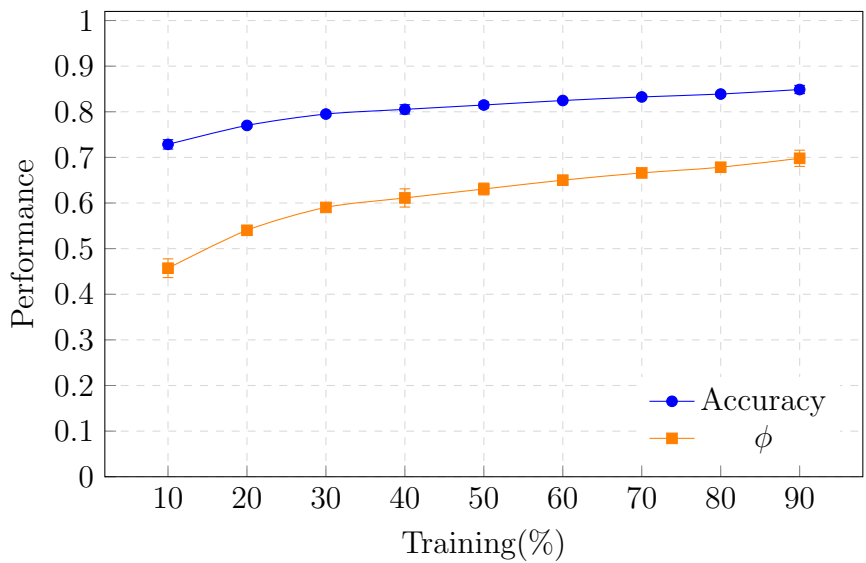

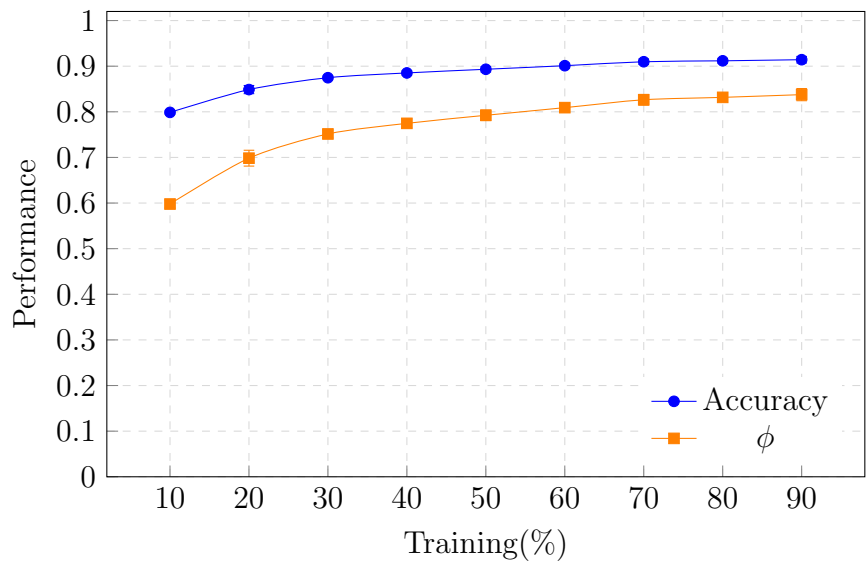
\begin{figure}[H]
	\centering
	\begin{tikzpicture}
	\begin{axis}[ymin=0, ymax=1.02,
	width=0.75\textwidth,
	height=0.5\textwidth,
	ytick={0,0.1,...,1.1}, ytick align=inside, ytick pos=left,
	xtick={0,10,...,100}, xtick align=inside, xtick pos=left,
	xlabel=Training(\%),
	ylabel=Performance,
	grid=major,
	grid style={dashed, gray!30},
	legend pos=south east,
	legend style={draw=none}]
	\addplot+[
	blue, mark options={blue, scale=1},
	smooth, 
	error bars/.cd, 
	y fixed,
	y dir=both, 
	y explicit
	] table [x=x, y=y,y error=error, col sep=comma] {
		x,  y,        error
		10, 0.79875 , 0.00432475
		20, 0.848793, 0.00894543
		30, 0.874771, 0.00423661
		40, 0.885258, 0.00489931
		50, 0.893295, 0.00579394
		60, 0.901112, 0.00524272
		70, 0.909794, 0.00599414
		80, 0.911813, 0.00527535
		90, 0.914298, 0.00719334
	};
	\addlegendentry{Accuracy}
	\addplot+[
	orange, mark options={orange, scale=1},
	smooth, 
	error bars/.cd, 
	y fixed,
	y dir=both, 
	y explicit
	] table [x=x, y=y,y error=error, col sep=comma] {
		x,  y,        error
		10, 0.597881, 0.00839482
		20, 0.698422, 0.0173697
		30, 0.751475, 0.00834473
		40, 0.774741, 0.00939326
		50, 0.792407, 0.0115808
		60, 0.809177, 0.00974662
		70, 0.826436, 0.0117047
		80, 0.831729, 0.00900419
		90, 0.837702, 0.0133518
	};
	\addlegendentry{$\phi$}
	\end{axis}
	\end{tikzpicture}
	\caption{Training and validating three classes: \textbf{Q4}, \textbf{Q5} and \textbf{Q6}. We generate (102+138+161) matrices. There are 11490 1's and 11449 0's. The method is NB. The Diophantine variables are included.}\label{threeclassdiophantinecurve}
\end{figure}

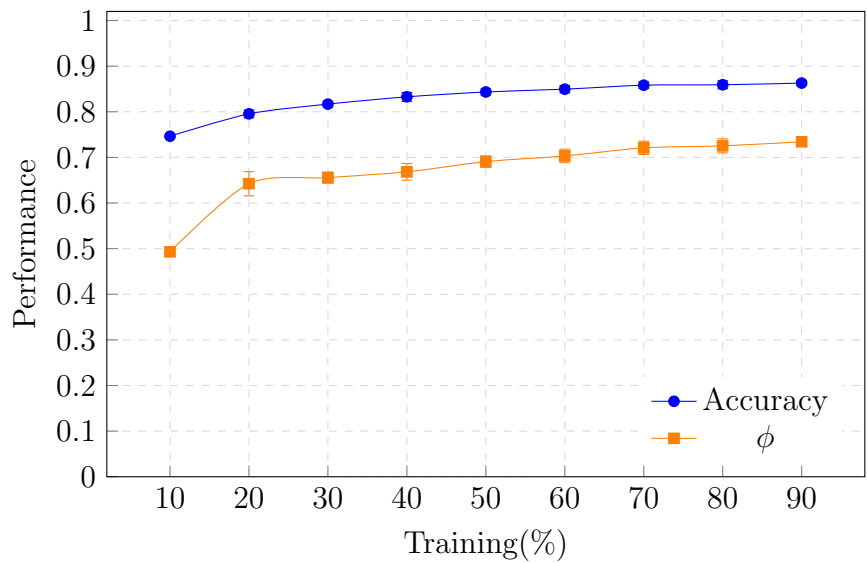
\begin{figure}[H]
	\centering
	\begin{tikzpicture}
	\begin{axis}[ymin=0, ymax=1.02,
	width=0.75\textwidth,
	height=0.5\textwidth,
	ytick={0,0.1,...,1.1}, ytick align=inside, ytick pos=left,
	xtick={0,10,...,100}, xtick align=inside, xtick pos=left,
	xlabel=Training(\%),
	ylabel=Performance,
	grid=major,
	grid style={dashed, gray!30},
	legend pos=south east,
	legend style={draw=none}]
	\addplot+[
	blue, mark options={blue, scale=1},
	smooth, 
	error bars/.cd, 
	y fixed,
	y dir=both, 
	y explicit
	] table [x=x, y=y,y error=error, col sep=comma] {
		x,  y,        error
		10, 0.746483, 0.00216391
		20, 0.795462, 0.00864077
		30, 0.816839, 0.0064732
		40, 0.832779, 0.00967043
		50, 0.843481, 0.00670754
		60, 0.849609, 0.00783483
		70, 0.858307, 0.00735097
		80, 0.859218, 0.0087428
		90, 0.862948, 0.00532504
	};
	\addlegendentry{Accuracy}
	\addplot+[
	orange, mark options={orange, scale=1},
	smooth, 
	error bars/.cd, 
	y fixed,
	y dir=both, 
	y explicit
	] table [x=x, y=y,y error=error, col sep=comma] {
		x,  y,        error
		10, 0.493014, 0.00431286
		20, 0.642437, 0.0265292
		30, 0.655636, 0.0121418
		40, 0.668381, 0.0182541
		50, 0.690738, 0.0124602
		60, 0.703203, 0.0149231
		70, 0.72096 , 0.0146171
		80, 0.72533 , 0.015437
		90, 0.734288, 0.0102563
	};
	\addlegendentry{$\phi$}
	\end{axis}
	\end{tikzpicture}
	\caption{Training and validating four classes: \textbf{Q4}, \textbf{Q5}, \textbf{Q6}, and \textbf{Q7}. We generate (102+138+161+102) matrices. There are 14099 1's and 14118 0's. The method is NB. The Diophantine variables are included.}\label{fourclassdiophantinecurve}
\end{figure}

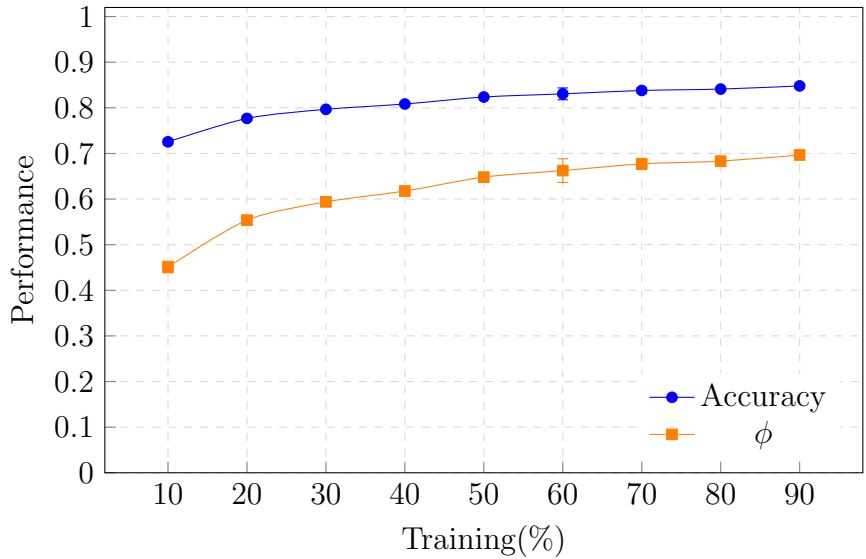
\begin{figure}[H]
	\centering
	\begin{tikzpicture}
	\begin{axis}[ymin=0, ymax=1.02,
	width=0.75\textwidth,
	height=0.5\textwidth,
	ytick={0,0.1,...,1.1}, ytick align=inside, ytick pos=left,
	xtick={0,10,...,100}, xtick align=inside, xtick pos=left,
	xlabel=Training(\%),
	ylabel=Performance,
	grid=major,
	grid style={dashed, gray!30},
	legend pos=south east,
	legend style={draw=none}]
	\addplot+[
	blue, mark options={blue, scale=1},
	smooth, 
	error bars/.cd, 
	y fixed,
	y dir=both, 
	y explicit
	] table [x=x, y=y,y error=error, col sep=comma] {
		x,  y,        error
		10, 0.725467, 0.00630823
		20, 0.776831, 0.00469516
		30, 0.796721, 0.00453842
		40, 0.808418, 0.00441331
		50, 0.823720, 0.0018998
		60, 0.830616, 0.012983
		70, 0.838043, 0.00618765
		80, 0.841075, 0.00197492
		90, 0.848055, 0.00472813
	};
	\addlegendentry{Accuracy}
	\addplot+[
	orange, mark options={orange, scale=1},
	smooth, 
	error bars/.cd, 
	y fixed,
	y dir=both, 
	y explicit
	] table [x=x, y=y,y error=error, col sep=comma] {
		x,  y,        error
		10, 0.450985, 0.0125998
		20, 0.553828, 0.00920577
		30, 0.593785, 0.00898658
		40, 0.617616, 0.00857504
		50, 0.648112, 0.00364185
		60, 0.662443, 0.0259115
		70, 0.677159, 0.0120218
		80, 0.683124, 0.00344858
		90, 0.696638, 0.00964367
	};
	\addlegendentry{$\phi$}
	\end{axis}
	\end{tikzpicture}
	\caption{Training and validating five classes: \textbf{Q4}, \textbf{Q5}, \textbf{Q6}, \textbf{Q7}, and \textbf{Q8}. We generate (102+138+161+102+161) matrices. There are 23273 1's and 24217 0's. The method is NB. The Diophantine variables are included.}\label{fiveclassdiophantinecurve}
\end{figure}

\addcontentsline{toc}{section}{References}
\bibliographystyle{JHEP}
\bibliography{references}

\providecommand{\href}[2]{#2}\begingroup\raggedright\begin{thebibliography}{10}

\bibitem{Seiberg:1994pq}
N.~Seiberg, \emph{{Electric - magnetic duality in supersymmetric Non-Abelian
  gauge theories}},
  \href{http://dx.doi.org/10.1016/0550-3213(94)00023-8}{\emph{Nucl. Phys.} {\bf
  B435} (1995) 129--146}, [\href{http://arxiv.org/abs/hep-th/9411149}{{\tt
  hep-th/9411149}}].

\bibitem{MR1887642}
S.~Fomin and A.~Zelevinsky, \emph{Cluster algebras. {I}. {F}oundations},
  \href{http://dx.doi.org/10.1090/S0894-0347-01-00385-X}{\emph{J. Amer. Math.
  Soc.} {\bf 15} (2002) 497--529}.

\bibitem{MR2004457}
S.~Fomin and A.~Zelevinsky, \emph{Cluster algebras. {II}. {F}inite type
  classification},
  \href{http://dx.doi.org/10.1007/s00222-003-0302-y}{\emph{Invent. Math.} {\bf
  154} (2003) 63--121}.

\bibitem{Feng:2000mi}
B.~Feng, A.~Hanany and Y.-H. He, \emph{{D-brane gauge theories from toric
  singularities and toric duality}},
  \href{http://dx.doi.org/10.1016/S0550-3213(00)00699-4}{\emph{Nucl. Phys. B}
  {\bf 595} (2001) 165--200}, [\href{http://arxiv.org/abs/hep-th/0003085}{{\tt
  hep-th/0003085}}].

\bibitem{Feng:2001bn}
B.~Feng, A.~Hanany, Y.-H. He and A.~M. Uranga, \emph{{Toric duality as Seiberg
  duality and brane diamonds}},
  \href{http://dx.doi.org/10.1088/1126-6708/2001/12/035}{\emph{JHEP} {\bf 12}
  (2001) 035}, [\href{http://arxiv.org/abs/hep-th/0109063}{{\tt
  hep-th/0109063}}].

\bibitem{Cachazo:2001sg}
F.~Cachazo, B.~Fiol, K.~A. Intriligator, S.~Katz and C.~Vafa, \emph{{A
  Geometric unification of dualities}},
  \href{http://dx.doi.org/10.1016/S0550-3213(02)00078-0}{\emph{Nucl. Phys. B}
  {\bf 628} (2002) 3--78}, [\href{http://arxiv.org/abs/hep-th/0110028}{{\tt
  hep-th/0110028}}].

\bibitem{Hanany:2005ve}
A.~Hanany and K.~D. Kennaway, \emph{{Dimer models and toric diagrams}},
  \href{http://arxiv.org/abs/hep-th/0503149}{{\tt hep-th/0503149}}.

\bibitem{Franco:2005rj}
S.~Franco, A.~Hanany, K.~D. Kennaway, D.~Vegh and B.~Wecht, \emph{{Brane dimers
  and quiver gauge theories}},
  \href{http://dx.doi.org/10.1088/1126-6708/2006/01/096}{\emph{JHEP} {\bf 01}
  (2006) 096}, [\href{http://arxiv.org/abs/hep-th/0504110}{{\tt
  hep-th/0504110}}].

\bibitem{Franco:2005sm}
S.~Franco, A.~Hanany, D.~Martelli, J.~Sparks, D.~Vegh and B.~Wecht,
  \emph{{Gauge theories from toric geometry and brane tilings}},
  \href{http://dx.doi.org/10.1088/1126-6708/2006/01/128}{\emph{JHEP} {\bf 01}
  (2006) 128}, [\href{http://arxiv.org/abs/hep-th/0505211}{{\tt
  hep-th/0505211}}].

\bibitem{Feng:2005gw}
B.~Feng, Y.-H. He, K.~D. Kennaway and C.~Vafa, \emph{{Dimer models from mirror
  symmetry and quivering amoebae}},
  \href{http://dx.doi.org/10.4310/ATMP.2008.v12.n3.a2}{\emph{Adv. Theor. Math.
  Phys.} {\bf 12} (2008) 489--545},
  [\href{http://arxiv.org/abs/hep-th/0511287}{{\tt hep-th/0511287}}].

\bibitem{Benvenuti:2006qr}
S.~Benvenuti, B.~Feng, A.~Hanany and Y.-H. He, \emph{{Counting BPS Operators in
  Gauge Theories: Quivers, Syzygies and Plethystics}},
  \href{http://dx.doi.org/10.1088/1126-6708/2007/11/050}{\emph{JHEP} {\bf 11}
  (2007) 050}, [\href{http://arxiv.org/abs/hep-th/0608050}{{\tt
  hep-th/0608050}}].

\bibitem{lauren}
S.~Fomin, L.~Williams and A.~Zelevinsky, \emph{{Introduction to Cluster
  Algebras}}.

\bibitem{Bourjaily:2016mnp}
J.~L. Bourjaily, S.~Franco, D.~Galloni and C.~Wen, \emph{{Stratifying On-Shell
  Cluster Varieties: the Geometry of Non-Planar On-Shell Diagrams}},
  \href{http://dx.doi.org/10.1007/JHEP10(2016)003}{\emph{JHEP} {\bf 10} (2016)
  003}, [\href{http://arxiv.org/abs/1607.01781}{{\tt 1607.01781}}].

\bibitem{Arkani-Hamed:2019plo}
N.~Arkani-Hamed, S.~He, T.~Lam and H.~Thomas, \emph{{Binary Geometries,
  Generalized Particles and Strings, and Cluster Algebras}},
  \href{http://arxiv.org/abs/1912.11764}{{\tt 1912.11764}}.

\bibitem{MR2567745}
V.~V. Fock and A.~B. Goncharov, \emph{Cluster ensembles, quantization and the
  dilogarithm}, \href{http://dx.doi.org/10.24033/asens.2112}{\emph{Ann. Sci.
  \'{E}c. Norm. Sup\'{e}r. (4)} {\bf 42} (2009) 865--930}.

\bibitem{Franco:2017lpa}
S.~Franco and G.~Musiker, \emph{{Higher Cluster Categories and QFT Dualities}},
  \href{http://dx.doi.org/10.1103/PhysRevD.98.046021}{\emph{Phys. Rev. D} {\bf
  98} (2018) 046021}, [\href{http://arxiv.org/abs/1711.01270}{{\tt
  1711.01270}}].

\bibitem{He:2017aed}
Y.-H. He, \emph{{Deep-Learning the Landscape}},
  \href{http://arxiv.org/abs/1706.02714}{{\tt 1706.02714}}.

\bibitem{He:2017set}
Y.-H. He, \emph{{Machine-learning the string landscape}},
  \href{http://dx.doi.org/10.1016/j.physletb.2017.10.024}{\emph{Phys. Lett. B}
  {\bf 774} (2017) 564--568}.

\bibitem{He:2018jtw}
Y.-H. He, \emph{{The Calabi-Yau Landscape: from Geometry, to Physics, to
  Machine-Learning}},  \href{http://arxiv.org/abs/1812.02893}{{\tt
  1812.02893}}.

\bibitem{He:2020lcy}
J.~Bao, Y.-H. He, E.~Hirst and S.~Pietromonaco, \emph{{Lectures on the
  Calabi-Yau Landscape}},  \href{http://arxiv.org/abs/2001.01212}{{\tt
  2001.01212}}.

\bibitem{Krefl:2017yox}
D.~Krefl and R.-K. Seong, \emph{{Machine Learning of Calabi-Yau Volumes}},
  \href{http://dx.doi.org/10.1103/PhysRevD.96.066014}{\emph{Phys. Rev.} {\bf
  D96} (2017) 066014}, [\href{http://arxiv.org/abs/1706.03346}{{\tt
  1706.03346}}].

\bibitem{Ruehle:2017mzq}
F.~Ruehle, \emph{{Evolving neural networks with genetic algorithms to study the
  String Landscape}},
  \href{http://dx.doi.org/10.1007/JHEP08(2017)038}{\emph{JHEP} {\bf 08} (2017)
  038}, [\href{http://arxiv.org/abs/1706.07024}{{\tt 1706.07024}}].

\bibitem{Carifio:2017bov}
J.~Carifio, J.~Halverson, D.~Krioukov and B.~D. Nelson, \emph{{Machine Learning
  in the String Landscape}},
  \href{http://dx.doi.org/10.1007/JHEP09(2017)157}{\emph{JHEP} {\bf 09} (2017)
  157}, [\href{http://arxiv.org/abs/1707.00655}{{\tt 1707.00655}}].

\bibitem{Betzler:2020rfg}
P.~Betzler and S.~Krippendorf, \emph{{Connecting Dualities and Machine
  Learning}}, \href{http://dx.doi.org/10.1002/prop.202000022}{\emph{Fortsch.
  Phys.} {\bf 68} (2, 2020) }, [\href{http://arxiv.org/abs/2002.05169}{{\tt
  2002.05169}}].

\bibitem{Krippendorf:2020gny}
S.~Krippendorf and M.~Syvaeri, \emph{{Detecting Symmetries with Neural
  Networks}},  \href{http://arxiv.org/abs/2003.13679}{{\tt 2003.13679}}.

\bibitem{Altman:2018zlc}
R.~Altman, J.~Carifio, J.~Halverson and B.~D. Nelson, \emph{{Estimating
  Calabi-Yau Hypersurface and Triangulation Counts with Equation Learners}},
  \href{http://dx.doi.org/10.1007/JHEP03(2019)186}{\emph{JHEP} {\bf 03} (2019)
  186}, [\href{http://arxiv.org/abs/1811.06490}{{\tt 1811.06490}}].

\bibitem{Demirtas:2018akl}
M.~Demirtas, C.~Long, L.~McAllister and M.~Stillman, \emph{{The Kreuzer-Skarke
  Axiverse}}, \href{http://dx.doi.org/10.1007/JHEP04(2020)138}{\emph{JHEP} {\bf
  04} (2020) 138}, [\href{http://arxiv.org/abs/1808.01282}{{\tt 1808.01282}}].

\bibitem{He:2015fif}
Y.-H. He, V.~Jejjala and L.~Pontiggia, \emph{{Patterns in Calabi--Yau
  Distributions}},
  \href{http://dx.doi.org/10.1007/s00220-017-2907-9}{\emph{Commun. Math. Phys.}
  {\bf 354} (2017) 477--524}, [\href{http://arxiv.org/abs/1512.01579}{{\tt
  1512.01579}}].

\bibitem{Cole:2019enn}
A.~Cole, A.~Schachner and G.~Shiu, \emph{{Searching the Landscape of Flux Vacua
  with Genetic Algorithms}},
  \href{http://dx.doi.org/10.1007/JHEP11(2019)045}{\emph{JHEP} {\bf 11} (2019)
  045}, [\href{http://arxiv.org/abs/1907.10072}{{\tt 1907.10072}}].

\bibitem{Hashimoto:2018ftp}
K.~Hashimoto, S.~Sugishita, A.~Tanaka and A.~Tomiya, \emph{{Deep learning and
  the AdS/CFT correspondence}},
  \href{http://dx.doi.org/10.1103/PhysRevD.98.046019}{\emph{Phys. Rev. D} {\bf
  98} (2018) 046019}, [\href{http://arxiv.org/abs/1802.08313}{{\tt
  1802.08313}}].

\bibitem{Anderson:2017aux}
L.~B. Anderson, X.~Gao, J.~Gray and S.-J. Lee, \emph{{Fibrations in CICY
  Threefolds}}, \href{http://dx.doi.org/10.1007/JHEP10(2017)077}{\emph{JHEP}
  {\bf 10} (2017) 077}, [\href{http://arxiv.org/abs/1708.07907}{{\tt
  1708.07907}}].

\bibitem{He:2019vsj}
Y.-H. He and S.-J. Lee, \emph{{Distinguishing elliptic fibrations with AI}},
  \href{http://dx.doi.org/10.1016/j.physletb.2019.134889}{\emph{Phys. Lett. B}
  {\bf 798} (2019) 134889}, [\href{http://arxiv.org/abs/1904.08530}{{\tt
  1904.08530}}].

\bibitem{Grimm:2019bey}
T.~W. Grimm, F.~Ruehle and D.~van~de Heisteeg, \emph{{Classifying Calabi-Yau
  threefolds using infinite distance limits}},
  \href{http://arxiv.org/abs/1910.02963}{{\tt 1910.02963}}.

\bibitem{Brodie:2019dfx}
C.~R. Brodie, A.~Constantin, R.~Deen and A.~Lukas, \emph{{Machine Learning Line
  Bundle Cohomology}},
  \href{http://dx.doi.org/10.1002/prop.201900087}{\emph{Fortsch. Phys.} {\bf
  68} (2020) 1900087}, [\href{http://arxiv.org/abs/1906.08730}{{\tt
  1906.08730}}].

\bibitem{Jejjala:2019kio}
V.~Jejjala, A.~Kar and O.~Parrikar, \emph{{Deep Learning the Hyperbolic Volume
  of a Knot}},
  \href{http://dx.doi.org/10.1016/j.physletb.2019.135033}{\emph{Phys. Lett. B}
  {\bf 799} (2019) 135033}, [\href{http://arxiv.org/abs/1902.05547}{{\tt
  1902.05547}}].

\bibitem{Mutter:2018sra}
A.~M\"utter, E.~Parr and P.~K. Vaudrevange, \emph{{Deep learning in the
  heterotic orbifold landscape}},
  \href{http://dx.doi.org/10.1016/j.nuclphysb.2019.01.013}{\emph{Nucl. Phys. B}
  {\bf 940} (2019) 113--129}, [\href{http://arxiv.org/abs/1811.05993}{{\tt
  1811.05993}}].

\bibitem{Deen:2020dlf}
R.~Deen, Y.-H. He, S.-J. Lee and A.~Lukas, \emph{{Machine Learning String
  Standard Models}},  \href{http://arxiv.org/abs/2003.13339}{{\tt 2003.13339}}.

\bibitem{Gal:2020dyc}
Y.~Gal, V.~Jejjala, D.~K. Mayorga~Pena and C.~Mishra, \emph{{Baryons from
  Mesons: A Machine Learning Perspective}},
  \href{http://arxiv.org/abs/2003.10445}{{\tt 2003.10445}}.

\bibitem{Ashmore:2019wzb}
A.~Ashmore, Y.-H. He and B.~A. Ovrut, \emph{{Machine learning Calabi-Yau
  metrics}},  \href{http://arxiv.org/abs/1910.08605}{{\tt 1910.08605}}.

\bibitem{He:2019nzx}
Y.-H. He and M.~Kim, \emph{{Learning Algebraic Structures: Preliminary
  Investigations}},  \href{http://arxiv.org/abs/1905.02263}{{\tt 1905.02263}}.

\bibitem{He:2020eva}
Y.-H. He, E.~Hirst and T.~Peterken, \emph{{Machine-Learning Dessins d'Enfants:
  Explorations via Modular and Seiberg-Witten Curves}},
  \href{http://arxiv.org/abs/2004.05218}{{\tt 2004.05218}}.

\bibitem{Alessandretti:2019jbs}
L.~Alessandretti, A.~Baronchelli and Y.-H. He, \emph{{Machine Learning meets
  Number Theory: The Data Science of Birch-Swinnerton-Dyer}},
  \href{http://arxiv.org/abs/1911.02008}{{\tt 1911.02008}}.

\bibitem{sagemath}
{The Sage Developers}, \emph{{S}ageMath, the {S}age {M}athematics {S}oftware
  {S}ystem ({V}ersion 9.0)}, 2019.

\bibitem{2011arXiv1102.4844M}
G.~{Musiker} and C.~{Stump}, \emph{{A compendium on the cluster algebra and
  quiver package in sage}},  \href{http://arxiv.org/abs/1102.4844}{{\tt
  1102.4844}}.

\bibitem{Kutasov:1995np}
D.~Kutasov and A.~Schwimmer, \emph{{On duality in supersymmetric Yang-Mills
  theory}}, \href{http://dx.doi.org/10.1016/0370-2693(95)00676-C}{\emph{Phys.
  Lett.} {\bf B354} (1995) 315--321},
  [\href{http://arxiv.org/abs/hep-th/9505004}{{\tt hep-th/9505004}}].

\bibitem{Kutasov:1995ve}
D.~Kutasov, \emph{{A Comment on duality in N=1 supersymmetric nonAbelian gauge
  theories}}, \href{http://dx.doi.org/10.1016/0370-2693(95)00392-X}{\emph{Phys.
  Lett.} {\bf B351} (1995) 230--234},
  [\href{http://arxiv.org/abs/hep-th/9503086}{{\tt hep-th/9503086}}].

\bibitem{Kapustin:1996nb}
A.~Kapustin, \emph{{The Coulomb branch of N=1 supersymmetric gauge theory with
  adjoint and fundamental matter}},
  \href{http://dx.doi.org/10.1016/S0370-2693(97)00209-8}{\emph{Phys. Lett.}
  {\bf B398} (1997) 104--109}, [\href{http://arxiv.org/abs/hep-th/9611049}{{\tt
  hep-th/9611049}}].

\bibitem{BFZ3}
A.~Berenstein, S.~Fomin and A.~Zelevinsky, \emph{Cluster algebras. {III}.
  {U}pper bounds and double {B}ruhat cells},
  \href{http://dx.doi.org/10.1215/S0012-7094-04-12611-9}{\emph{Duke Math. J.}
  {\bf 126} (2005) 1--52}.

\bibitem{SherZel}
P.~Sherman and A.~Zelevinsky, \emph{Positivity and canonical bases in rank 2
  cluster algebras of finite and affine types},
  \href{http://dx.doi.org/10.17323/1609-4514-2004-4-4-947-974}{\emph{Mosc.
  Math. J.} {\bf 4} (2004) 947--974, 982}.

\bibitem{Gabriel}
P.~Gabriel, \emph{Unzerlegbare {D}arstellungen. {I}},
  \href{http://dx.doi.org/10.1007/BF01298413}{\emph{Manuscripta Math.} {\bf 6}
  (1972) 71--103; correction, ibid. 6 (1972), 309}.

\bibitem{2008arXiv0811.1703F}
A.~{Felikson}, M.~{Shapiro} and P.~{Tumarkin}, \emph{{Skew-symmetric cluster
  algebras of finite mutation type}}, {\emph{arXiv e-prints} (Nov, 2008)
  arXiv:0811.1703}, [\href{http://arxiv.org/abs/0811.1703}{{\tt 0811.1703}}].

\bibitem{DerkOwen}
H.~Derksen and T.~Owen, \emph{New graphs of finite mutation type},
  {\emph{Electron. J. Combin.} {\bf 15} (2008) Research Paper 139, 15}.

\bibitem{Complete}
M.~Alim, S.~Cecotti, C.~C\'{o}rdova, S.~Espahbodi, A.~Rastogi and C.~Vafa,
  \emph{B{PS} quivers and spectra of complete {${N}=2$} quantum field
  theories}, \href{http://dx.doi.org/10.1007/s00220-013-1789-8}{\emph{Comm.
  Math. Phys.} {\bf 323} (2013) 1185--1227}.

\bibitem{2006math......8367F}
S.~{Fomin}, M.~{Shapiro} and D.~{Thurston}, \emph{{Cluster algebras and
  triangulated surfaces. Part I: Cluster complexes}}, {\emph{arXiv Mathematics
  e-prints} (Aug, 2006) math/0608367},
  [\href{http://arxiv.org/abs/math/0608367}{{\tt math/0608367}}].

\bibitem{Kac}
V.~G. Kac, \emph{Infinite root systems, representations of graphs and invariant
  theory}, \href{http://dx.doi.org/10.1007/BF01403155}{\emph{Invent. Math.}
  {\bf 56} (1980) 57--92}.

\bibitem{Feng:2002kk}
B.~Feng, A.~Hanany, Y.-H. He and A.~Iqbal, \emph{{Quiver theories, soliton
  spectra and Picard-Lefschetz transformations}},
  \href{http://dx.doi.org/10.1088/1126-6708/2003/02/056}{\emph{JHEP} {\bf 02}
  (2003) 056}, [\href{http://arxiv.org/abs/hep-th/0206152}{{\tt
  hep-th/0206152}}].

\bibitem{Franco:2003ja}
S.~Franco, A.~Hanany, Y.-H. He and P.~Kazakopoulos, \emph{{Duality walls,
  duality trees and fractional branes}},
  \href{http://arxiv.org/abs/hep-th/0306092}{{\tt hep-th/0306092}}.

\bibitem{tensorflow2015-whitepaper}
M.~Abadi et~al., \emph{{TensorFlow}: Large-scale machine learning on
  heterogeneous systems},  2015.

\bibitem{Benvenuti:2004dw}
S.~Benvenuti and A.~Hanany, \emph{{New results on superconformal quivers}},
  \href{http://dx.doi.org/10.1088/1126-6708/2006/04/032}{\emph{JHEP} {\bf 04}
  (2006) 032}, [\href{http://arxiv.org/abs/hep-th/0411262}{{\tt
  hep-th/0411262}}].

\bibitem{Hanany:2012mb}
A.~Hanany, Y.-H. He, C.~Sun and S.~Sypsas, \emph{{Superconformal Block Quivers,
  Duality Trees and Diophantine Equations}},
  \href{http://dx.doi.org/10.1007/JHEP11(2013)017}{\emph{JHEP} {\bf 11} (2013)
  017}, [\href{http://arxiv.org/abs/1211.6111}{{\tt 1211.6111}}].

\bibitem{Franco:2002mu}
S.~Franco and A.~Hanany, \emph{{Toric duality, Seiberg duality and
  Picard-Lefschetz transformations}},
  \href{http://dx.doi.org/10.1002/prop.200310091}{\emph{Fortsch. Phys.} {\bf
  51} (2003) 738--744}, [\href{http://arxiv.org/abs/hep-th/0212299}{{\tt
  hep-th/0212299}}].

\bibitem{Franco:2020ijt}
S.~Franco, A.~Hasan and X.~Yu, \emph{{On the Classification of Duality Webs for
  Graded Quivers}},  \href{http://arxiv.org/abs/2001.08776}{{\tt 2001.08776}}.

\bibitem{Gadde:2013lxa}
A.~Gadde, S.~Gukov and P.~Putrov, \emph{{(0, 2) trialities}},
  \href{http://dx.doi.org/10.1007/JHEP03(2014)076}{\emph{JHEP} {\bf 03} (2014)
  076}, [\href{http://arxiv.org/abs/1310.0818}{{\tt 1310.0818}}].

\bibitem{Franco:2016nwv}
S.~Franco, S.~Lee and R.-K. Seong, \emph{{Brane brick models and 2d (0, 2)
  triality}}, \href{http://dx.doi.org/10.1007/JHEP05(2016)020}{\emph{JHEP} {\bf
  05} (2016) 020}, [\href{http://arxiv.org/abs/1602.01834}{{\tt 1602.01834}}].

\bibitem{Franco:2016tcm}
S.~Franco, S.~Lee, R.-K. Seong and C.~Vafa, \emph{{Quadrality for
  Supersymmetric Matrix Models}},
  \href{http://dx.doi.org/10.1007/JHEP07(2017)053}{\emph{JHEP} {\bf 07} (2017)
  053}, [\href{http://arxiv.org/abs/1612.06859}{{\tt 1612.06859}}].

\bibitem{Smith}
J.~H. Smith, \emph{{Some properties of the spectrum of a graph}},  in
  \emph{Combinatorial Structures and their Applications (Proc. Calgary
  Internat. Conf., Calgary, Alta., 1969)}, pp.~403--406, New York: Gordon and
  Breach, 1970.

\bibitem{Mathematica}
W.~R. Inc., ``Mathematica, {V}ersion 12.0.''

\bibitem{NaiveBayes}
H.~Zhang, \emph{The {O}ptimality of {N}aive {B}ayes},  in \emph{Proceedings of
  the Seventeenth International Florida Artificial Intelligence Research
  Society Conference, FLAIRS 2004}, vol.~2, 01, 2004.

\bibitem{DBLP:journals/corr/OSheaN15}
K.~O'Shea and R.~Nash, \emph{An introduction to convolutional neural networks},
  {\emph{CoRR} (2015) }, [\href{http://arxiv.org/abs/1511.08458}{{\tt
  1511.08458}}].

\bibitem{2014arXiv1412.6980K}
D.~P. {Kingma} and J.~{Ba}, \emph{{Adam: A Method for Stochastic
  Optimization}}, {\emph{arXiv e-prints} (Dec, 2014) arXiv:1412.6980},
  [\href{http://arxiv.org/abs/1412.6980}{{\tt 1412.6980}}].

\bibitem{Kohavi95astudy}
R.~Kohavi, \emph{A study of cross-validation and bootstrap for accuracy
  estimation and model selection},  in \emph{IJCAI’95: Proceedings of the
  14th International Joint Conference on Artificial Intelligence - Volume 2},
  pp.~1137--1143, Morgan Kaufmann, 1995.

\bibitem{Chicco2020TheAO}
D.~Chicco and G.~Jurman, \emph{The advantages of the matthews correlation
  coefficient (mcc) over f1 score and accuracy in binary classification
  evaluation},  in \emph{BMC Genomics}, vol.~21, 2020.

\end{thebibliography}\endgroup

\end{document}